\documentclass[twocolumn,tighten]{aastex62}

\usepackage{graphicx}
\usepackage{bm}
\usepackage{xspace}
\usepackage{ulem}

\usepackage{amsmath}
\usepackage{subfigure}

\usepackage{afterpage}

\renewcommand{\figurename}{Fig.}
\renewcommand{\tablename}{Table}

\newcommand{\maxi}{{MAXI}\xspace}
\newcommand{\swift}{\textit{Swift}\xspace}
\newcommand{\uhuru}{\textit{Uhuru}\xspace}
\newcommand{\rxte}{\textit{RXTE}\xspace}
\newcommand{\xmm}{\textit{XMM-Newton}\xspace}
\newcommand{\rosat}{\textit{ROSAT}\xspace}

\newcommand{\ergcms}{erg cm$^{-2}$ s$^{-1}$}
\newcommand{\lognlogs}{$\log N$--$\log S$\xspace}

\begin{document}
\title{
The 7-year \maxi/GSC X-ray Source Catalog in the High Galactic-Latitude Sky (3MAXI)
}

\author{T. Kawamuro}
\altaffiliation{JSPS fellow (PD)}
\affil{National Astronomical Observatory of Japan, Osawa, Mitaka, Tokyo 181-8588, Japan}
\author{Y. Ueda}
\affil{Department of Astronomy, Kyoto University, Kitashirakawa-Oiwake-cho, Sakyo-ku, Kyoto 606-8502, Japan}
\author{M. Shidatsu}
\affil{{\it MAXI} team, RIKEN, 2-1, Hirosawa, Wako-shi, Saitama 351-0198, Japan}
\author{T. Hori}
\affil{Department of Astronomy, Kyoto University, Kitashirakawa-Oiwake-cho, Sakyo-ku, Kyoto 606-8502, Japan}
\author{M. Morii}
\affil{The Institute of Statistical Mathematics, 10-3 Midori-cho, Tachikawa, Tokyo 190-8562, Japan}
\author{S. Nakahira}
\affil{{\it MAXI} team, RIKEN, 2-1, Hirosawa, Wako-shi, Saitama 351-0198, Japan}
\author{N. Isobe}
\affil{Institute of Space and Astronautical Science, Japan Aerospace Exploration Agency, 3-1-1 Yoshinodai, Chuo, Sagamihara, Kanagawa 252-5210, Japan}
\author{N. Kawai}
\affil{Department of Physics, Tokyo Institute of Technology, 2-12-1 Ookayama, Meguro-ku, Tokyo 152-8551, Japan}
\author{T. Mihara}
\affil{{\it MAXI} team, RIKEN, 2-1, Hirosawa, Wako-shi, Saitama 351-0198, Japan}
\author{M. Matsuoka}
\affil{{\it MAXI} team, RIKEN, 2-1, Hirosawa, Wako-shi, Saitama 351-0198, Japan}
\author{T. Morita}
\affil{Department of Astronomy, Kyoto University, Kitashirakawa-Oiwake-cho, Sakyo-ku, Kyoto 606-8502, Japan}
\author{M. Nakajima}
\affil{School of Dentistry at Matsudo, Nihon University, 2-870-1, Sakaecho-nishi, Matsudo, 271-8587, Chiba, Japan}
\author{H. Negoro}
\affil{Department of Physics, Nihon University, 1-8 Kanda-Surugadai, Chiyoda-ku, Tokyo 101-8308, Japan}
\author{S. Oda}
\affil{Department of Astronomy, Kyoto University, Kitashirakawa-Oiwake-cho, Sakyo-ku, Kyoto 606-8502, Japan}
\author{T. Sakamoto}
\affil{College of Science and Engineering, Department of Physics and Mathematics, Aoyama Gakuin University, 5-10-1 Fuchinobe, Chuo-ku, Sagamihara, Kanagawa 252-5258, Japan}
\author{M. Serino}
\affil{College of Science and Engineering, Department of Physics and Mathematics, Aoyama Gakuin University, 5-10-1 Fuchinobe, Chuo-ku, Sagamihara, Kanagawa 252-5258, Japan}
\author{M. Sugizaki}
\affil{{\it MAXI} team, RIKEN, 2-1, Hirosawa, Wako-shi, Saitama 351-0198, Japan}
\author{A. Tanimoto}
\affil{Department of Astronomy, Kyoto University, Kitashirakawa-Oiwake-cho, Sakyo-ku, Kyoto 606-8502, Japan}
\author{H. Tomida}
\affil{Institute of Space and Astronautical Science, Japan Aerospace Exploration Agency, 3-1-1 Yoshinodai, Chuo, Sagamihara, Kanagawa 252-5210, Japan}
\author{Y. Tsuboi}
\affil{Department of Physics, Faculty of Science \& Engineering, Chuo University, 1-13-27 Kasuga, Bunkyo, Tokyo 112-8551, Japan}
\author{H. Tsunemi}
\affil{Department of Earth and Space Science, Graduate School of Science, Osaka University, 1-1 Machikaneyama-cho, Toyonaka, Osaka 560-0043, Japan}
\author{S. Ueno}
\affil{Institute of Space and Astronautical Science, Japan Aerospace Exploration Agency, 3-1-1 Yoshinodai, Chuo, Sagamihara, Kanagawa 252-5210, Japan}
\author{K. Yamaoka}
\affil{Institute for Space-Earth Environmental Research (ISEE), Nagoya University, Furo-cho, Chikusa-ku, Nagoya, Aichi 464-8601, Japan}
\author{S. Yamada}
\affil{Department of Astronomy, Kyoto University, Kitashirakawa-Oiwake-cho, Sakyo-ku, Kyoto 606-8502, Japan}
\author{A. Yoshida}
\affil{College of Science and Engineering, Department of Physics and Mathematics, Aoyama Gakuin University, 5-10-1 Fuchinobe, Chuo-ku, Sagamihara, Kanagawa 252-5258, Japan}
\author{W. Iwakiri}
\affil{{\it MAXI} team, RIKEN, 2-1, Hirosawa, Wako-shi, Saitama 351-0198, Japan}
\author{Y. Kawakubo}
\affil{College of Science and Engineering, Department of Physics and Mathematics, Aoyama Gakuin University, 5-10-1 Fuchinobe, Chuo-ku, Sagamihara, Kanagawa 252-5258, Japan}
\author{Y. Sugawara}
\affil{Institute of Space and Astronautical Science, Japan Aerospace Exploration Agency, 3-1-1 Yoshinodai, Chuo, Sagamihara, Kanagawa 252-5210, Japan}
\author{S. Sugita}
\affil{Department of Physics, Tokyo Institute of Technology, 2-12-1 Ookayama, Meguro-ku, Tokyo 152-8551, Japan}
\author{Y. Tachibana}
\affil{Department of Physics, Tokyo Institute of Technology, 2-12-1 Ookayama, Meguro-ku, Tokyo 152-8551, Japan}
\author{T. Yoshii}
\affil{Department of Physics, Tokyo Institute of Technology, 2-12-1 Ookayama, Meguro-ku, Tokyo 152-8551, Japan}

\correspondingauthor{Taiki Kawamuro}
\email{taiki.kawamuro@nao.ac.jp}


\begin{abstract} 

  We present the third \maxi/GSC catalog in the high Galactic-latitude sky ($|b| > 10^\circ$) based 
  on the 7-year data from 2009 August 13 to 2016 July 31, complementary to that in the low Galactic-latitude 
  sky \citep[$|b| < 10^\circ$; ][]{Hor18}. We compile 682 sources  detected at significances of 
  $s_{\rm D,4-10~keV} \geq 6.5$ in the 4--10 keV band. A two-dimensional image fit based on the Poisson
  likelihood algorithm ($C$-statistics) is adopted for the detections and constraints on their fluxes
  and positions. The 4--10 keV sensitivity reaches $\approx 0.48$ mCrab, or $\approx 5.9 \times 10^{-12}$
  \ergcms, over the half of the survey area. 
Compared with the 37-month catalog \citep{Hir13}, 
which adopted a threshold of $s_{\rm D,4-10~keV} \geq 7$, 
the source number increases by a factor of $\sim$1.4. 
The fluxes in the 3--4 keV and 10--20 keV bands are further estimated, and 
  hardness ratios (HRs) are calculated using the 3--4 keV, 4--10 keV, 3--10 
 keV, and 10--20 keV band fluxes.
  We also make the 4--10 keV lightcurves in one year bins
for all the sources and characterize their variabilities with an index based on a likelihood function and the excess
 variance. Possible counterparts are found from five major X-ray survey catalogs by \swift, \uhuru, \rxte, \xmm,
 and \rosat, and an X-ray galaxy-cluster catalog (MCXC). Our catalog provides the fluxes, positions, detection significances,
HRs, one-year bin lightcurves, variability indices, and counterpart candidates.
  
\end{abstract}

\section{INTRODUCTION}\label{intro}

The Monitor of All-sky X-ray Image \cite[MAXI; ][]{Mat09} onboard the International
Space Station (ISS) has been successfully monitoring the X-ray sky since its launch
in 2009 August. The Gas Slit Camera (GSC) on the MAXI \citep{Mih11}, covering the 2--30 keV band, 
surveys the sky with two instantaneous fields-of-view of 3$^\circ$.0$\times$160$^\circ$
separated by 84 degrees. Rotating its fields-of-view with a period of 92
minutes according to the orbital motion of the ISS, the GSC eventually
covers a large fraction of the sky (95\%) in one day \citep{Sug11}.

As an all sky X-ray survey mission, the \maxi/GSC has achieved the best sensitivity
in the 4--10 keV band, which is not covered by the 
{\it ROSAT} all sky survey \citep[$<$ 2 keV; ][]{Tru82} and hard X-ray
($>$ 10 keV) surveys performed with the {\it Swift}/BAT \citep{Geh04} and
{\it INTEGRAL}/IBIS \citep{Win03}. The 4--10 keV X-rays have good
penetrating power up to a column density of $\log N_{\rm H}/{\rm
cm}^{-2} \sim$ 23, providing samples less biased against the absorption
than the {\it ROSAT} catalogs. Also, this energy band has an advantage
in detecting sources with ``soft'' spectra above 4 keV (e.g., a steep
power law), compared with hard X-rays above 10 keV. 
Thus, the \maxi/GSC survey is complementary to other 
X-ray surveys conducted in different energy bands. 

A variety of X-ray sources have been detected by accumulating the
\maxi/GSC data on different time scales. The \maxi Alert system
automatically informs us of variabilities within a few days
\citep{Neg12,Neg16}. As a result, number of new X-ray transients
have been discovered \citep[e.g., ][]{Mor13,Ser15,Shi17}, 
and recurrent activities of known sources have been captured \citep[e.g.,][]{Iso15}.
Most of them are Galactic objects. Also, systematic surveys focusing
on a few month variability provided unique samples 
including objects undetected in shorter or longer time-scale surveys 
\cite[e.g., tidal disruption events; ][]{Kaw16}. 
To extract the best sensitivities for most of sources, 
high and low Galactic latitude sky catalogs
were produced by integrating all the \maxi/GSC
data available then (7 months to 7 years; \citealt{Hir11,Hir13,Hor18}).
Particularly, the high latitude catalogs have been 
used as a basis to study extragalactic objects,
mainly active galactic nuclei \citep[AGNs; e.g.,
][]{Ued11,Ued14,Ter15,Ino16,Iso16,Abe17}.

In this paper, we present the third \maxi/GSC X-ray source catalog in
the high Galactic-latitude sky ($|b| >10^\circ$) constructed from the
first 7-year data. This is an extension of the 7-month and 37-month
catalogs \citep{Hir11,Hir13}, and is complementary to the 7-year low
Galactic-latitude catalog \citep{Hor18}. The catalog consists of sources
detected in the 4--10 keV data, to be consistent with the previous works. 
We also estimate the fluxes in the 3--4 keV and 10--20 keV bands of
each source, and calculate the hardness ratios (HRs) among the four 
bands (3--4 keV, 4--10 keV, 3--10 keV, and 10--20 keV) to obtain 
spectral information. The 10--20 keV band flux is new information
that was unavailable in the previous high Galactic-latitude catalogs.
Also, by analyzing time-sliced images, we make the 4--10 keV lightcurves
in one-year bins for all the detected sources. Transiently brightened
sources undetected in the 7-year image are found in these time-sliced
data, which are summarized as a transient event catalog in Appendix.

This paper is organized as follows. Section~\ref{sec:img_ana} describes
analysis procedures from data reduction to source detection, and provides the source catalog. 
In Section~\ref{sec:time_ana} we make the one-year bin lightcurves and search for transient events. 
Cross-matching with other X-ray catalogs is performed in Section~\ref{sec:cross-match}. Then, we 
discuss the statistical properties of the cataloged sources in Section~\ref{sec:result} 
based on the fluxes, HRs, and the time variability index. 
The $\log N$--$\log S$ relation is also shown in Section~\ref{sec:lognlogs}. 
We summarize our work in Section~\ref{sec:sum}.

\begin{figure*}
\includegraphics[scale=0.23,angle=-90]{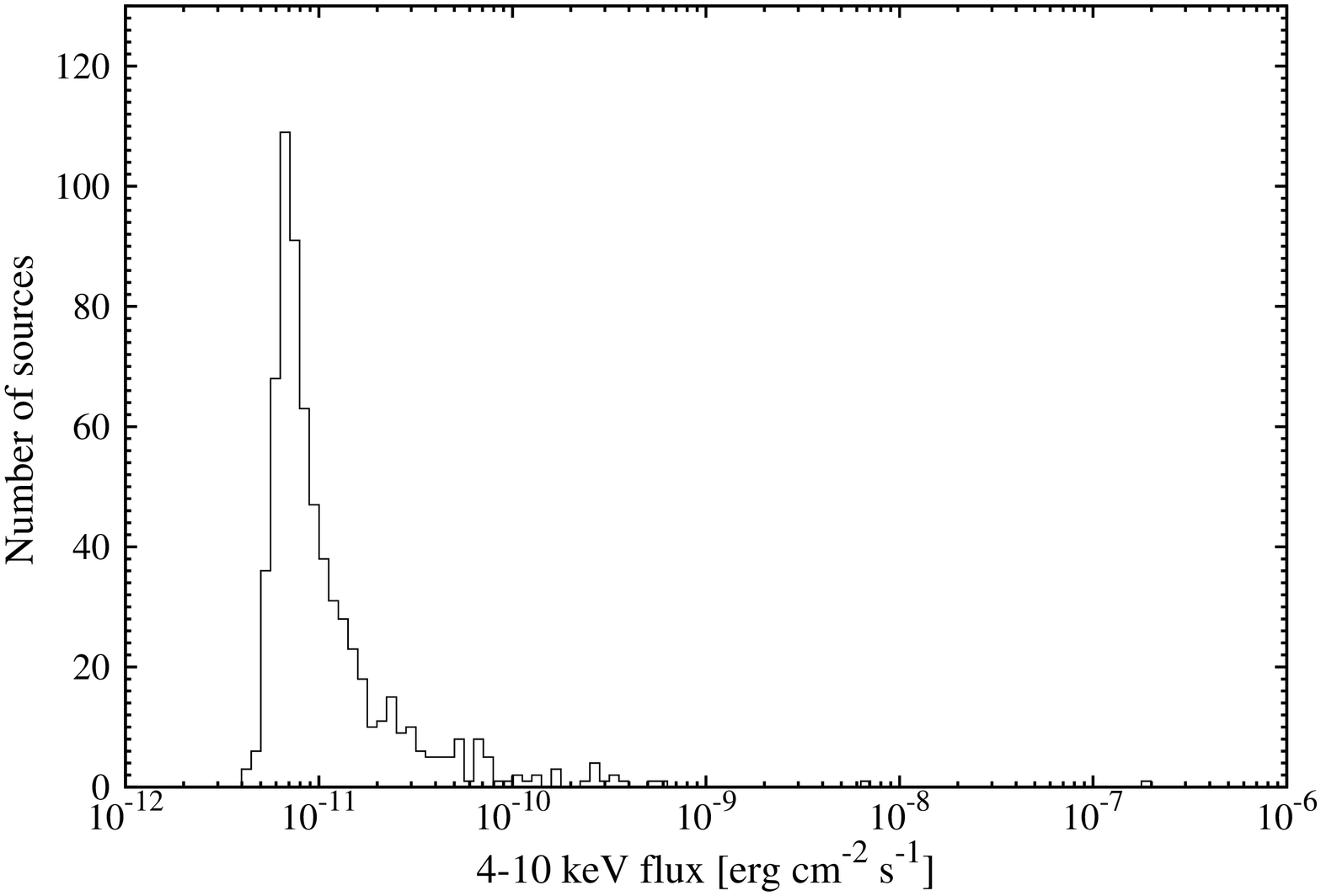} 
\includegraphics[scale=0.23,angle=-90]{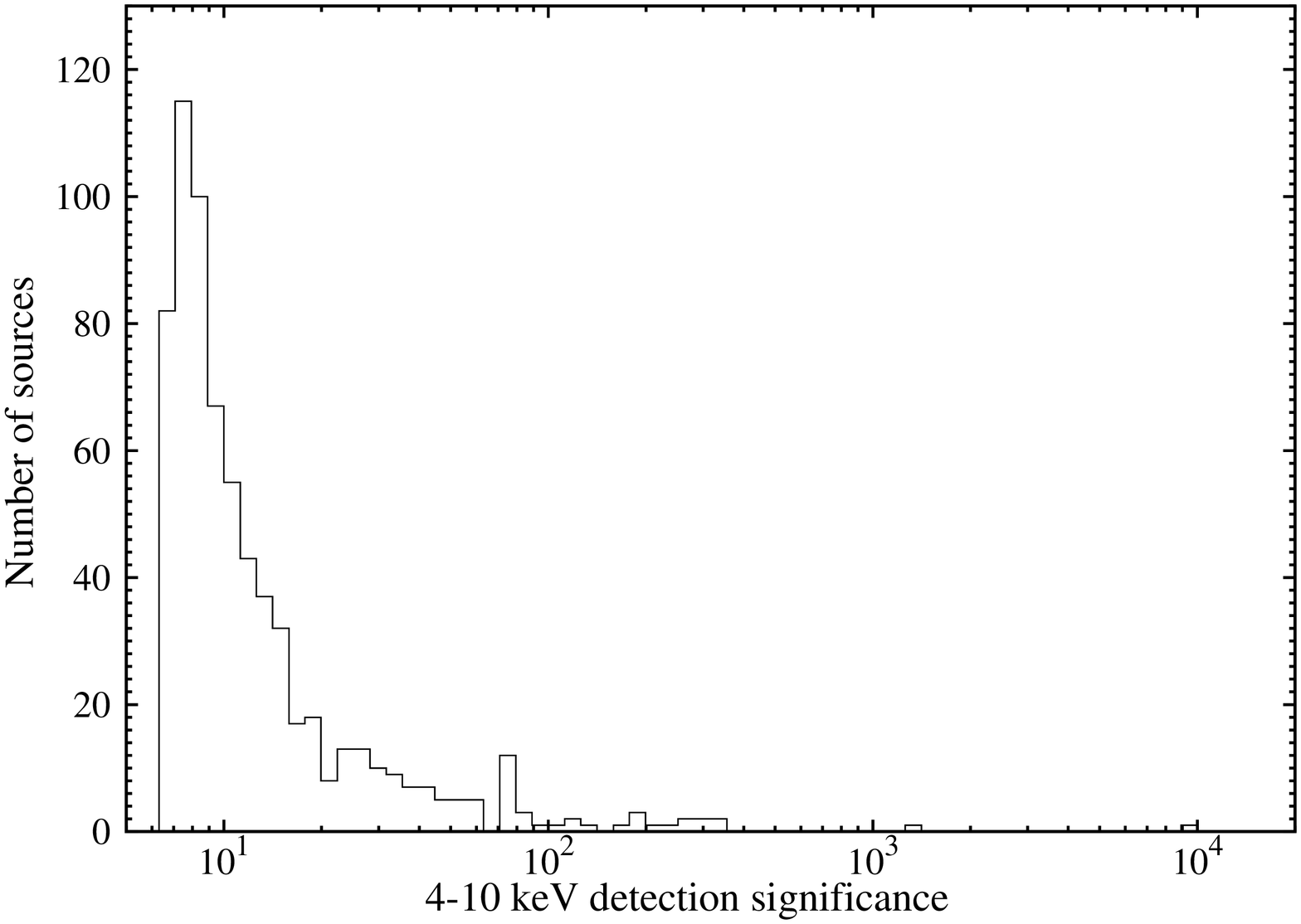}
\includegraphics[scale=0.23,angle=-90]{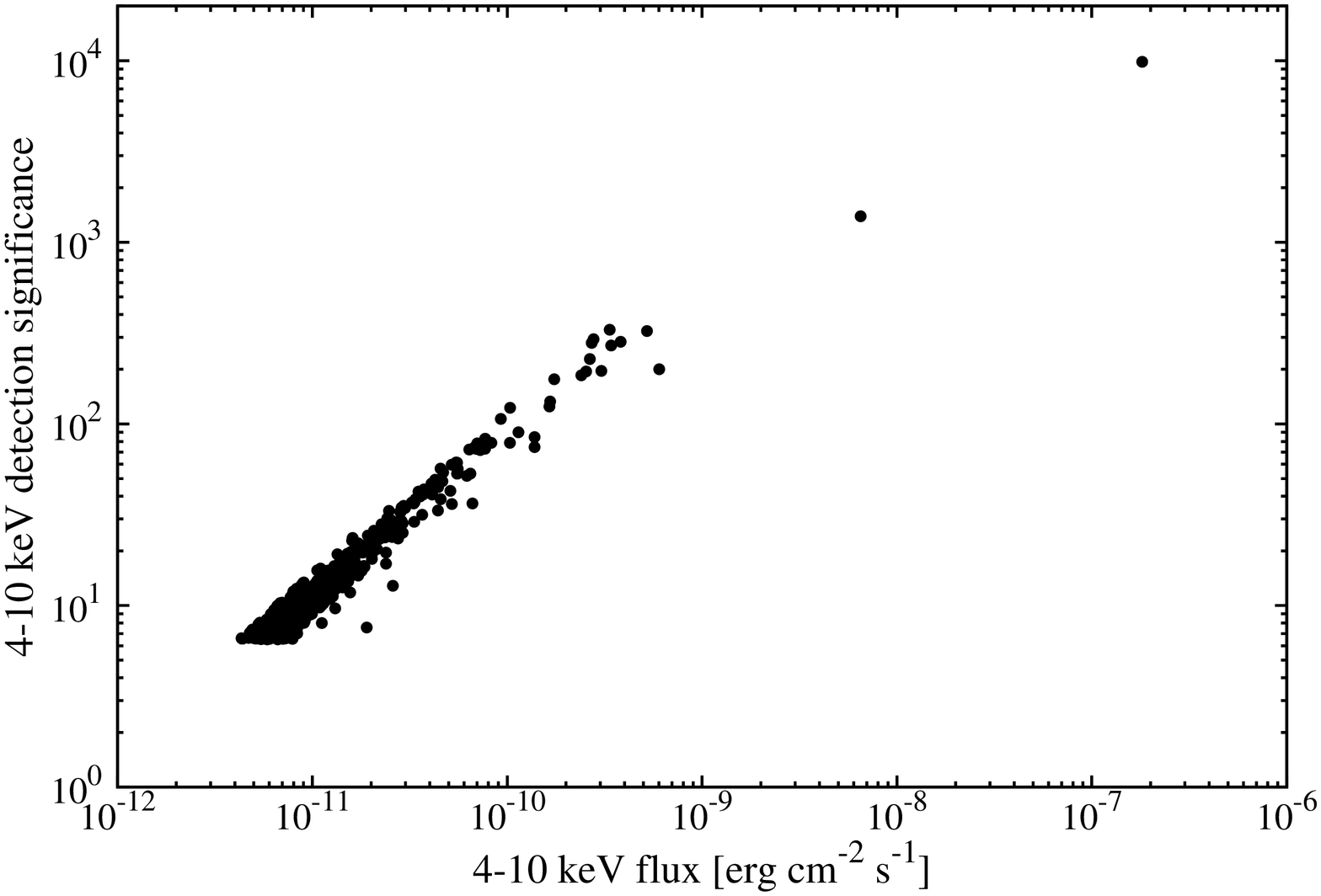}
\caption{\label{fig:basic_info_1}
Histogram of 
the 4--10 keV flux (left), that of the detection significance (middle), 
and their correlation plot (right).  
}
\end{figure*}

\section{Catalog Construction}\label{sec:img_ana}

\subsection{Overview of the \maxi/GSC Data}\label{sec:data}

We analyze the \maxi/GSC event data from 2009 August 13 to 2016 July 31 
provided by RIKEN (rev. 1.8)\footnote{
  The official day of the MAXI first light is 2009 August 15 
  \texttt{(http://iss.jaxa.jp/en/kiboexp/ef/maxi/maxi\_first\_light.html)}, 
  when both the GSCs and Solid-state Slit Cameras \citep{Tom11} started the operation. 
  However, because the activation of all the GSCs was already completed on 2009 
  August 13, the GSC data taken earlier than the first-light day are available.
},
focusing on those at high Galactic latitudes ($|b|>10^\circ$).
The same data screening criteria as adopted by \citet{Hor18} 
(see their Section~2) are applied. We exclude the data taken 
with four GSCs (GSC3, GSC6, GSC9, and GSCa) out of all the 12 GSCs 
from the analysis. GSC6, GSC9, and GSCa were operated only for $<$1 year, which 
is too short to establish a reliable background model (see below).
The GSC3 data are highly contaminated by background
counts due to defects of the veto counters \citep{Sug11}, and are also 
subject to large calibration uncertainties in the position determination.
Apart from them, the GSC0 data since 2013 June 15 are not used 
because they are affected by gas leakage.
We select the 4--10 keV band for source detection by
considering the high quantum efficiency and low background rate
\citep{Mih11}.

Here we summarize main differences of the third \maxi/GSC high
Galactic-latitude catalog compared with the second one \citep{Hir13}.
A major advantage is the increased photon statistics (37 months
to 84 months). Others are improved calibration of the background
(the non X-ray background plus the cosmic X-ray background) and
the point spread function (PSF), as detailed in \citet{Hor18}. 
These are key components in our image analysis (Section~\ref{sec:step1}). 
It was found  that the background event rates well correlate with 
the rates of simultaneous events in the carbon-anode cells and the
tungsten-anode cells for particle veto \citep[i.e., VC-counts;
][]{Mih11,Sug11}. This correlation was already used to reproduce the
background events in \citet{Hir13}. In the new background model, we also 
take into account the direction of movement of the ISS with respect to
the Earth (to the north or south), on which the background-to-VC-count 
correlation is found to largely depend (see Figure~2 of \citealt{Hor18}; \citealt{Shi17a}). 
In return, the 10--20 keV data can be now reliably utilized, which
were not analyzed in the previous work. Also, an updated PSF database
as a function of energy and detector position of each counter (see Appendix
of \citealt{Hor18}) is implemented to the \maxi event simulator,
``\textit{maxisim}''\citep{Egu09}.

\subsection{Image Analysis}\label{sec:step1}

Here we briefly describe the image analysis for source 
detection, which is essentially the same as that adopted in \citet{Hir13}. 
In the event files, each photon has 
information of sky position (RA and Decl.), which 
is converted from an estimated position in the detector 
along the anode direction
and the pointed direction of the collimator at the detected time. 
We first divide the all-sky image into 768 tangentially projected 
images of $14.^\circ0\times14.^\circ0$ size (binned by $0.^\circ1\times0.^\circ1$), 
whose central coordinates are determined by the HEALPix system \citep{Gor05}.
Then, we apply the processes described in the following subsections 
to each projected image. 
It consists of two steps: (1) making a tentative source-candidate list
based on simple photon statistics, and 
(2) determining the flux, significance, and position by image fitting 
with a model composed of the background and PSFs. 
To model the background image, we produce simulated background
events that have 10 times more counts than those of real data, using the 
\maxi/GSC background generator \citep{Egu09}. 

\subsubsection{Source Finding}\label{sec:step1}

We first make a source-candidate list to be used as the input to the image-fit process.
  Combining the real data (consisting of source and background events)
and the simulated background data, 
  we calculate ``nominal significances''
as (real$-$background counts)/$\sqrt{{\rm real~counts}}$, 
where those counts are integrated within a radius of $1^\circ$ 
around each image bin. From this ``significance map'', we identify 
a bin showing the highest significance as the position of the first source 
candidate. After masking a $3^\circ$ radius circular region centered on
the source-candidate position, we again search for another
candidate. This is iterated until no peak is found with a significance
larger than 15$\sigma$. Subsequently, to find fainter sources, we adjust
the normalization of the background model so that the counts of the
masked real and masked background data become identical, assuming that 
the contribution from remaining fainter sources to the observed counts 
is ignorable. Such fine tuning of the background (typically by $<5$\%) 
is required due to systematic uncertainties in the background model and 
fluctuation of the level of the cosmic X-ray background.  
Then, we repeat the same source finding procedures 
above by lowering the detection threshold to 5.0$\sigma$.
Merging the source-candidate lists from all the images,
we sometimes find multiple sources within one degree, which is smaller than a typical 
size (FWHM) of the \maxi/GSC PSF ($\sim 1^\circ.5$). 
This happens because there are some overlap area among the projected images.
In such cases, we only leave the source candidate located closest to the 
center of the analyzed image.

\subsubsection{Image Fit}\label{sec:step2}

\begin{figure}
  \begin{center}
    \vspace{.3cm}
\includegraphics[scale=0.25,angle=-90]{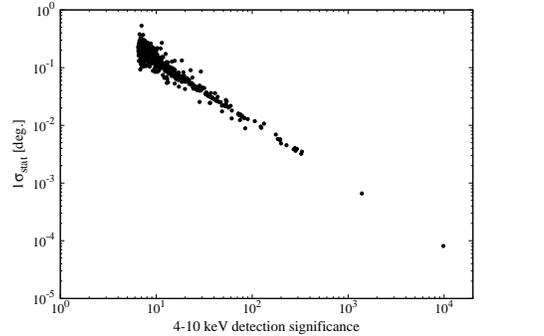} 
\caption{\label{fig:basic_info_2}
Positional error (1$\sigma_{\rm stat}$) as a function of 
the 4--10 keV detection significance ($s_{\rm D,4-10~keV}$). 
}
\end{center}
\end{figure}

\begin{figure*}
  \hspace{-0.7cm} 
\includegraphics{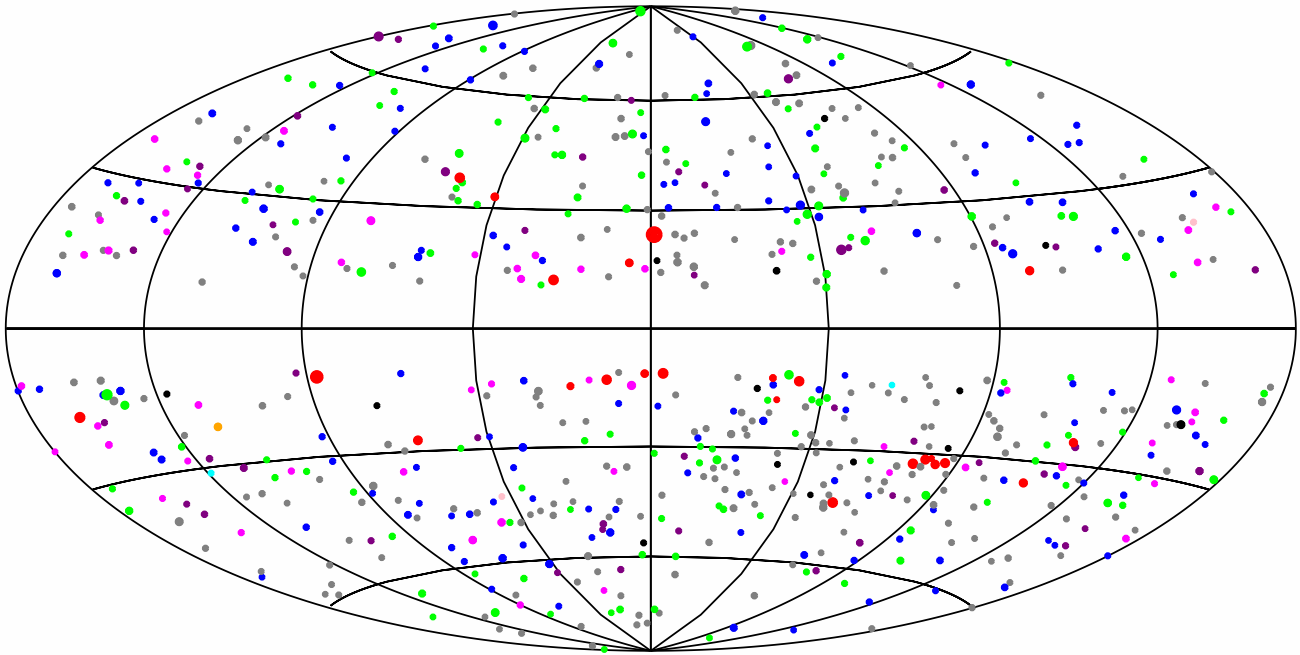}
\caption{\label{fig:allsky-map}
Spatial distribution of the detected 682 X-ray sources in Galactic
coordinates. The center corresponds to 
($l$, $b$) = ($0^\circ.0$, $0^\circ.0$). The radii of the circles are proportional to the logarithm 
of the 4--10 keV flux. 
According to the results of catalog cross-matching
(Section~\ref{sec:cross-match}),
a color is assigned to each source type: 
unidentified X-ray sources in BAT105
(black); galaxies (cyan); galaxy clusters (green); Seyfert galaxies
(blue); blazars (purple); cataclysmic variables/stars (magenta); X-ray
binaries (red); pulsars (pink); confused sources (orange); and sources
without any counterparts in the reference catalogs (grey).}
\end{figure*}

The PSFs for the source candidates are constructed from simulated photon events 
produced by the \maxi simulator \citep{Egu09} with the exactly the 
same observing conditions as for the real data. 
We assume a point source having the same spectrum as the Crab Nebula. 
Throughout this paper, we assume that Crab Nebula has a power law with a
photon index of 2.1 and a normalization of 10 photons cm$^{-2}$ s$^{-1}$ 
keV$^{-1}$ at 1 keV, absorbed with a hydrogen column density of
$2.6\times10^{21}$ cm$^{-2}$. The fluxes are 3.96$\times$10$^{-9}$, 1.21$\times$10$^{-8}$,
8.51$\times$10$^{-9}$, and 1.61$\times$10$^{-8}$ erg cm$^{-2}$ s$^{-1}$ in the 3--4 keV,
4--10 keV, 10--20 keV, and 3--10 keV respectively, which correspond to ``1 Crab''.

Then, we fit the PSF and background models to the the real image.
Here we restrict the fitting region to an inner
$11^\circ.0\times11.^\circ0$ area, because the outer region may be
affected by sources located just outside the edge of the 
14$^\circ.0\times$14$^\circ.0$ region.
The fluxes and peak positions of the PSFs and the normalization of the
background are left as free parameters.
We determine these parameters by 
Poisson maximum-likelihood (ML) algorithm ($C$-statistics), 
using the MINUIT software package. The likelihood function is defined as
\begin{eqnarray}
C & & \equiv 2 \sum_{i,j} \mathcal{L} (i,j), 
\end{eqnarray}
where
\begin{footnotesize}
\[
\mathcal{L} (i,j) = \begin{cases}
  M (i,j) - D (i,j)+ D (i,j) \{ \ln D (i,j) - \ln M (i,j) \}  \\
  \hspace{5cm}  (D_{i,j} > 0) \\ 
  M (i,j)   \hspace{4.08cm} (D_{i,j} = 0). 
\end{cases}
\]
\end{footnotesize}
The observed data and model at the $(i, j)$ bin are denoted with $D(i, j)$ and $M(i, j)$,
respectively. The 1$\sigma$ statistical error of each parameter is estimated by finding a
value when the likelihood function ($C$) is increased from its minimum by unity. The detection
significance is defined as (best-fit flux) / (its lower-side 1$\sigma$ error).
We confirm that the lower- and upper-side errors are generally consistent
with each other. 
This is expected because the number of counts is large enough to follow
a Gaussian distribution.
Since we measure a relative count rate to that that 
would be expected from the Crab Nebula spectrum, the flux 
is given in the ``Crab'' unit.

\subsubsection{Iteration of Source Finding and Image Fit}\label{sec:step3}

Some sources located near bright sources could be missed in the first
source-candidate list because we masked out the 3 degree circular
region around each source candidate. To make our source finding
complete, we repeat the source finding process (Section~\ref{sec:step1})
by using a new significance map (smoothed residual map) where the best-fit 
model (background plus PSFs) obtained in the above image-fit process
(Section~\ref{sec:step2}) is adopted as the background. Then, including
those new source candidates detected with nominal significances of $\geq$ 
5.5$\sigma$, we perform image fit again. This provides the final source
catalog.

\ifnum0=1
We try to salvage sources undetected in the previous significance maps,
which were based on the simple estimate of the background level. The
non-detection is likely because such are located around brighter
sources. Hence, new significance maps are made using the real data and
the best-fit total model instead of the background model only. Then, we
repeat the source finding as done in Sections~\ref{sec:step1}.  The PSFs
for those detected at $\geq$ 5.5$\sigma$ are simulated. The more strict
detection criterion is adopted to reduce the false source
detection. After adding the new candidates to the initial list, we
perform the image fitting to finalize the third \maxi/GSC catalog. 
\fi

\subsection{Catalog}\label{sec:step3}

We detect 682 sources with 4--10 keV detection significances of 
$s_{\rm D,4-10~keV} \geq 6.5$
at high Galactic latitudes ($|b| > 10^\circ$).
Table~\ref{tab:catalog1} gives our source catalog. 
  Considering the improved calibration of the background and PSF models (Section~\ref{sec:data}) 
  and following the 7-year ``low''-Galactic latitude catalog \citep{Hor18}, 
we have decided to lower the detection threshold from $s_{\rm
D,4-10~keV} \geq 7$, which was adopted in the 37-month catalog
\citep{Hir13}, to $s_{\rm D,4-10~keV} \geq 6.5$ in our new catalog.
The number of the cataloged sources increases by a factor of 
$\approx 1.4$ compared with \citet{Hir13}, where 500 sources were listed. If 
the same threshold of 
$s_{\rm D,4-10~keV} \geq 7$ as in \citet{Hir13}
is adopted, the source number of our catalog becomes 615.
  The counterpart candidate is 
listed 
in Table~\ref{tab:catalog2}, as detailed in Section~\ref{sec:cross-match}.
The sensitivity 
limit achieved for 50\% of the survey area is $\sim 0.48$ mCrab, or 
$\sim$ 5.9 $\times$ 10$^{-12}$ \ergcms 
(Section~\ref{sec:lognlogs}). Figure~\ref{fig:basic_info_1} shows the
distribution of the flux and detection significance in the 4--10 keV
band, and their correlation. Figure~\ref{fig:basic_info_2} plots the 
statistical errors in the position (1$\sigma_{\rm stat}$) against the
detection significance. The spatial distribution in the Galactic
coordinates is shown in Figure~\ref{fig:allsky-map}. The
classification of sources in this figure is detailed in
Section~\ref{sec:cross-match}.

\subsubsection{Spurious Source Fraction}\label{sec:fal_src}

The detection threshold should be determined to maximize the number of
detected sources while suppressing that of spurious sources caused by
statistical and/or systematic fluctuations of the data to an acceptable
level. We estimate the number of spurious sources in our catalog by counting
that of negative signals with the same detection threshold, assuming that
their significance distribution is symmetrical in positive and negative sides 
around the background level. We first search the significance maps produced
from the data and the best-fit models for negative-signal candidates.
Then, we create their PSF models through simulation and perform image
fitting by allowing their normalizations to be negative. We find that
our catalog may contain 26 spurious sources with $s_{\rm D,4-10~keV} \geq 6.5$, 
which corresponds to 4\% of the total 682 sources. If we adopt
$s_{\rm D,4-10~keV} \geq 6$ as the threshold, the number of spurious sources
drastically increases to 58 sources, 8\% of the total 766 sources. Thus, we
select $s_{\rm D,4-10~keV} = 6.5$ as the threshold. Depending on the scientific
purpose, catalog users can choose a more conservative threshold; the estimated
fraction of spurious sources becomes 3\% (17/615) for $s_{\rm D,4-10~keV} \geq 7$
and 0.6\% (3/477) for $s_{\rm D,4-10~keV} \geq 8$.


\subsubsection{Hardness Ratios}\label{sec:hr} 

To calculate HRs of the detected sources, 
we determine the fluxes in the 3--4 keV and 10--20 keV
bands in the following way.
Tangentially projected images in these bands 
are produced from the event files. 
We prepare the input source list for the image fit
from our catalog produced above.
For completeness, here we also include the source candidates
whose detection significances were less than $6.5\sigma$ 
in the 4--10 keV band.
We then perform image fit to the image in the 3--4 keV or 10--20 keV band
using this source list. 
During the fitting process, the positions are fixed at those determined 
in the 4--10 keV band. 
This process is repeated by adding 
new source candidates found in the smoothed residual map with 
the nominal significances (see Section~\ref{sec:step1}) above 5.5$\sigma$. 
Eventually, we find no new sources 
detected with significances above $6.5\sigma$ 
in the 3--4 keV or 10--20 keV band but not included 
in the above catalog based on the 4--10 keV band survey.
The 3--10 keV fluxes are also calculated from 
the 3--4 keV and 4--10 keV fluxes obtained above by assuming
the Crab-like spectrum. 
We define the HR as 
\begin{equation}
{\rm HR} = \frac{f_{\rm H}-f_{\rm S}}{f_{\rm H}+f_{\rm S}}, 
\end{equation} 
where $f_{\rm H}$ and $f_{\rm S}$ correspond to the hard and soft band
fluxes, respectively, in the Crab units. 
The catalog lists three HRs: HR1 (3--4 keV, 4--10 keV), 
HR2 (4--10 keV, 10--20 keV), and HR3 (3--10 keV, 10--20 keV).

\begin{figure*}
\includegraphics[scale=0.36,angle=-90]{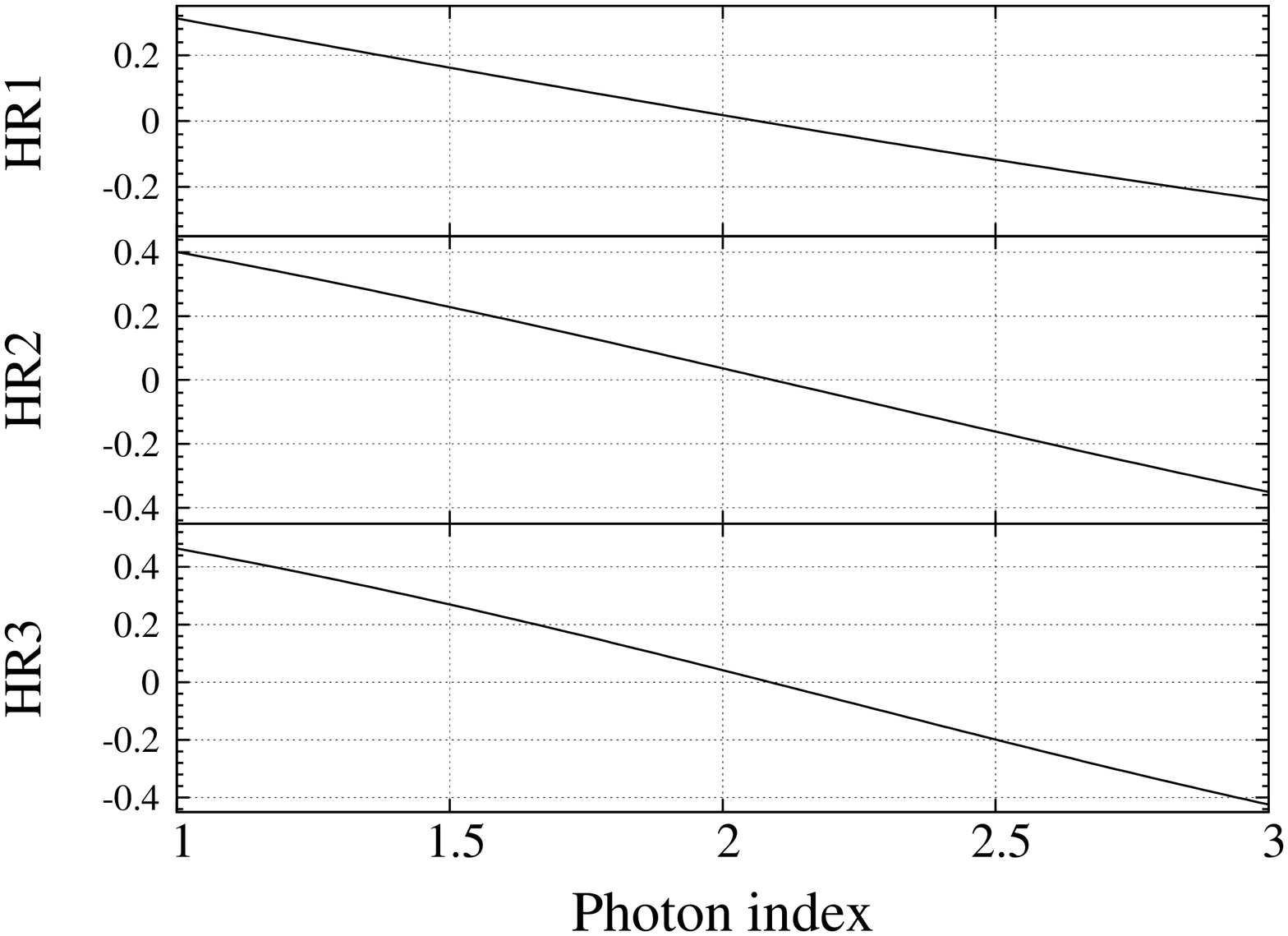}\hspace{-.5cm}
\includegraphics[scale=0.36,angle=-90]{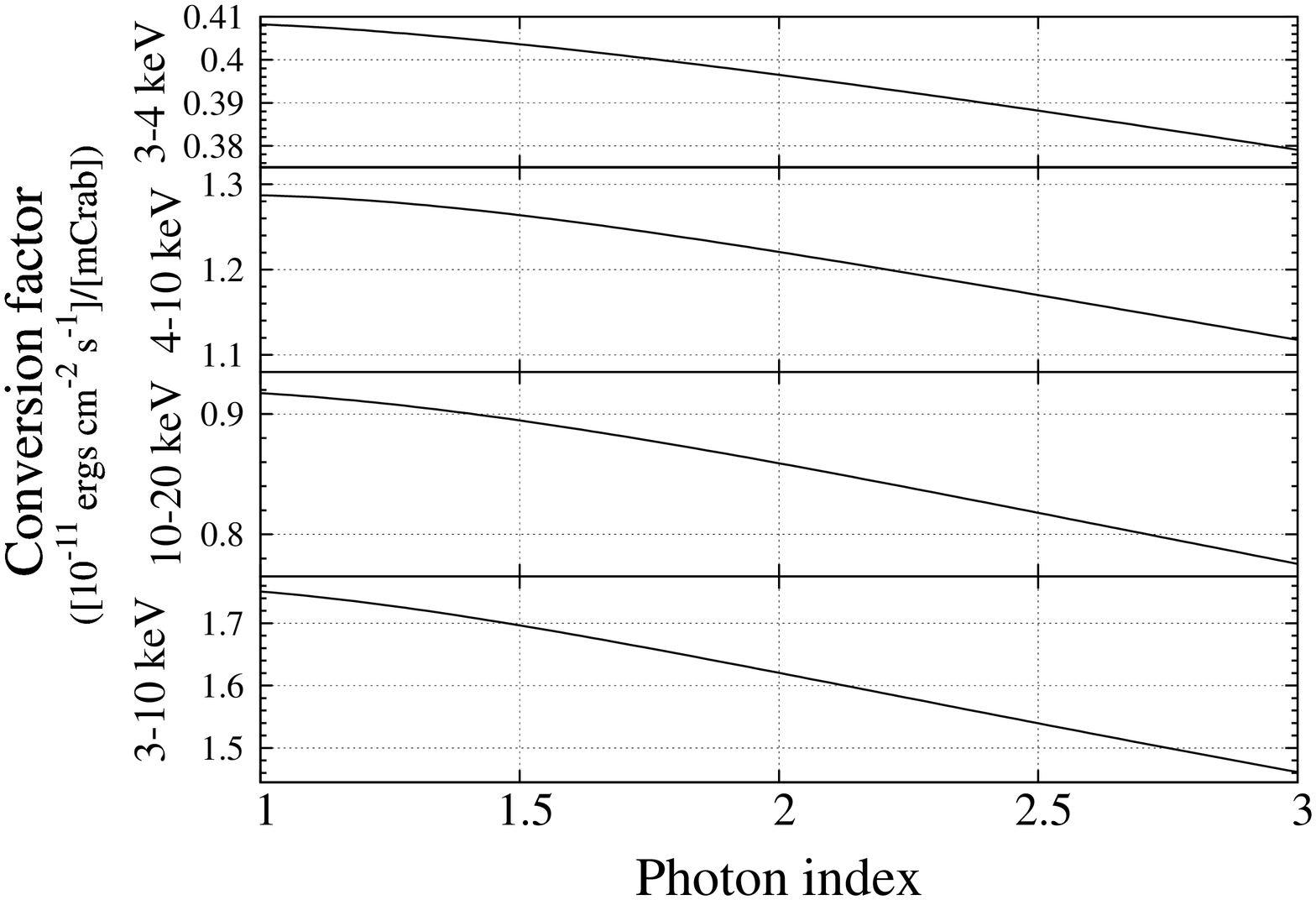}\\
\includegraphics[scale=0.36,angle=-90]{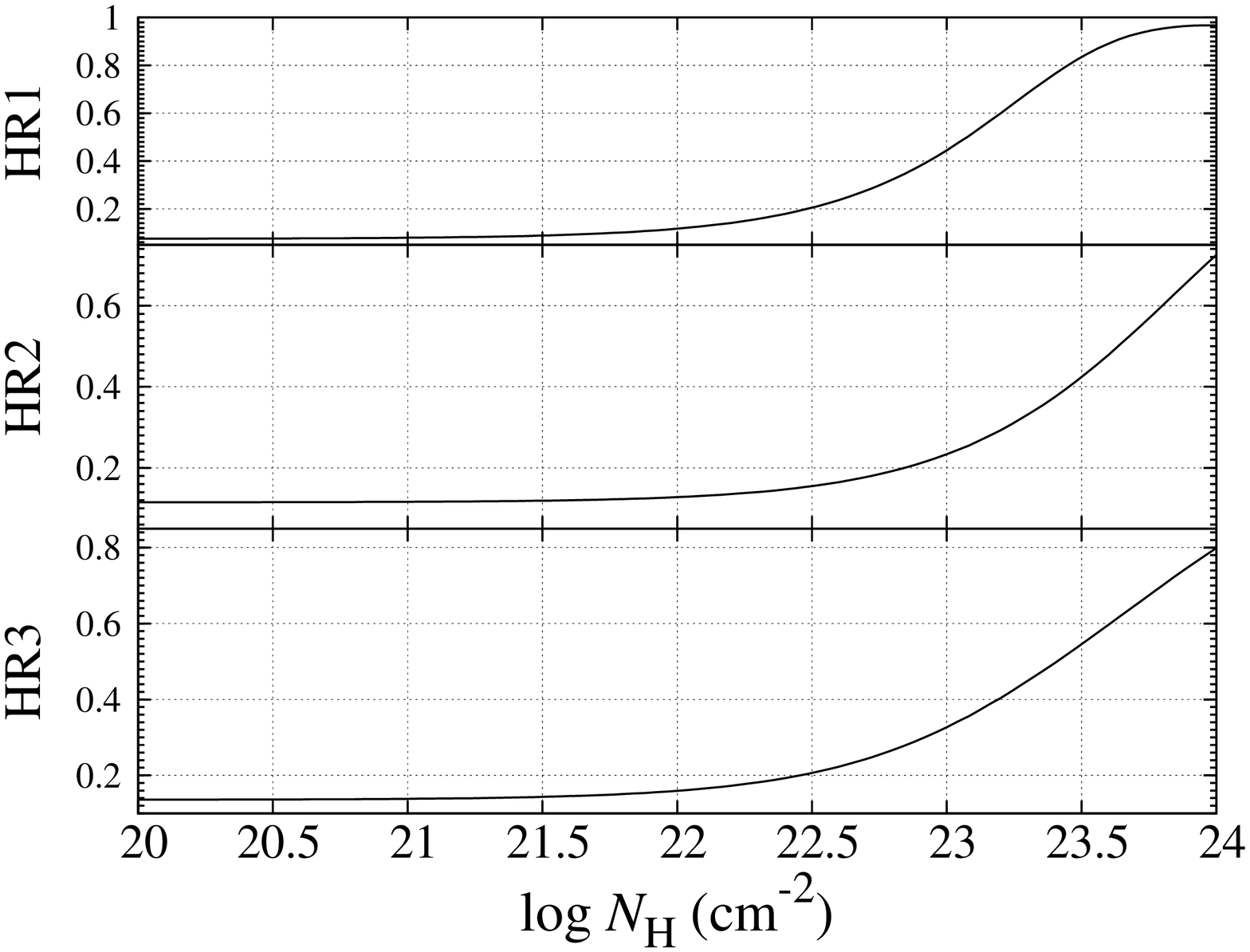}\hspace{-.5cm}
\includegraphics[scale=0.36,angle=-90]{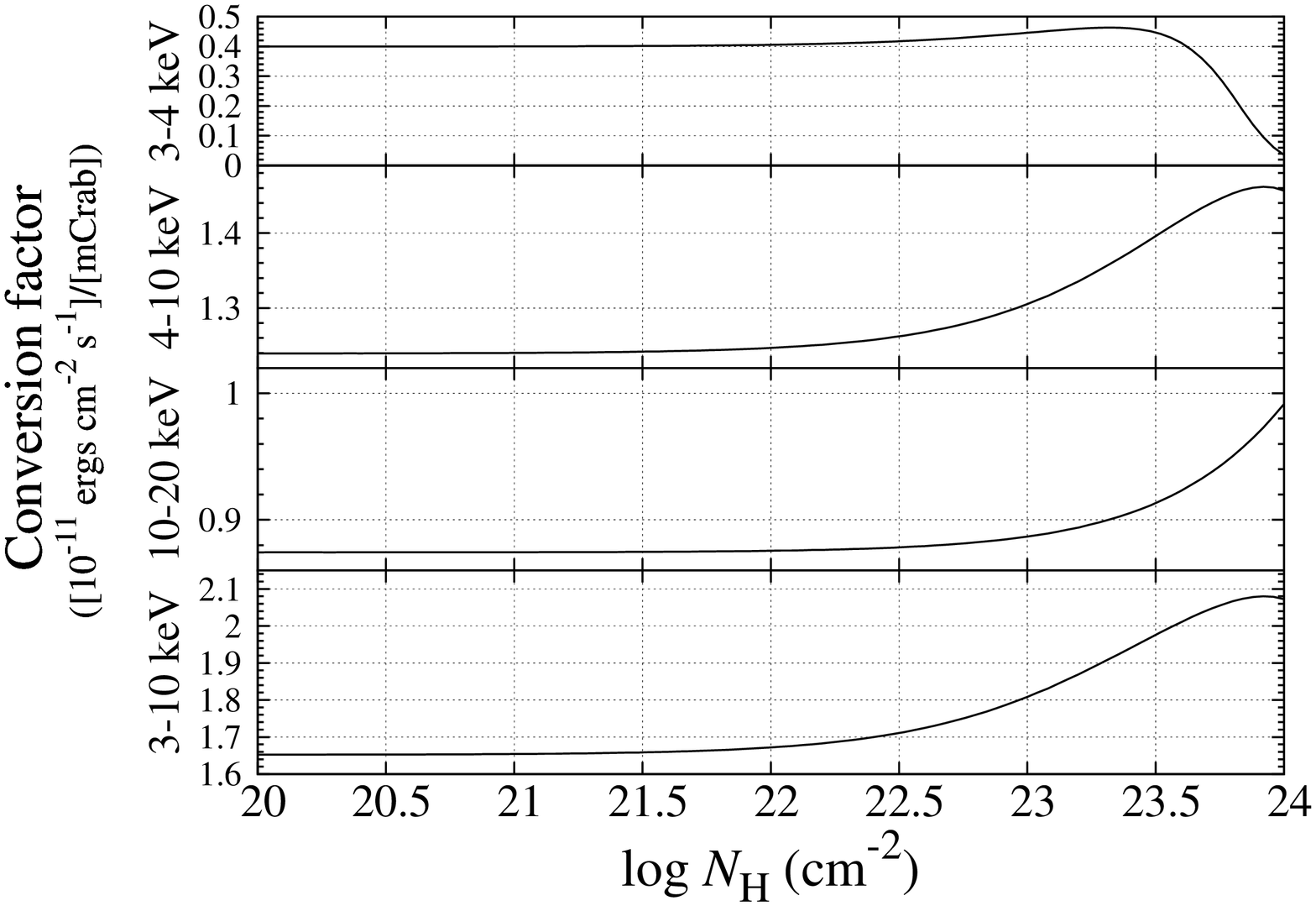}
\caption{\label{fig:conversions} 
Upper left: Relation between photon index and HRs: HR1 (3--4 keV, 
4--10 keV), HR2 (4--10 keV, 10--20 keV), and HR3 (3--10 keV, 10--20  keV). 
No absorption other than the Galactic one ($N^{\rm Gal}_{\rm H} = 2.5\times10^{20}$ cm$^{-2}$) is considered. 
Upper right: Conversion factor in units of $10^{-11}$  \ergcms~mCrab$^{-1}$ as a function of photon index. 
Lower left: Relation between hydrogen column density and HRs for an absorbed power-law model with a photon index of 1.8.
  Lower right: Conversion factor in units of $10^{-11}$  \ergcms~mCrab$^{-1}$ as a function of hydrogen column density.
}
\end{figure*}

To deduce the spectral shape from the HRs, we calculate the relation between the
HRs and spectra parameters by assuming an absorbed power law model. The parameters
are a photon index ($\Gamma$) and an intrinsic hydrogen column density ($N_{\rm H}$).
We also take into account a representative Galactic absorption of $N^{\rm Gal}_{\rm H}
= 2.5\times10^{20}$ cm$^{-2}$ in addition to the intrinsic one.
  This column density roughly corresponds to the minimum value over the high Galactic
  latitude sky \citep[$|b|>10^\circ$; ][]{Dal72}, and should be considered as a lower limit
  for extragalactic objects. We confirmed, however, that the spatial variation in the
  total Galactic column ($N^{\rm Gal}_{\rm H} < 2\times10^{21}$ cm$^{-2}$ for $|b|>10^\circ$) 
  has little effect on the following results. 
The source redshift is not considered for simplicity. We generate, through 
the \maxi on-demand tool\footnote{{\tt http://maxi.riken.jp/mxondem/}}, the 7-year averaged
\maxi/GSC response file, and calculate the predicted count rates in these bands for given
spectral parameters. We consider two cases: a power-law model with and without an
intrinsic absorption. 
The Galactic absorption is taken into account in both cases.
In the former case, we fix $\Gamma = 1.8$ and vary $\log N_{\rm H}$/cm$^{-2}$ in a range of
20--24, whereas in the latter case, we fix $N_{\rm H} = 0$ and 
vary $\Gamma$ in a range of 1--3. 
The results are plotted in the left panels of Figure~\ref{fig:conversions}. We also
calculate the conversion factors from the flux in the Crab units to that
in the CGS system (\ergcms), which are presented in the right panels of
Figure~\ref{fig:conversions}. For an unabsorbed power law with $\Gamma = 1.8$,
HR1 = 0.0749, HR2 = 0.115, and HR3 = 0.136, and the conversion factors are 
$4.00\times10^{-12}$, 
$1.24\times10^{-11}$, 
$1.65\times10^{-11}$, and 
$8.74\times10^{-12}$ \ergcms~mCrab$^{-1}$
in the 3--4 keV, 4--10 keV, 10--20 keV, and 3--10 keV, respectively. 

\section{Time Sliced Image Analysis}\label{sec:time_ana} 

\subsection{4--10 keV lightcurves}\label{sec:lc} 

Flux variability provides us with a key to deduce the nature of X-ray sources; 
for example, we can use the fact that many of compact objects 
(e.g., X-ray binaries and AGN) are transients, whereas extended 
sources like galaxy clusters are more persistent. 

Here, we produce the 4--10 keV lightcurves in 1-year binning. 
We split the real, background, and PSF data used in the catalog 
construction into each year, and perform image fitting for each time-sliced dataset.
We adopt the same observation period as those in the catalog construction, 
and therefore the time bins in 2009 and 2016 have smaller sizes than the other bins. 
In addition to the cataloged sources, there may be transient sources that
can be detected at significances above 6.5$\sigma$ in a 1-year image 
but not in the 7-year accumulated data. 
Thus, we systematically search for such sources (see Section~\ref{sec:trans_obj}) 
and incorporate their PSFs in the image fitting. 
The positions of the cataloged sources are fixed at those determined from 
the whole 7-year data, while the positions of the newly detected transients 
are at those determined in the time bin where their detection significances 
are the highest. Then, we construct the lightcurves of the individual sources 
by fitting the normalizations of their PSFs and the background model. 

\begin{figure*}
\hspace{0cm}
\includegraphics[scale=0.36,angle=-90]{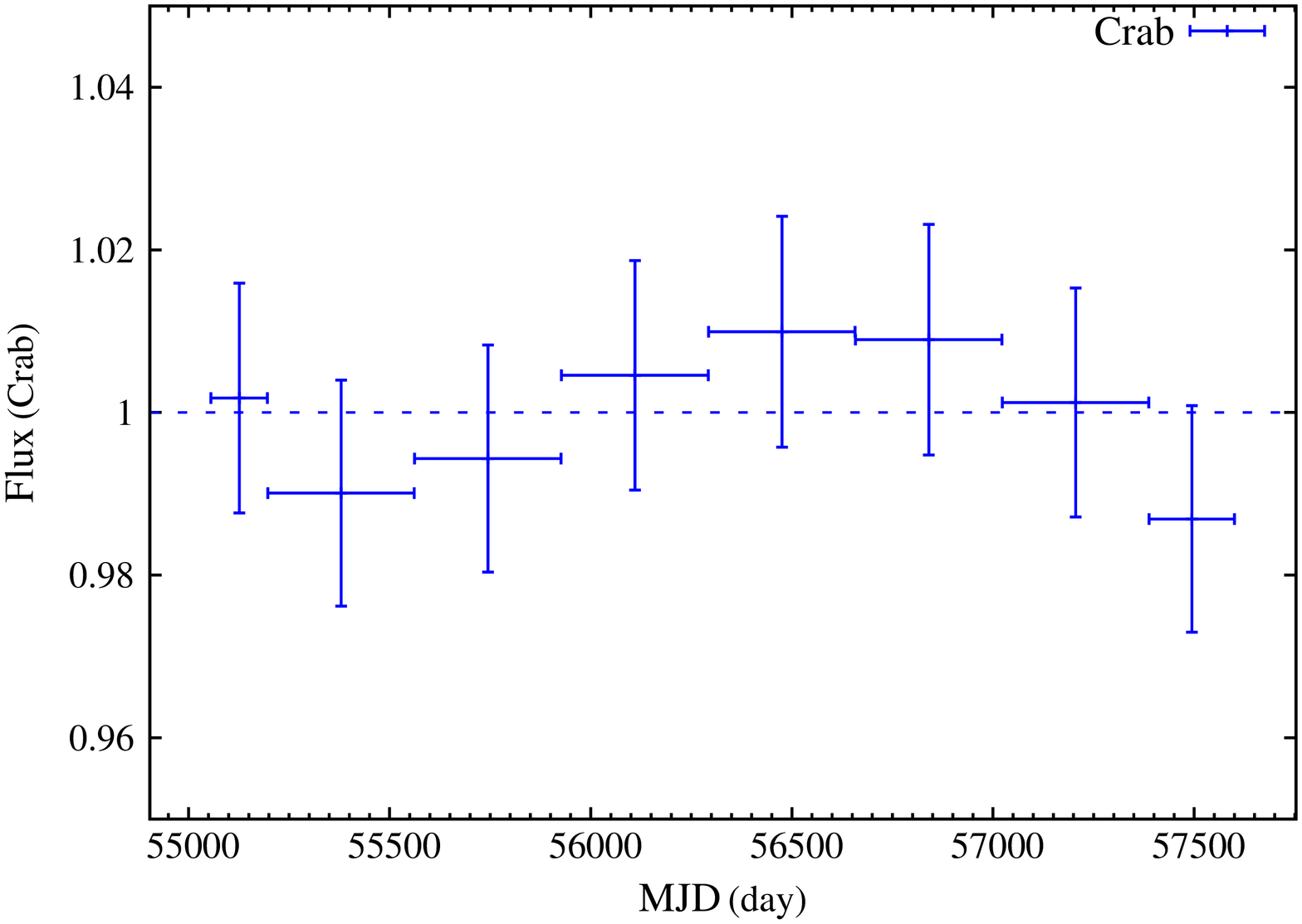} \hspace{-.2cm} 
\includegraphics[scale=0.36,angle=-90]{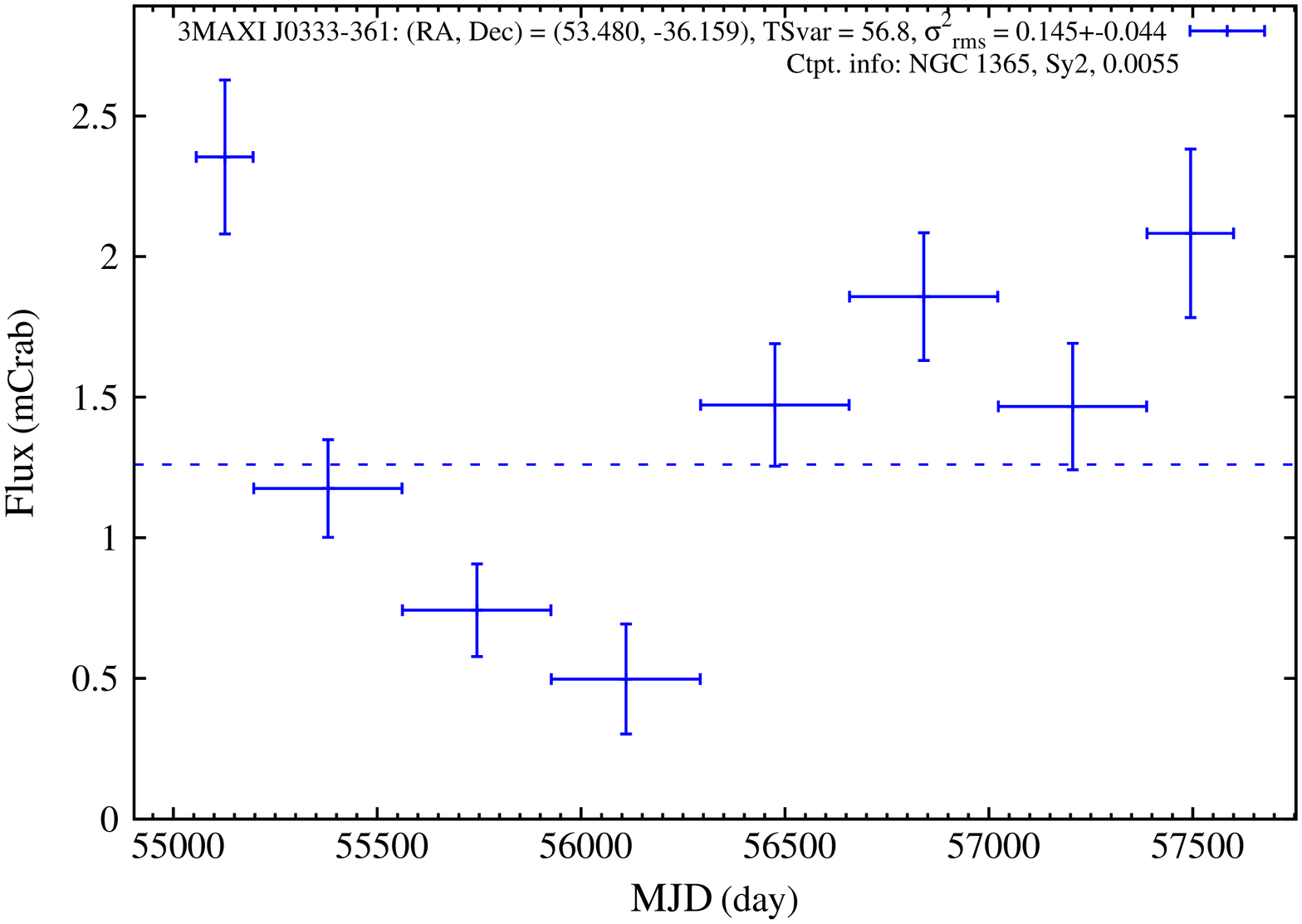}
\caption{\label{fig:lcs}
Left: One-year-bin lightcurve of the Crab Nebula in the 4--10 keV band. 
The dashed line denotes the the 7-year-averaged Crab flux, which is defined as 1 Crab. 
Right: The same as the left but for NGC 1365 (AGN) in a unit of mCrab. 
The counterpart information (name, type, and redshift) is listed in the legend. 
Both lightcurves take account of a systematic flux error of 1.4\% (see text). 
}
\label{detection_ex}
\end{figure*}

In addition to the statistical flux errors, we derive the systematic error, which may remain because 
we do not consider possible time variation of the PSF 
profile in our PSF database. We simply determine 
the systematic error so that the Crab lightcurve is consistent with 1 Crab over the 7 years. 
This is a conservative way to estimate the systematic errors, since the
X-ray flux of Crab is not perfectly constant \citep{Wil11} 
and its variability is included in the estimation. 
As shown in the left panel of Figure~\ref{fig:lcs}, adding 1.4\% error is found to be sufficient. This value is taken
into account in all the lightcurves as the systematic errors.
This systematic uncertainty is comparable to that derived in \citet{Hor18}, 1.6\%, and much reduced compared with 
that in the 37-month data \citep[$\sim$10\%; ][]{Hir13,Kaw16}. This improvement can be ascribed to the sophisticated PSF
model. Among the lightcurves (Figure~\ref{fig:all_lcs}), we present, as an interesting case, the NGC 1365 
(AGN) one in the right panel of Figure~\ref{fig:lcs}, which shows the strong variability by a factor $\sim$5 within the 7
years. The lightcurve data are published in online.

We characterize the lightcurves in two steps. First, to statistically examine whether a lightcurve is constant or varies, 
we adopt a likelihood-based method as used in \cite{Nol12}. This is more
appropriate than the simple least chi-squared method when the photon
statistics follow the Poisson distribution, and therefore can be used
for faint sources as well as for bright ones. 
We calculate a test statistics defined as 
\begin{equation}
{\rm TS}_{\rm var} = 2 \sum_{i} \frac{\Delta F^2_i}{\Delta F_i^2 + f^2F^2_{\rm const}}  [\log \mathcal{L}'_i (F_i) - \log \mathcal{L}'_i(F_{\rm const})], 
\end{equation}
where $F_i$, $\Delta F_i$, $F_{\rm const}$, and $f$ denote the best-fit flux, statistical uncertainty, 7-year 
averaged flux, and systematic error (i.e., 1.4\%), respectively, at the $i$-th time bin. TS$_{\rm var}$ follows the 
chi-squared distribution when the number of bins is large. If TS$_{\rm var} > 18.48$, we can reject the null hypothesis that the flux is constant at the 99\% confidence level for 7 degrees of freedom. Next, we derive the 
excess variance ($\sigma_{\rm rms}^2$) following \citet{Vau03}, or Equations (3) and (4) in \citet{Hor18}, to investigate the strength of flux variations. 
TS$_{\rm var}$ and $\sigma_{\rm rms}$ are listed in Table~\ref{tab:catalog1}.

\subsection{Transient Event Search}\label{sec:trans_obj}

We systematically search for transient events that are missed
in the catalog based on the 7-year integrated images.
First, we create the significance maps in the individual years 
in the the same manner as described in Section~\ref{sec:step1},
using the best-fit model only including the cataloged sources. 
We then identify residuals with nominal significances 
above 5.5$\sigma$ as the candidates of transients. Then, we 
perform the image fitting by including their PSFs 
to the model in addition to those of the catalog sources. 
Four transient sources are found to show significances of $s_{\rm D,4-10~keV} \geq$ 6.5 in a time bin.
Table~\ref{tab:trans_catalog} and Figure~\ref{fig:trans_lc} present their X-ray 
properties and the lightcurves, respectively. Their coordinates are estimated from a period where the
source shows the highest significance among all the time bins. Note that circular regions with a $3^\circ$ 
radius around the very bright sources Sco X-1, Cyg X-2, and Crab are 
excluded from this
analysis to avoid fake source detection due to systematic errors in the PSF calibration.

\section{Counterparts of Detected X-ray Sources}\label{sec:cross-match}

\subsection{Positional Cross-matching with Other X-ray Catalogs}\label{sec:cross-match}

Identification of the X-ray sources is an important task for the statistical use of the catalog. 
To find possible counterparts of our catalog sources including the transients (hereafter we call
"the third \maxi sources"), we spatially cross-match them to those in other X-ray catalogs in the
following order; the \swift/BAT 105-month catalog \citep[BAT105; ][]{Oh18}, the \uhuru fourth catalog
\citep[4U; ][]{For78}, the RXTE All-Sky Monitor long-term observed source table
(XTEASMLONG\footnote{{\tt https://heasarc.gsfc.nasa.gov/W3Browse/xte/xteasmlong.html}}), Meta-Catalog 
of X-Ray Detected Clusters of Galaxies \citep[MCXC; ][]{Pif11}, the \xmm slew survey catalog (XMMSL2 
\footnote{{\tt https://www.cosmos.esa.int/web/xmm-newton/xmmsl2-ug}}), and the \rosat all-sky survey 
bright source catalog \citep[1RXS; ][]{Vog99}. To reduce false identifications, we only take account
of XMMSL2 and 1RXS sources brighter than $>1.0\times10^{-12}$ erg cm$^{-1}$ s$^{-1}$ and $>$ 0.30
cts~s$^{-1}$, respectively. This is a flux corresponding to 0.3 mCrab when the Crab spectrum is assumed.
The statistical positional errors of the third \maxi sources　($\sigma_{\rm stat}$) are 
evaluated as $\sqrt{dx^2+dy^2}$, where $dx$ and $dy$ are one-dimensional spatial errors
in perpendicular directions at the 1$\sigma$ level. 
A systematic error ($\sigma_{\rm sys}$) is further　
added to the statistical ones as $\sigma_{\rm pos} = \sqrt{\sigma_{\rm stat}^2 + \sigma_{\rm sys}^2}$.
We adopt $\sigma_{\rm sys}=0.^\circ03$, determined to self-consistently match the cross-matching result
(see Section~\ref{sec:sys_err} and Figure~\ref{fig:ang}). Here, we identify a counterpart if its position
falls within the $2.5 \sigma_{\rm pos}$ of a third \maxi source.
Using several representative sources, we have verified that 
the radius of $2.5 \sigma_{\rm stat}$ typically corresponds to
that of the 95\% confidence contour for the source positions
where the $C$ value in Equation~(1) is increased from the best value by 6.0.
The positional errors in the reference catalogs are ignored. Note that, if we find 
more than one source as counterpart candidates, we list the object nearest to the \maxi position. 
The cross-matching result is summarized in Table~\ref{tab:catalog2}, and the numbers of the 
positionally matched MAXI sources are listed in Table~\ref{tab:match_result}. The breakdown of the
source types is seen in Table~\ref{tab:type_dist}. The counterparts of the transients (Section~\ref{sec:trans_obj})
are also identified from the {\it Swift}/BAT Hard X-ray Transient Monitor catalog\footnote{{\tt
    https://swift.gsfc.nasa.gov/results/transients/}} in the same manner, but with the $3.0\sigma$ positional
errors, because the probability of false matching is much lower than in the other cases (see Section~\ref{sec:chan_coin}).

In some cases, we suspect that a matched counterpart is likely
incorrect, judging from the X-ray properties (i.e., the time
variability, hardness ratios, and observed fluxes).  Also, some bright
galaxy clusters are not identified from the automatic 
cross-matching. This could be because our PSF model is developed so that 
it reproduces a point source and would not properly fit such extended
sources. More plausible counterparts for them are listed in the catalog, 
with the associated comments in the (9) column. 


\begin{deluxetable*}{lccccccccc}
  \tablefontsize{\footnotesize}
  \tabletypesize{\scriptsize}
  \tablecaption{Numbers of 
\maxi\ Sources Positionally Matched with Other Catalogs.
\label{tab:match_result}} 
  \tablewidth{0pt}
  \tablehead{
   \colhead{Catalog name} & \colhead{BAT105} & \colhead{4U} & \colhead{XTEASMLONG} & \colhead{MCXC} & \colhead{XMMSL2} & \colhead{1RXS} &  \colhead{One catalog only$^{a}$} 
  }
  \startdata
    BAT105     & 239(243)$^{b}$   & 10           & 77     & 10       & 97       & 133      & 75 \\
    4U         & ...              & 17(17)$^{b}$ & 12     &  4       &  8       &  11      & 2  \\
    XTEASMLONG & ...              & ...          & 94(95)$^{b}$ &  1       & 43       &  61      & 5  \\
    MCXC       & ...              & ...          & ...    & 152(164)$^{b}$ & 16       &  66      & 82 \\
    XMMSL2     & ...              & ...          & ...    & ...      & 198(226)$^{b}$ &  98      & 77 \\ 
    1RXS       & ...              & ...          & ...    & ...      & ...     & 250(251)$^{b}$ & 48 \\ \vspace{-3mm} 
    \enddata
    \tablecomments{The numbers of \maxi sources matched
 with both catalogs given in the first column and first row are listed.
      (a) Numbers of the \maxi sources matched only with one catalog
given in the first column. 
      (b) Numbers in parentheses represent those of total matched sources in each catalog. 
    } 
\end{deluxetable*}

\begin{deluxetable}{ccc}
  \tablefontsize{\footnotesize}
  \tabletypesize{\scriptsize}
  \tablecaption{Numbers of Sources in Each Category. \label{tab:type_dist}}
  \tablewidth{0pt}
  \tablehead{
    \colhead{Category} & \colhead{Number of sources}
  }
  \startdata
   unknown$^{a}$       & 14  \\
   galaxy clusters     & 135 \\ 
   galaxies            & 2   \\ 
   Seyfert galaxies    & 142 \\ 
   blazars             & 48  \\ 
   cataclysmic variables/stars           & 54  \\ 
   pulsars             & 2   \\  
   X-ray binaries      & 24  \\ 
   confused            & 1   \\ 
   unmatched$^{b}$     & 260 \\ \hline 
   total               & 682 \\ \vspace{-3.5mm}
  \enddata
\tablenotetext{}{NOTE.--- 
(a) Sources that have possible counterparts in at least one of 
the cross-matched catalogs but have no classification.
(b) Sources that have no counterparts in the cross-matched catalogs.
}
\end{deluxetable}

\subsection{Probability of Chance Coincidence}\label{sec:chan_coin}

The above cross-matching is based only on the separation angles between the \maxi and other 
reference sources, and some counterparts are incorrectly identified because they just coincidentally 
fall within the positional error of a third \maxi source. We derive the expected number of such events by 
combining integrated positional errors of the \maxi sources used for each cross-matching and 
source number densities (i.e., source number divided by our survey area) of the reference catalogs.
We find that 10, 1, 2, 11, 11, and 7 \maxi sources may be misidentified in referring to BAT105, 4U, XTEASMLONG,
MCXC, XMMSL2, and 1RXS, respectively. Note that due to incomplete all-sky survey of XMMSL2, its 
derived value should be regarded as a lower limit. 
If we use only MCXC and \swift/BAT 105-month catalogs, similarly to the
case of \citet{Hir13}, who used MCXC and the \swift/BAT 70-month catalog \citep{Bau13}, 
the expected fraction of the false identification is $\sim$ 0.03. This remains at the same level as 
in the 37-month catalog, in spite of the increased source number in the 7-year \maxi and 
\swift/BAT 105-month catalogs. This small value is likely because the systematic errors 
of the PSF model are reduced in our catalog.

\subsection{Systematic Positional Error}\label{sec:sys_err}

\begin{figure}
\includegraphics[scale=0.3,angle=-90]{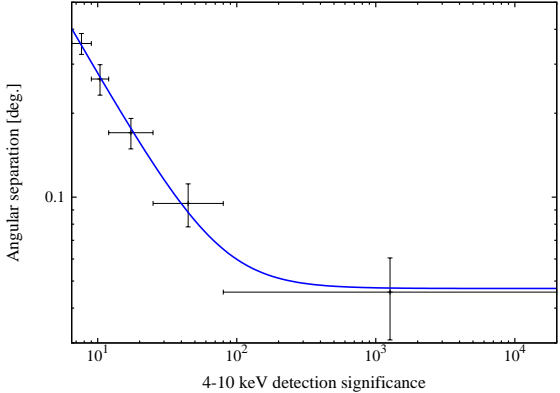}
\caption{\label{fig:ang}
Correlation between the detection significance ($s_{\rm D,4-10~keV}$) 
and the 90\% confidence limit of the separation angles
between the \maxi\ sources and their possible counterparts.
The data (black crosses) can be described as 
$\sqrt{(A/s_{\rm D,4-10~keV})^B + C^2}$ (blue solid line), where 
$A = 2.27\pm0.05$, $B=1.74\pm0.02$, and $C=0.047\pm0.001$.
}
\end{figure}

Using the results in Section~\ref{sec:cross-match}, we examine 
the separation angles between the \maxi sources and their possible
counterparts as a function of the detection significance ($s_{\rm
D,4-10~keV}$). Figure~\ref{fig:ang} plots the 90\% confidence limit of
the separation angles, estimated in each significance bin. 
For consistency, we fit the data with the same equation as that in \citet{Hir13}, $\sqrt{(A/s_{\rm D,4-10~keV})^B + C^2}$.
The best-fit parameters are $A = 2.27\pm0.05$, $B=1.74\pm0.02$, and $C=0.047\pm0.001$, suggesting 
the systematic error at a 90\% confidence limit of $\approx0.^\circ047$. 
The corresponding 1$\sigma$ error is $\sim 0.^\circ03$. 
The systematic error is successfully 
reduced from the previous works \citep[$\approx0.^\circ05$; ][]{Hir11,Hir13}.


\section{Statistical Properties of Our Catalog}\label{sec:result}

\subsection{Correlations among X-ray Properties}\label{sec:cor}

\begin{figure*}
\includegraphics[scale=0.24,angle=-90]{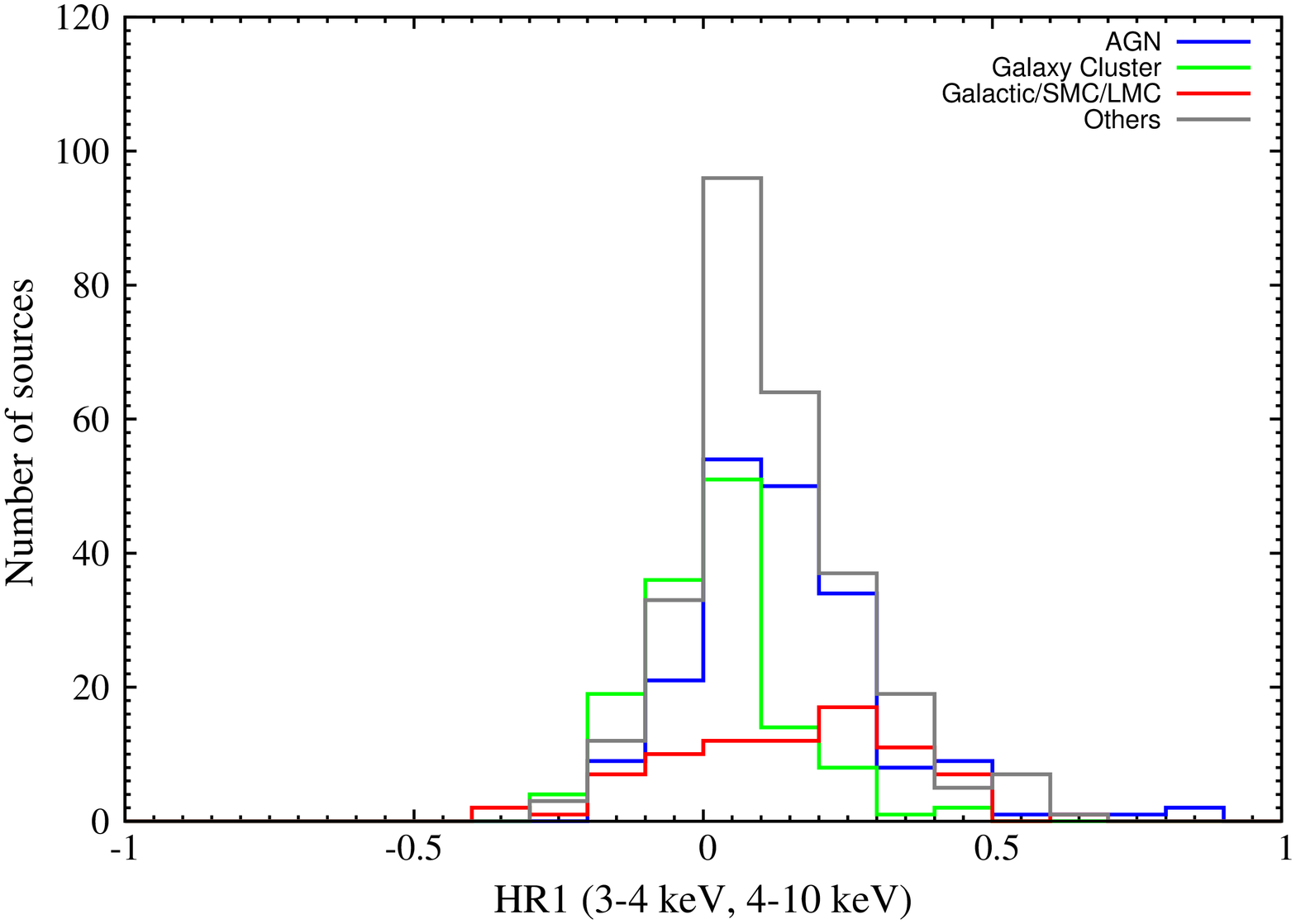}\hspace{-0.1cm}
\includegraphics[scale=0.24,angle=-90]{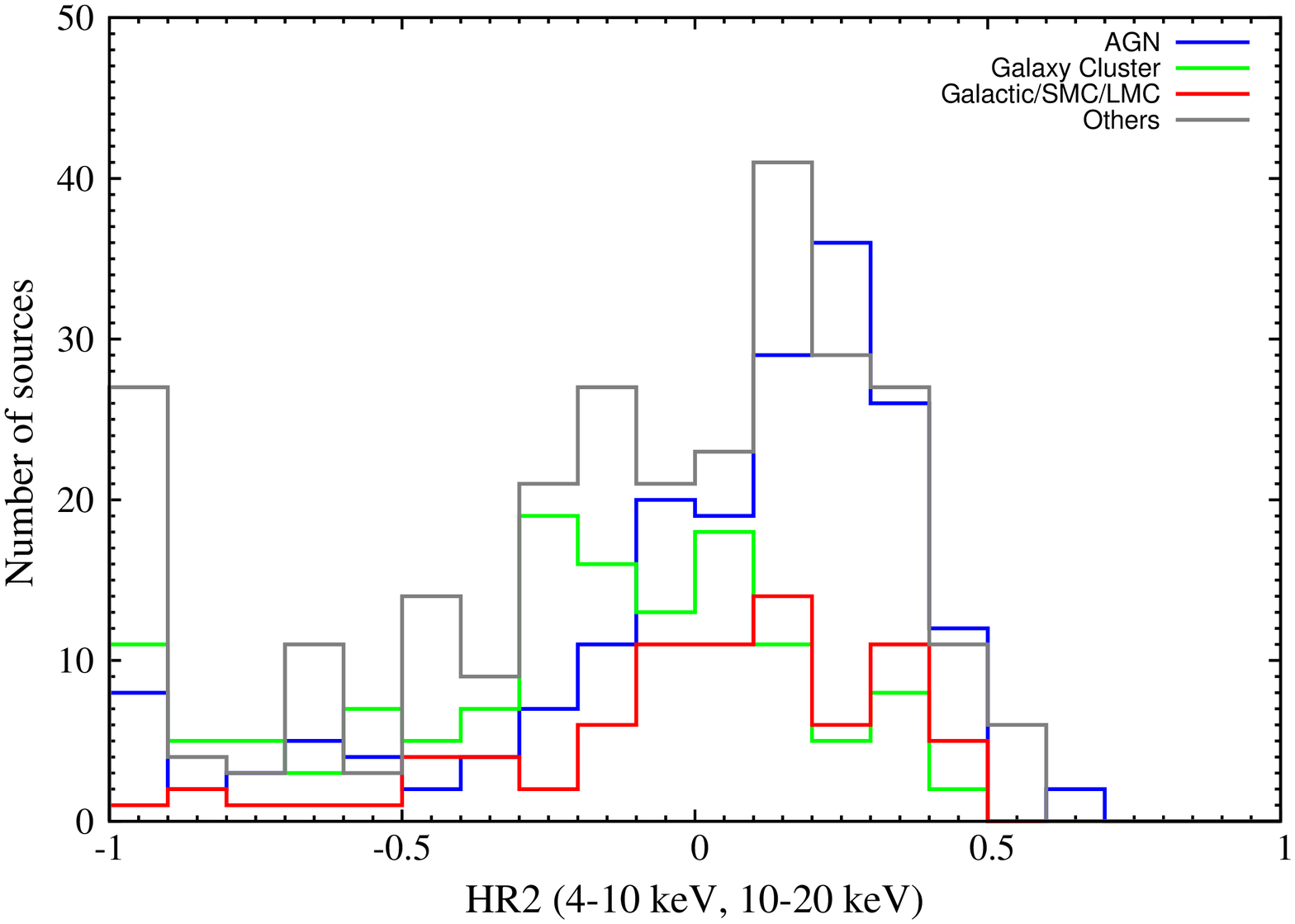}\hspace{-0.1cm}
\includegraphics[scale=0.24,angle=-90]{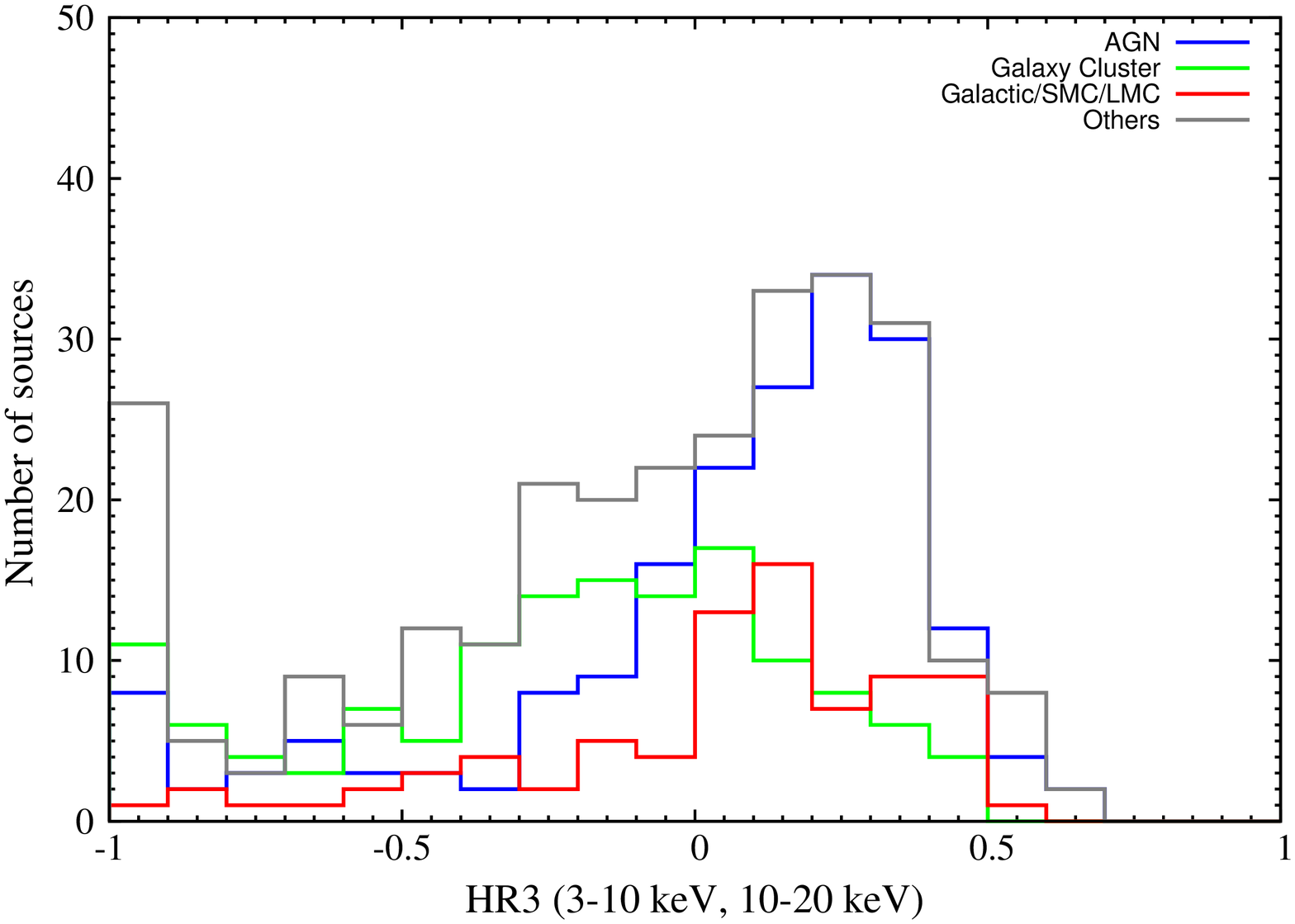}
\caption{\label{fig:hr_hist}
Histograms of the HRs. Here the \maxi sources are categorized into four types: 
AGN (blue), galaxy cluster (green), Galactic/SMC/LMC source (red), and others (gray). 
}
\end{figure*}

\begin{figure*}
\includegraphics[scale=0.24,angle=-90]{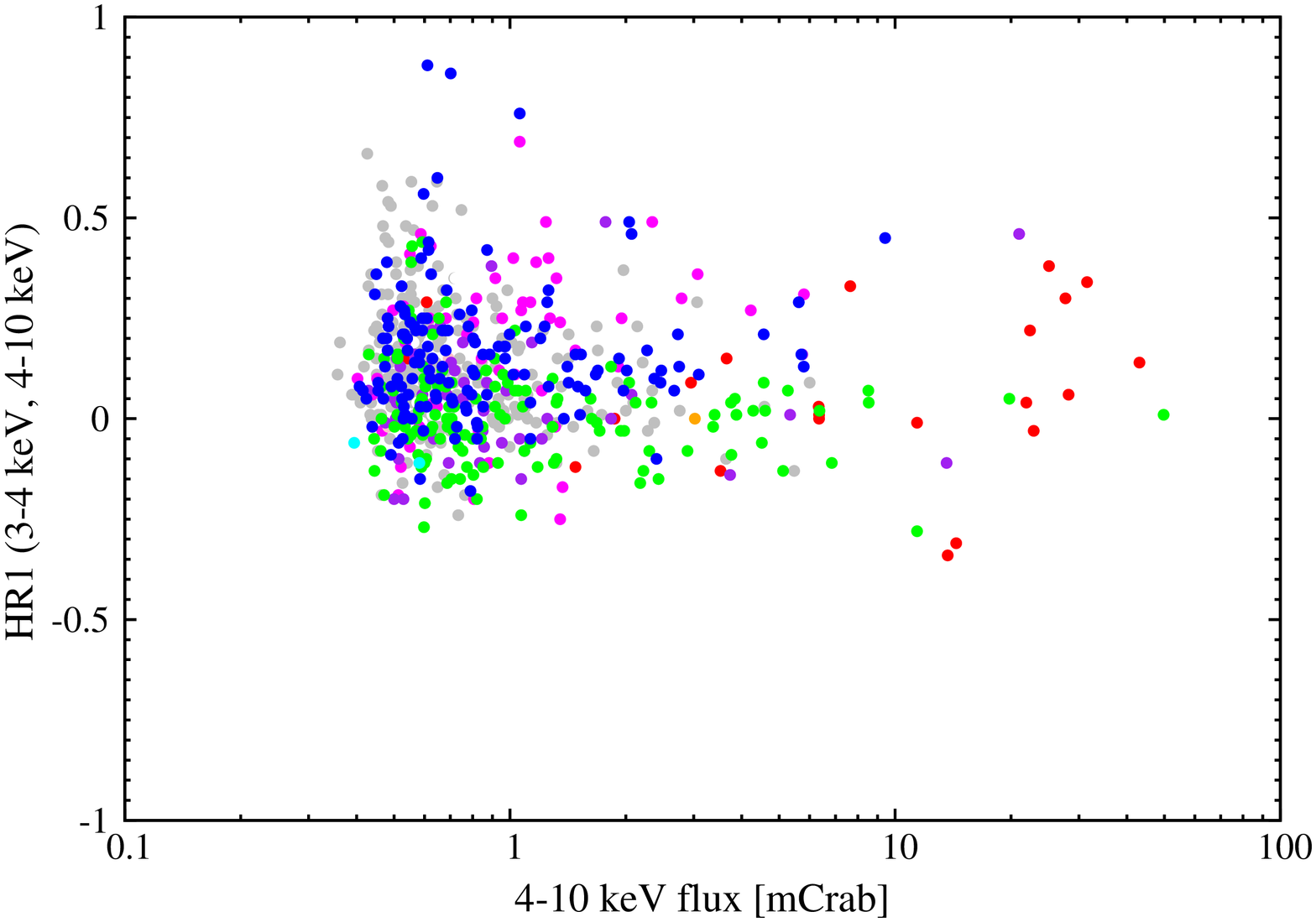}\hspace{-0.1cm}
\includegraphics[scale=0.24,angle=-90]{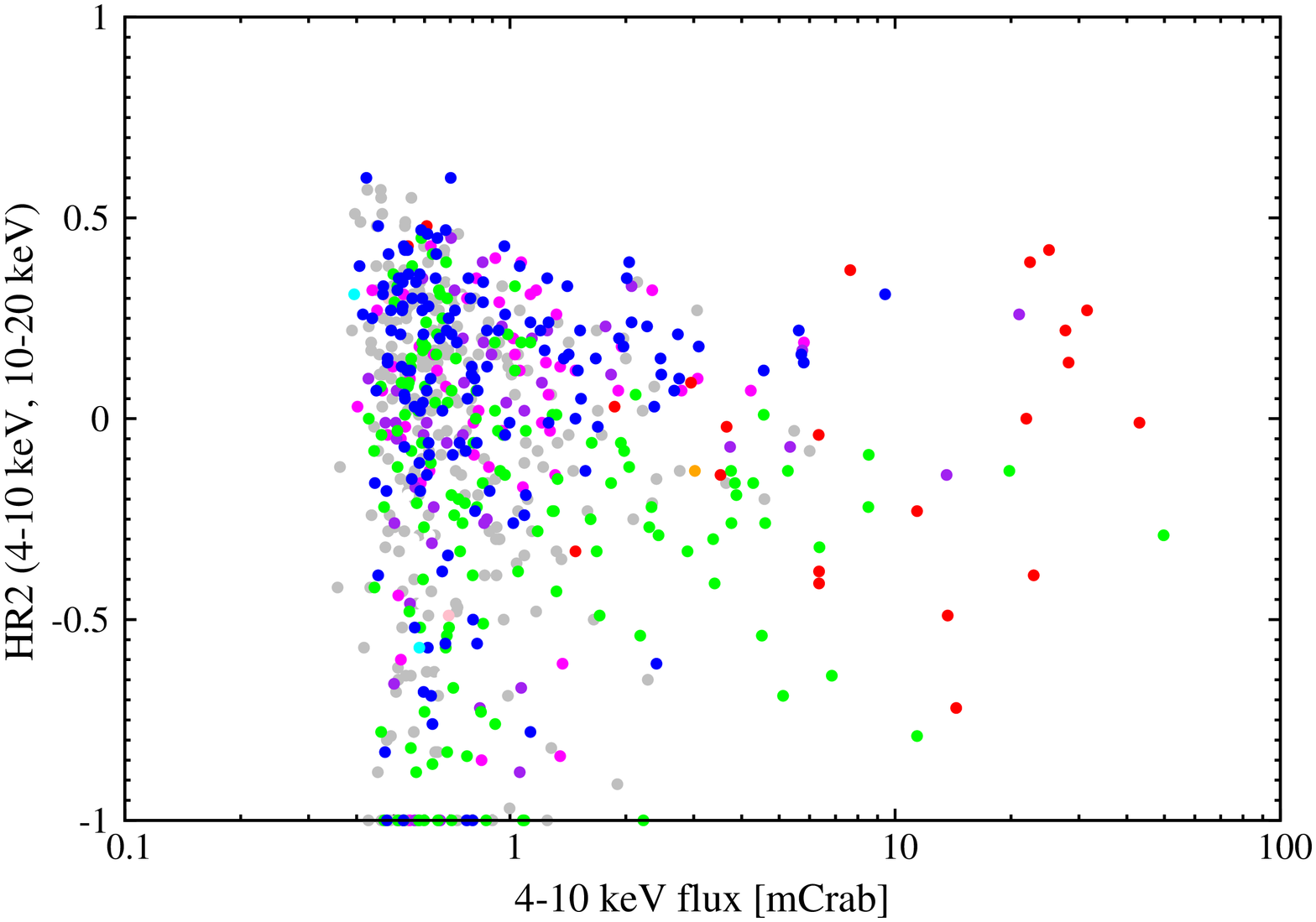}\hspace{-0.1cm}
\includegraphics[scale=0.24,angle=-90]{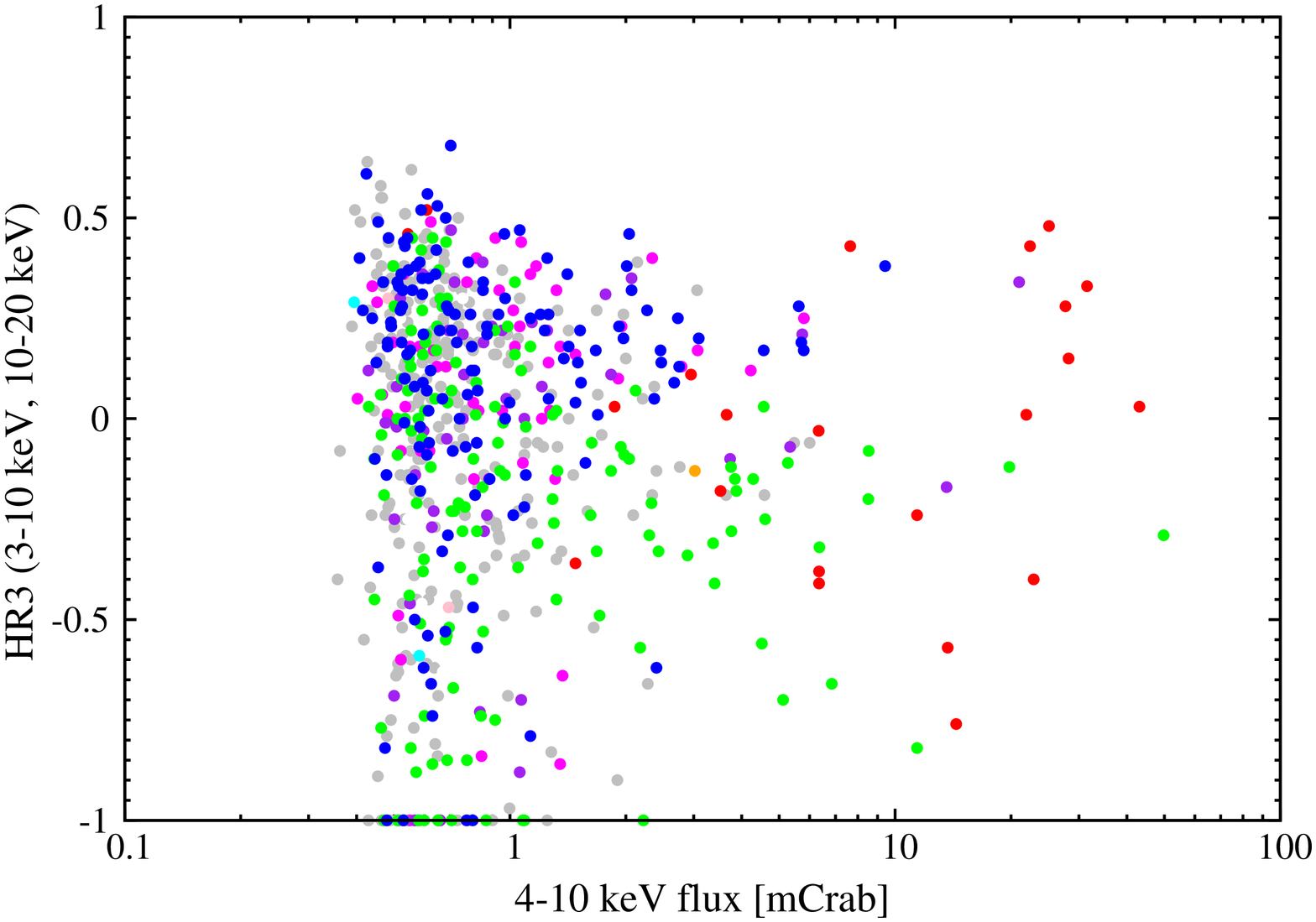}
\caption{\label{fig:f410_vs_hr}
Scatter plots between the HRs and the 4--10 keV flux. 
The color definition is the same as in Figure~\ref{fig:allsky-map}. 
}
\end{figure*}

\begin{figure*}
\includegraphics[scale=0.24,angle=-90]{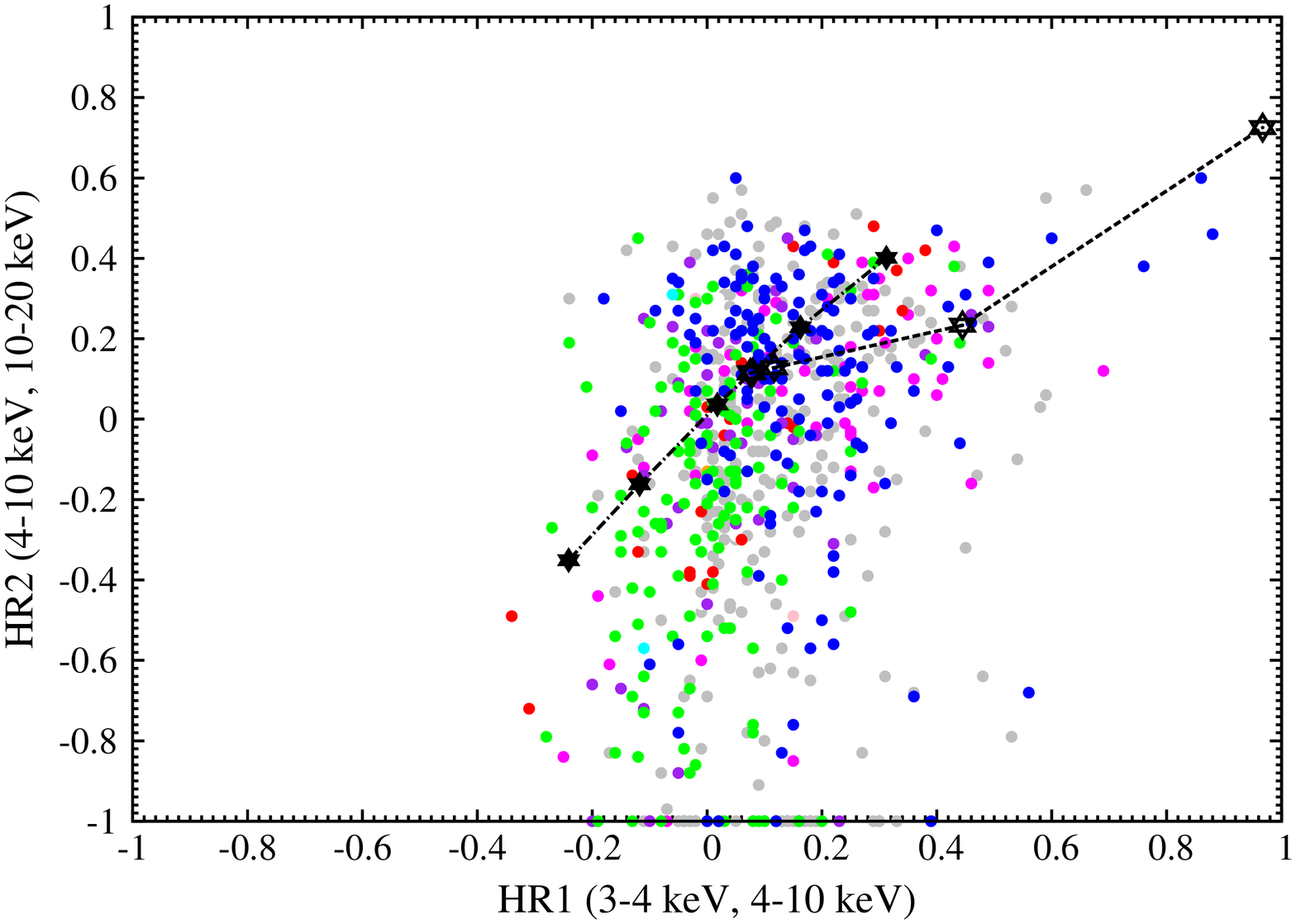}\hspace{-0.2cm}
\includegraphics[scale=0.24,angle=-90]{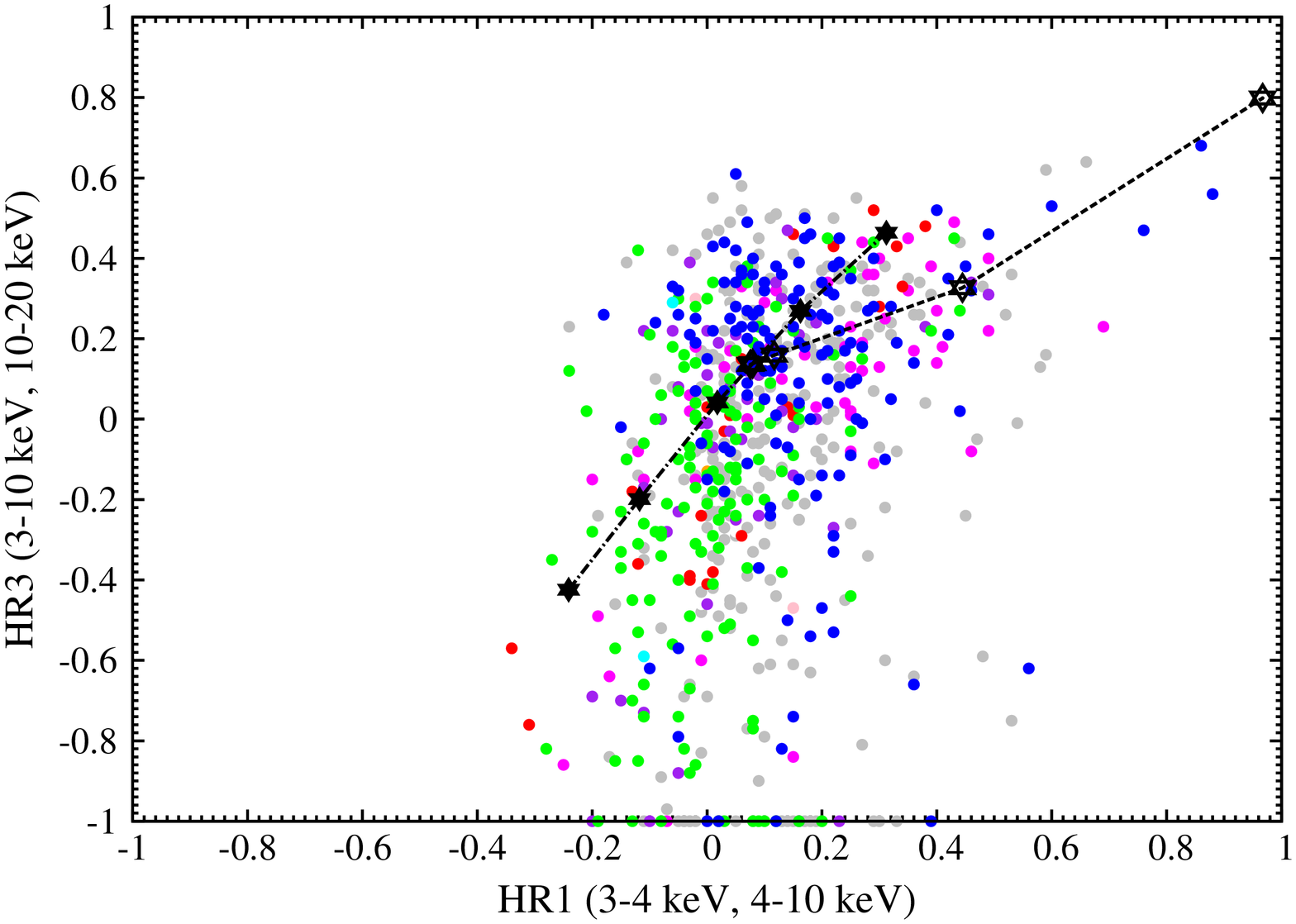}\hspace{-0.2cm}
\includegraphics[scale=0.24,angle=-90]{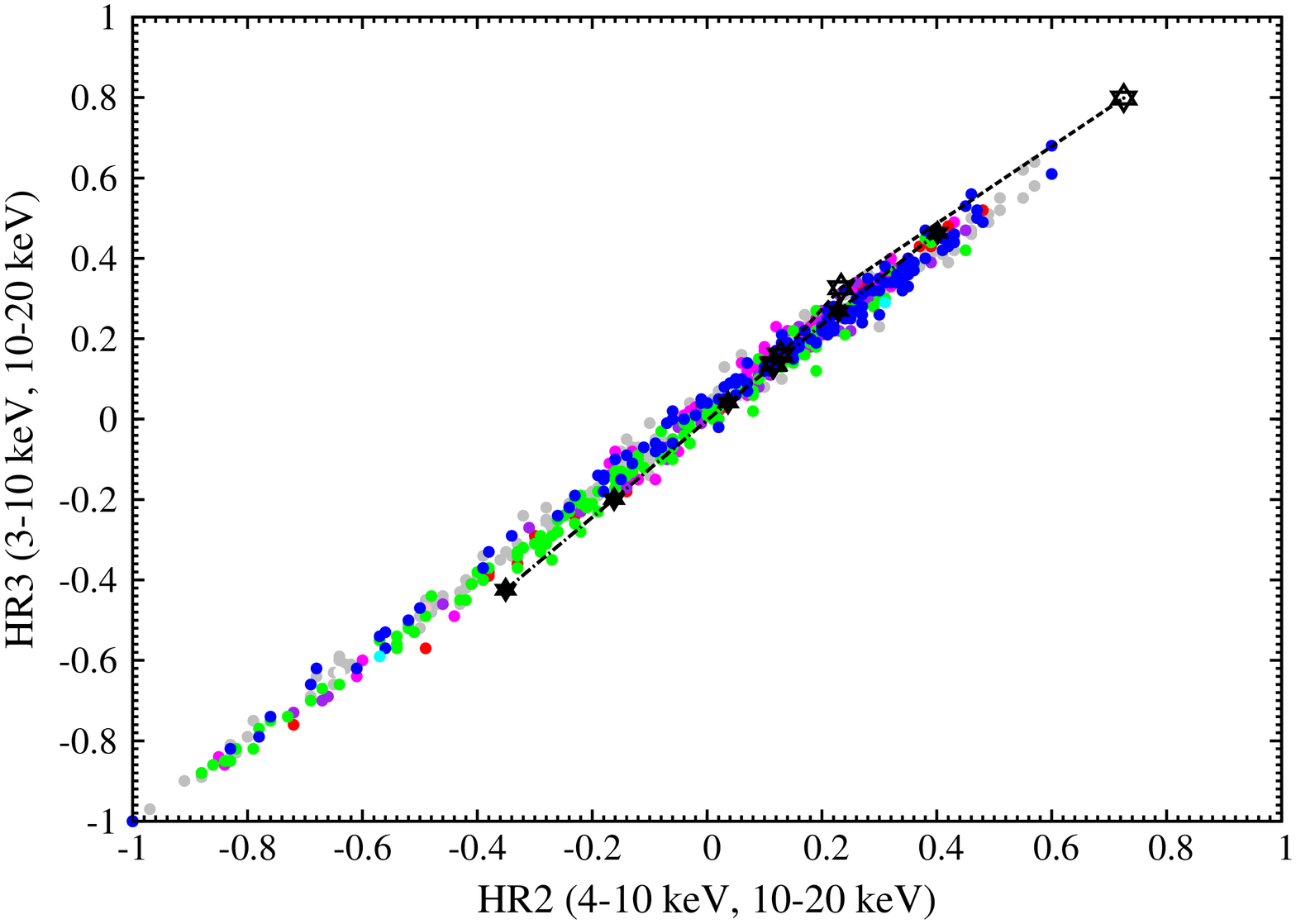}
\caption{\label{fig:colcol}
X-ray color-color plots among the three HRs. The filled stars and
dot-dashed lines illustrate the HR track for 
a power-law model without intrinsic absorption. 
The stars from upper right to lower left correspond to
photon indices from 1.0 to 3.0 with a step of 0.5.
The open stars and dashed lines present the track for 
an absorbed power-law model with a photon index of 1.8.  The
open stars from lower left to upper right correspond to $\log N_{\rm H}$/cm$^{-2}$ from
20.0 to 24.0 with a step of 1.0
The color definition is the same as those in Figure~\ref{fig:allsky-map}. 
}
\end{figure*}

\begin{figure*}
\begin{center}
\includegraphics[scale=0.32,angle=-90]{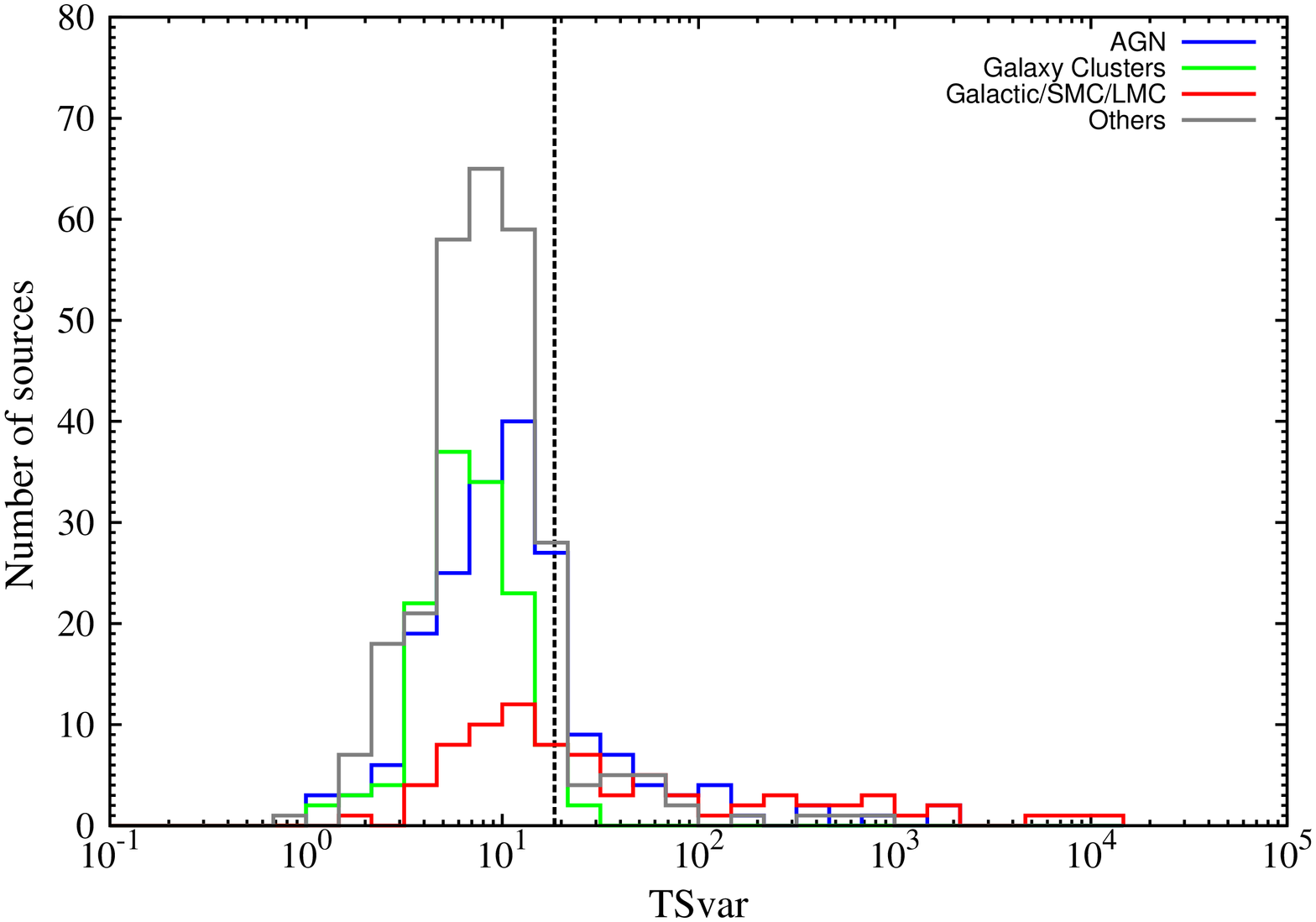}
\includegraphics[scale=0.32,angle=-90]{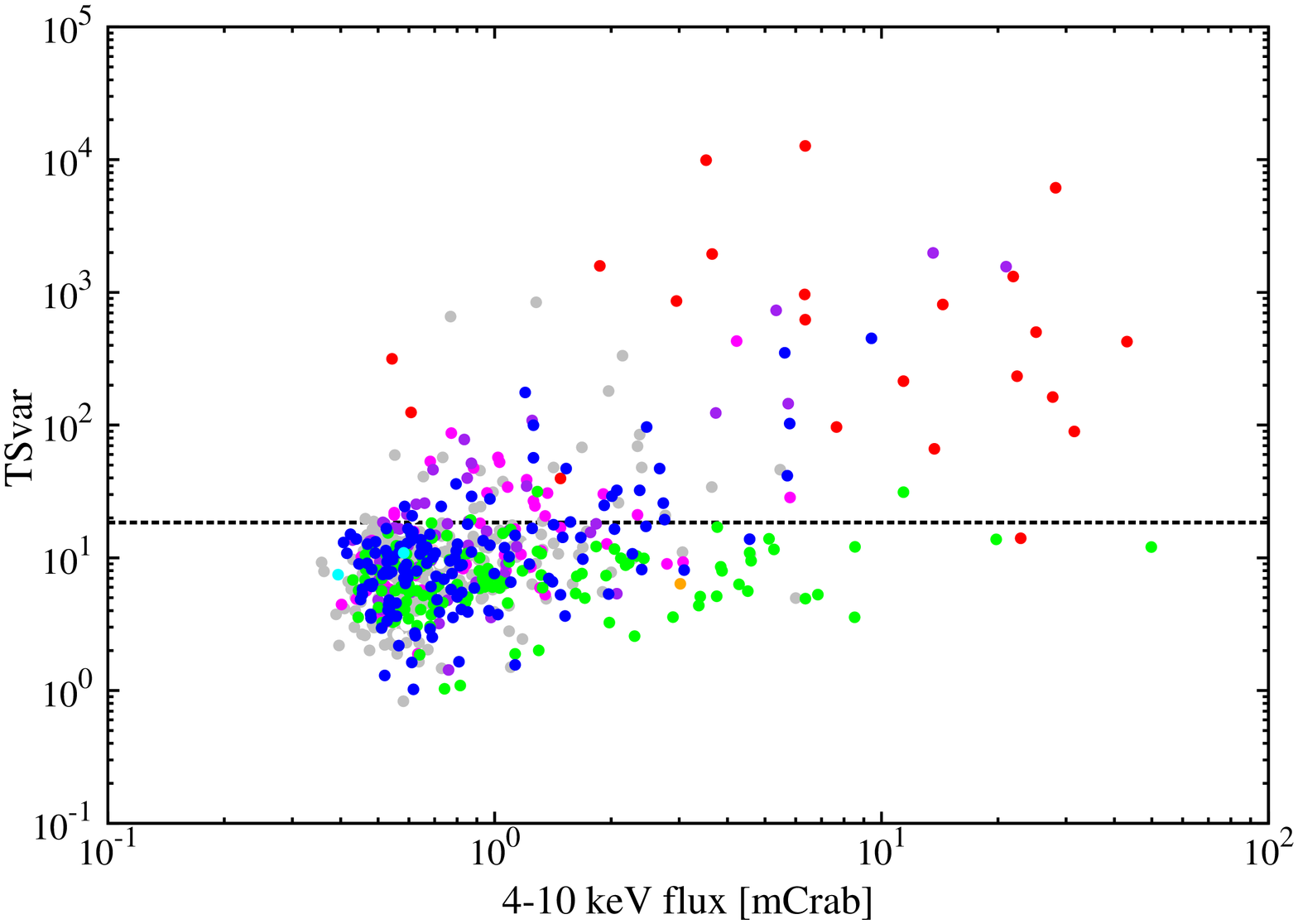}
\caption{\label{fig:tsvar}
Left:
Histogram of the time variability index (TS$_{\rm var}$).
The color definition is the same as that in Figure~\ref{fig:hr_hist}. 
Right: Scatter plot between the 4--10 keV flux and TS$_{\rm var}$.
The colors are the same as in Figure~\ref{fig:allsky-map}. 
In both figures, variable sources are located in 
the area of TS$_{\rm var} > 18.48$ (dashed line), 
corresponding to a 99\% confidence limit. 
}
\end{center}
\end{figure*}

\begin{figure*}
\includegraphics[scale=0.24,angle=-90]{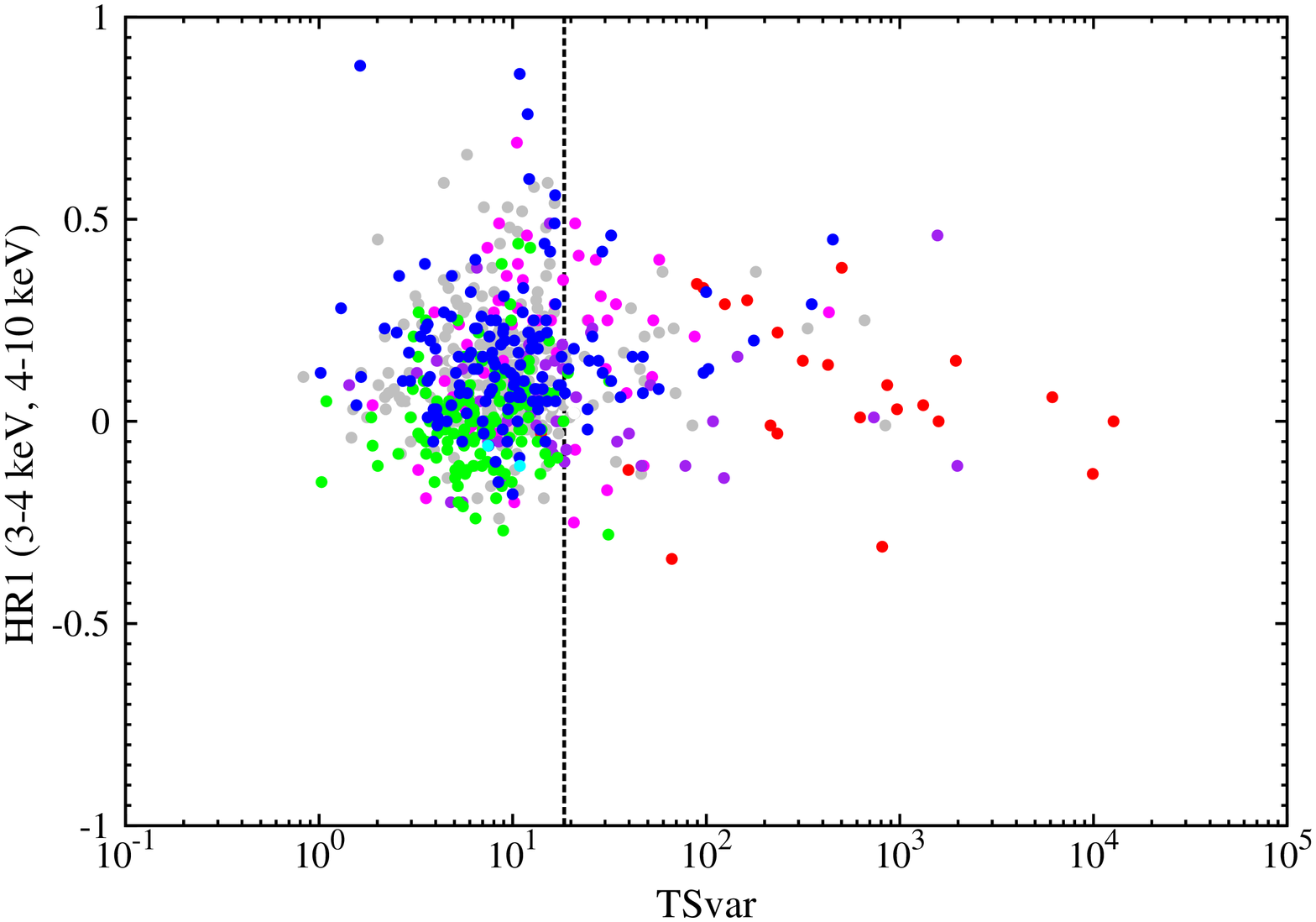} 
\includegraphics[scale=0.24,angle=-90]{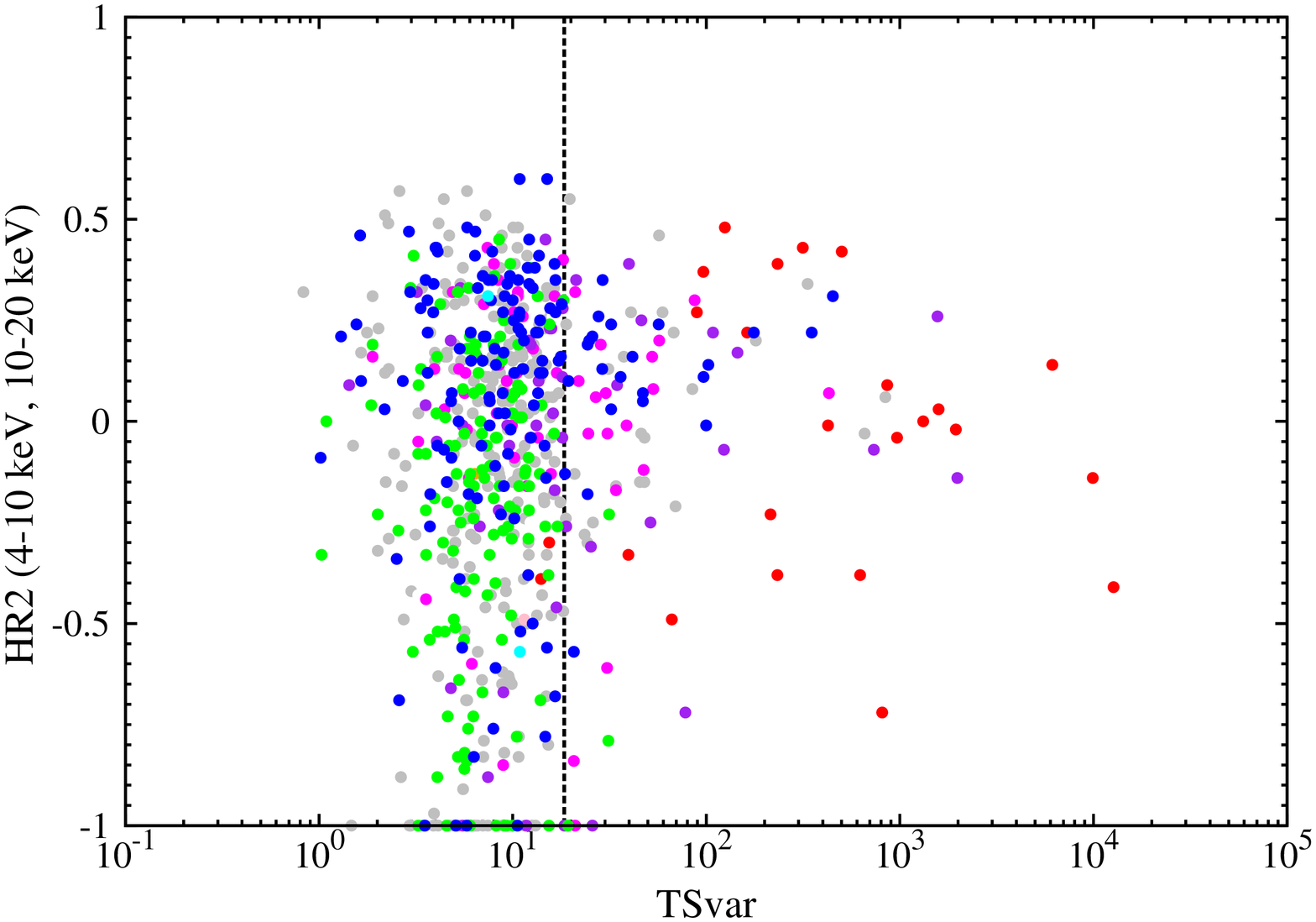}
\includegraphics[scale=0.24,angle=-90]{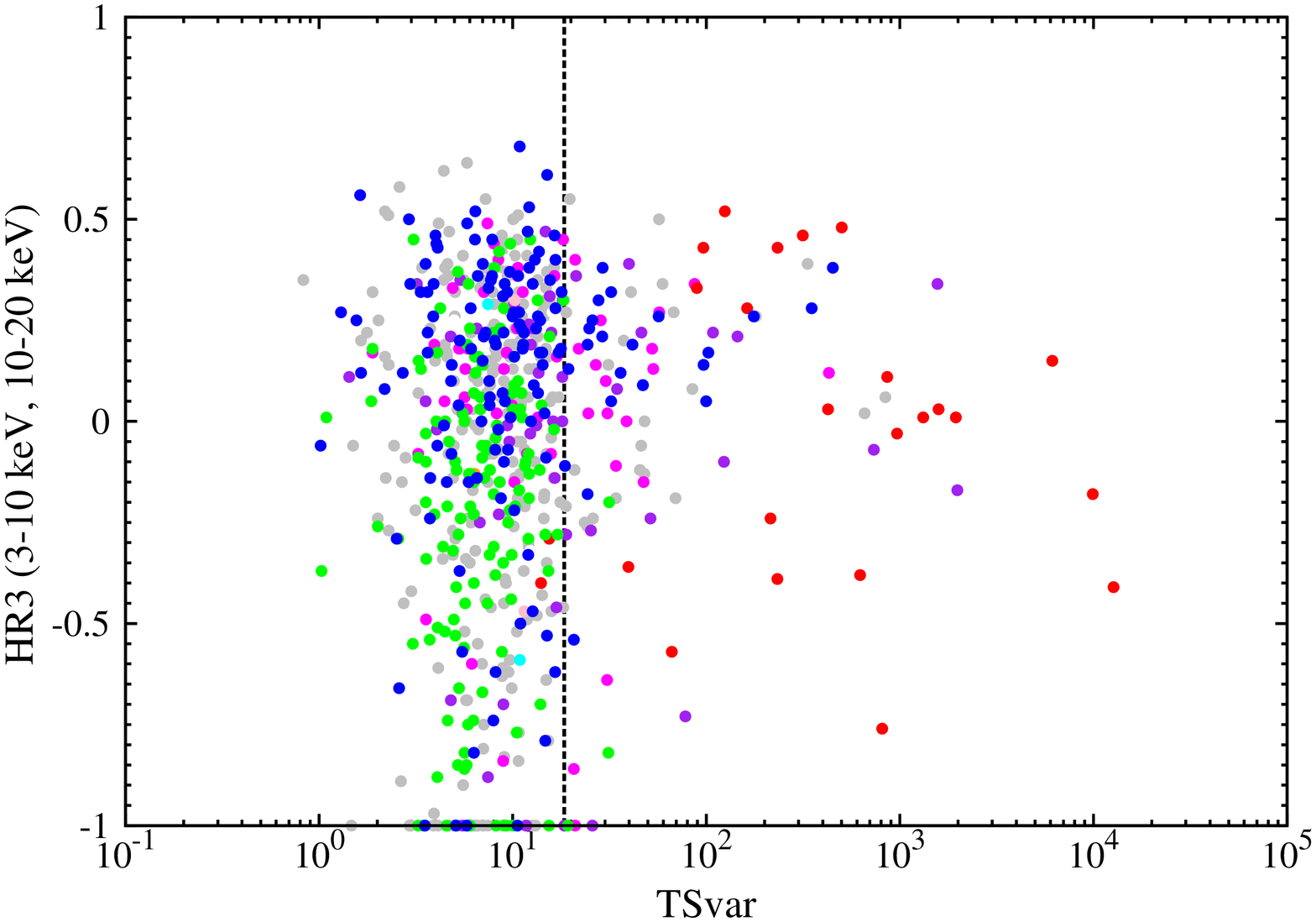}
\caption{\label{fig:tsvar2hr}
Scatter plots between the HRs and TS$_{\rm var}$. 
The color are the same as in Figure~\ref{fig:allsky-map}. 
}
\end{figure*}

In this section, we describe basic properties of our catalog based on the parameters constrained 
so far: the fluxes, the HRs, and the time variability index (TS$_{\rm var}$). Here, we 
  simply categorize the third \maxi sources into four types: AGN (i.e., Seyfert galaxies and blazars), 
  galaxy cluster, Galactic/Small/Large Magellanic Cloud (SMC/LMC) source (i.e., cataclysmic variables/stars,
  X-ray binaries, and pulsars), and the others.
The HR distributions are shown in Figure~\ref{fig:hr_hist}, 
and the scatter plots between the HRs and the 4--10 keV fluxes in Figure~\ref{fig:f410_vs_hr}. A noticeable
point is that the HRs of AGNs are typically harder than those of the galaxy clusters. The difference is more 
clearly seen in the HRs using the 10--20 keV band. This is likely to be a natural consequence of the 
difference in their typical spectra; galaxy 
clusters are bright in soft X-ray bands by optically-thin thermal
emission, whereas AGNs exhibit harder spectral profiles, roughly
described with a power-law component with a photon index of $\sim
1.4$--$2.4$ \citep[e.g.,][]{Ued14,Kaw16b}. Strong absorption, often seen
in AGNs, even more hardens their spectrum, and thereby higher HRs are
expected. A more detailed view can be obtained from X-ray color-color
plots in Figure~\ref{fig:colcol}, which overplots HR tracks predicted
from a single (absorbed) power-law model based on
Figure~\ref{fig:conversions}.

Figure~\ref{fig:tsvar} shows the histogram of the time variability index (TS$_{\rm var}$) and its scatter plot against 
the 4--10 keV flux. Also, Figure~\ref{fig:tsvar2hr} plots the HRs with respect to TS$_{\rm var}$. 
Most of the galaxy clusters have no significant flux variation, with TS$_{\rm var} < $ 18.48, but 
there are two exceptions showing moderate TS$_{\rm var}$ values of $\approx 30$. 
One is 3MAXI J1231$+$122 identified as Virgo cluster, and this apparent variability may be due to 
uncertainties associated with our PSF model, which assumes a point source, and/or the presence of 
AGN in M87 \citep{Won17}. 
The other source, 3MAXI J1702$+$339 is located near the bright, variable source Her X-1, and 
the part of its flux could be contaminated by Her X-1, causing the fake variability. 

%

\subsection{Comparison with \swift/BAT 105-month Catalog}\label{sec:cor}

\begin{figure}
\includegraphics[scale=0.32,angle=-90]{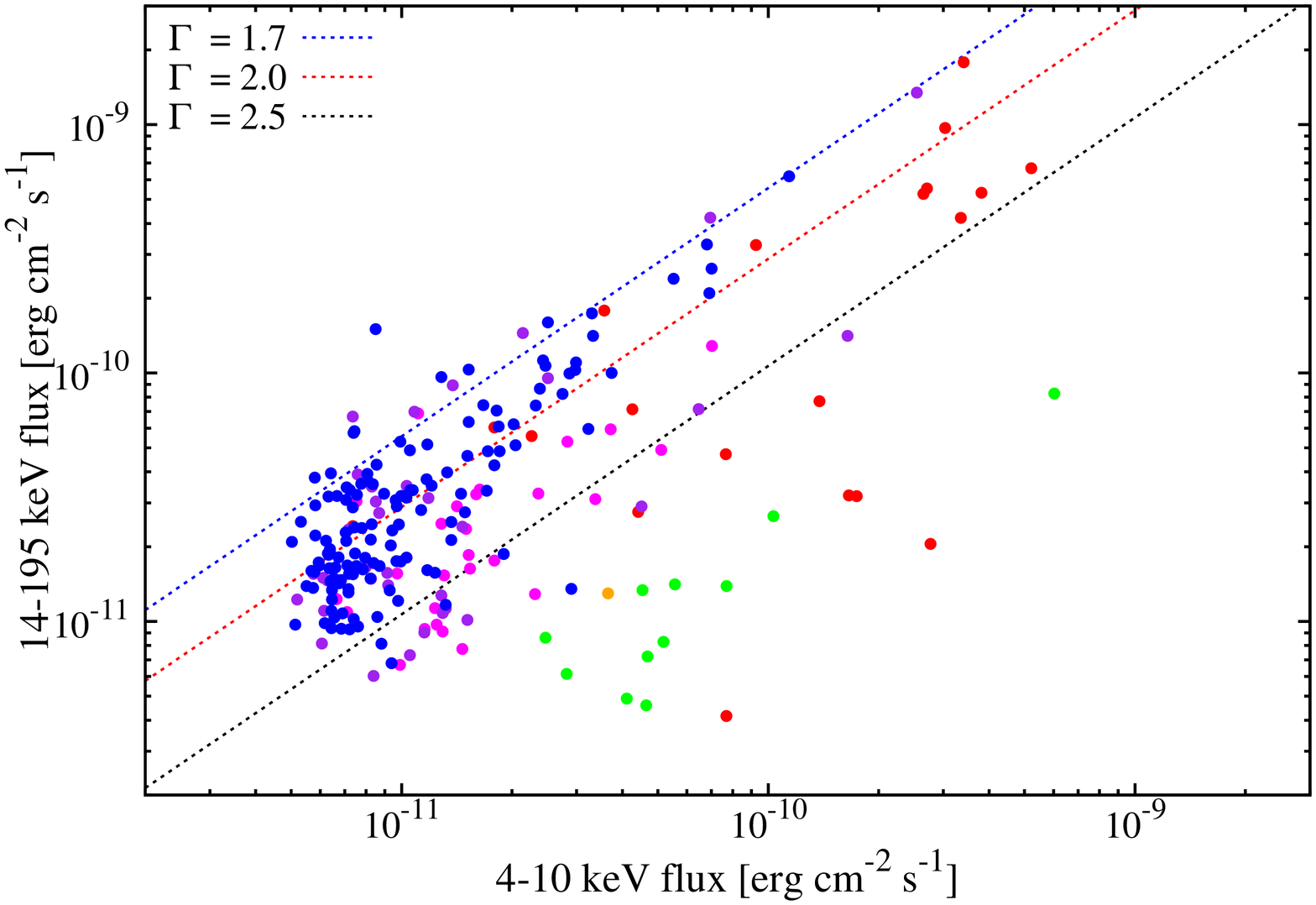}
\caption{\label{fig:f410_vs_f14195}
Scatter plot between the 4--10 keV and 14--195 keV fluxes 
for \maxi\ sources having a single counterpart in 
the \swift/BAT 105-month catalog.
The 4--10 keV fluxes are calculated 
with a conversion factor $1.21\times10^{-11}$ \ergcms mCrab$^{-1}$, 
while the 14--195 keV ones are retrieved from the \swift/BAT 105-month catalog \citep{Oh18}. Dashed 
lines present the relations
for an unabsorbed power-law spectrum with different photon indices:
$\Gamma = $ 2.5 (black), $\Gamma = $ 2.0 (red), and $\Gamma = $ 1.7 (blue). 
}
\end{figure}

Here we briefly view the nature of the third \maxi sources that have a single counterpart in the BAT105. 
Figure~\ref{fig:f410_vs_f14195} plots the correlation between the 4--10 keV and 14--195 keV 
fluxes, together with the flux ratios predicted from unabsorbed power-law models with three 
different photon indices ($\Gamma =$ 1.7, 2.0, and 2.5). Note that we use the 4--10 keV fluxes
estimated through the conversion factor of $1.21\times10^{-11}$ \ergcms mCrab$^{-1}$,  
while the 14--195 keV fluxes are retrieved from \citet{Oh18}. 
As seen in Fig.~\ref{fig:f410_vs_f14195}, most of the \maxi sources 
have soft spectra, corresponding to photon indices larger than 1.7. 

\subsection{$\log$N--$\log$S Relation}\label{sec:lognlogs}

\begin{figure}
  \begin{center}
\includegraphics[scale=0.68,angle=-0]{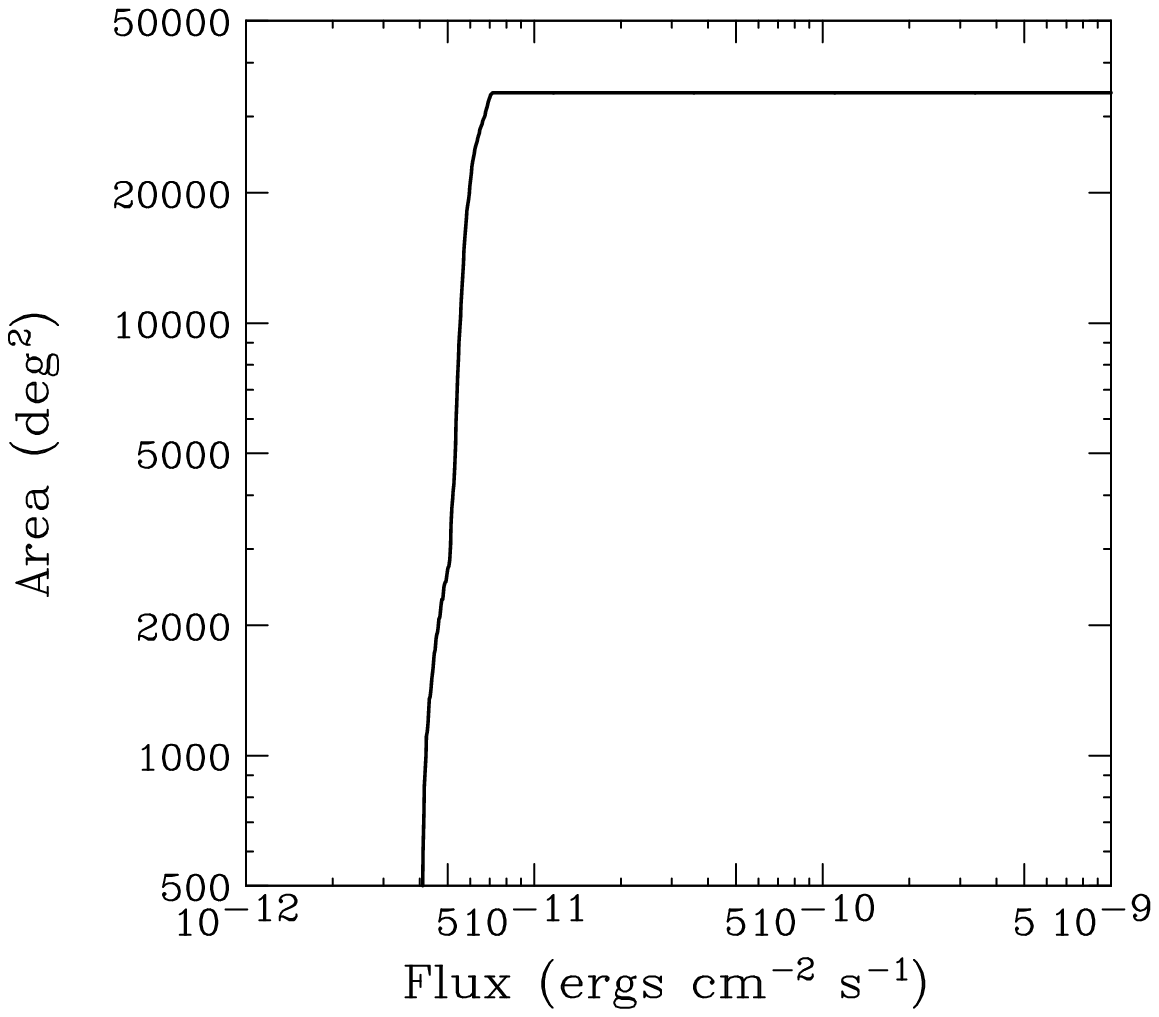}\\ \vspace{.3cm}
\includegraphics[scale=0.68,angle=-0]{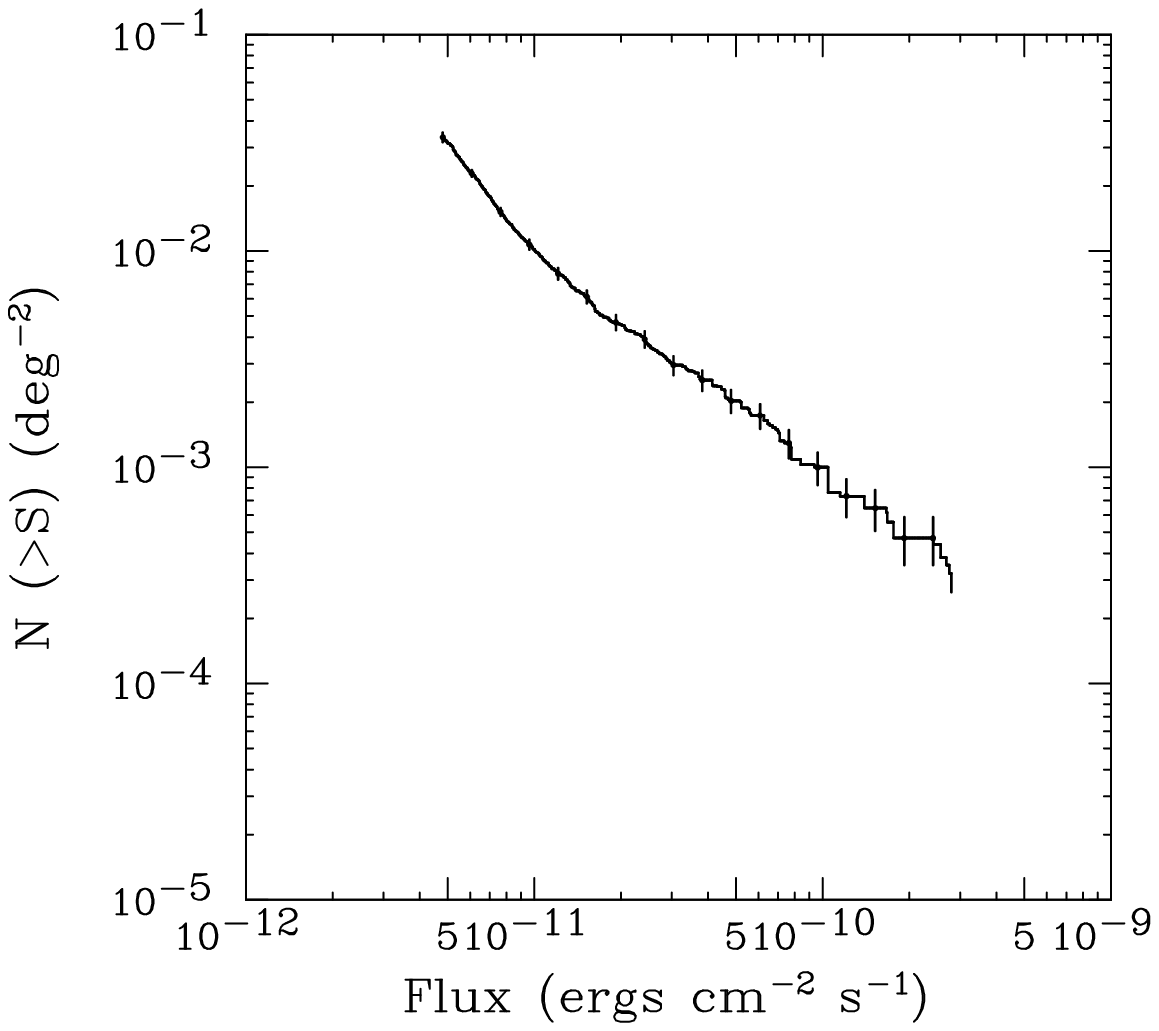}
\caption{\label{fig:lognlogs}
(Upper panel) Survey area plotted against 4--10 keV flux for the MAXI/GSC 7-year survey at $|b|>10^\circ$. 
(Lower panel) \lognlogs relation in the integral form. 
}
\end{center}
\end{figure}

We present the \lognlogs relation, the cumulative flux distributions of the sources detected 
at $s_{\rm D,4-10~keV} \geq $ 6.5. 
To derive it, we estimate the sensitivity at every sky position by
combining the background counts and the effective exposure map, 
given in Figure~1 of \citet{Hor18}.
The effective exposure map, in units of cm$^2$ s, is 
estimated via simulation of the cosmic X-ray background, with uniform brightness 
(see \citet{Hor18} for more details). 
The survey area as a function of the sensitivity 
(i.e., the area curve) is displayed in the upper panel of Figure~\ref{fig:lognlogs}. 
From this curve, we can find that our survey is complete, in the half of the high Galactic-latitude 
sky ($|b|> 10^\circ$), down to $5.9\times10^{-12}$ \ergcms, superior to the 37-month catalog 
\citep[i.e., $7.5\times10^{-12}$ \ergcms ;][]{Hir13}. The deepest sensitivity of $\sim 4\times10^{-12}$ 
\ergcms~is consistent with the lowest flux among the actually detected sources. 

We divide the number of sources with fluxes of $S \sim S+dS$
by the survey area at $S$, and obtain the \lognlogs
relation in the differential form. 
By integrating it, we obtain the 
number density $N (>S)$ of total sources having fluxes above $S$
(the lower panel of Figure~\ref{fig:lognlogs}).
At fluxes lower than $8\times10^{-11}$ ergs cm$^{-2}$
s$^{-1}$, the \lognlogs relation has a slightly steeper slope than that expected for sources uniformly distributed in
a static Euclidean universe (i.e., $N (>S) \propto S^{-1.5}$). 
This may be partially caused by the Eddington bias
(i.e., fainter sources than the sensitivity limit emerge due to
statistical fluctuation).

\section{Summary}\label{sec:sum}

We have provided the third \maxi/GSC X-ray source catalog in the
high-latitude sky ($b>$10$^\circ$) utilizing the first 7-year (2009
August 13 to 2016 July 31) data in the 4--10 keV band. The catalog
contains 682 X-ray sources detected at $s_{\rm D,4-10~keV} \geq
6.5$. Over the half of the survey area, the 4--10 keV band sensitivity
reaches $\sim 5.9\times 10^{-12}$ \ergcms, which is the highest ever
achieved as an all-sky X-ray survey in a similar energy band. 
The source number has increased by a factor of $\approx$1.4 compared
with the 37-month \maxi/GSC catalog \citep{Hir13}, which adopted 
$s_{\rm D,4-10~keV} \geq 7$ as the detection threshold.
We have cross-matched the cataloged sources with those in the other
X-ray catalogs (BAT105, 4U, XTEASMLONG, MCXC, XMMSL2, and 1RXS), and
found the counterparts of 422 sources.

In addition to the basic properties in the 4--10 keV band, we provide the 7-year averaged fluxes in 
the 3--4 keV, 10--20 keV, and 3--10 keV bands, and also their HRs. The HRs, particularly the ones 
using the 10--20 keV fluxes, seem to be the most helpful to discriminate between AGNs and 
galaxy clusters. We have also built the database of the 1-year-bin 4--10 keV lightcurves, with 
their variability index (TS$_{\rm var}$) and the excess variance ($\sigma_{\rm rms}^2$). 
We find that the compact objects, such as X-ray binaries and AGNs, show strong variability 
(TS$_{\rm var} > $18.48), demonstrating that the variability is a key to uncover the nature 
of unidentified X-ray sources. 
Future follow-up observations with a higher spatial and timing resolution are encouraged to
associate the unidentified sources in our catalog with those in other catalogs, for improving 
the completeness of the catalog so that it can be used in statistical studies.

We have searched the 1-year time-sliced data for transient sources
(Section~\ref{sec:trans_obj}), and found four objects that are not detected from 
the 7-year accumulated data (Table~\ref{tab:trans_catalog}). They are indeed 
classified to transient compact objects (i.e., X-ray binary, pulsar, and AGN/tidal 
disruption event). 

\acknowledgments

We thank Kyuseok Oh for providing the Swift/BAT 105-m catalog prior to the official
publication. Part of this work was financially supported by the Grant-in-Aid for JSPS
Fellows for young researchers (TK and TH), for Scientific Research
17H06362 (YU, NK, TM, HN, and AY) and 17K05384 (YU), and for Young
Scientists (B) 16K17672 (MS). MS acknowledges support by the Special
Postdoctoral Researchers Program at RIKEN. This research has made use of
{\it MAXI} data provided by RIKEN, JAXA and the {\it MAXI} team.

\appendix
\setcounter{table}{0}
\setcounter{figure}{0}
\def\thesection{\Alph{section}}
\def\thetable{\Alph{table}}
\def\thefigure{\Alph{figure}}
\setlength{\topmargin}{1.3cm}
 \begin{longrotatetable}
  \clearpage
\renewcommand{\arraystretch}{1.1}
\begin{deluxetable*}{p{0.2cm}cccp{-0.5cm}cp{-0.5cm}cp{-0.5cm}cp{-0.5cm}ccccp{-0.6cm}ccccccccccc}
  \thispagestyle{empty}
  \tablefontsize{\footnotesize}
  \tabletypesize{\scriptsize}
  \tablecaption{X-ray properties of sources in the third MAXI/GSC catalog. \label{tab:catalog1}}
  \tablehead{ \colhead{(1)} 
& \colhead{(2)} & \colhead{(3)} & \colhead{(4)} & \colhead{(5)} & \colhead{(6)} & \colhead{(7)} & \colhead{(8)} & \colhead{(9)}   & \colhead{(10)} & \colhead{(11)} & \colhead{(12)} & \colhead{(13)} & \colhead{(14)} & \colhead{(15)} & \colhead{(16)} & \colhead{(17)} \\ 
\colhead{No.} & \colhead{3MAXI} & \colhead{R.A.} & \colhead{Decl.} & \colhead{$\sigma_{\rm stat}$} & 
\colhead{$s_{\rm D, 4-10 keV}$} & \colhead{$f_{\rm 4-10 keV}$} & \colhead{$s_{\rm D, 3-4 keV}$} & \colhead{$f_{\rm 3-4 keV}$} & 
\colhead{$s_{\rm D, 10-20 keV}$} & \colhead{$f_{\rm 10-20 keV}$} & \colhead{$f_{\rm 3-10 keV}$} & 
\colhead{HR1} & \colhead{HR2} & \colhead{HR3} & \colhead{TS$_{\rm var}$} & \colhead{$\sigma^2_{\rm rms}$} 
  }
  \startdata
1 & J0001$-$684 & $0.251$ & $-68.463$ & 0.215 & $11.1$ & 7.74$\pm$0.70 & $9.8$ & 2.19$\pm$0.23 & $1.2$ & 3.00$\pm$2.60 & 9.93$\pm$0.74 & $0.07\pm0.07$ & $-0.29\pm0.40$ & $-0.27\pm0.40$ & 2.29 & $-0.045\pm0.037$ \\
2 & J0004$-$348 & $1.037$ & $-34.897$ & 0.161 & $7.8$ & 7.36$\pm$0.94 & $11.2$ & 2.87$\pm$0.26 & $2.0$ & 6.67$\pm$3.41 & 10.23$\pm$0.98 & $-0.09\pm0.08$ & $0.13\pm0.26$ & $0.10\pm0.26$ & 4.98 & $-0.047\pm0.059$ \\
3 & J0006+727 & $1.602$ & $72.742$ & 0.097 & $12.5$ & 10.39$\pm$0.83 & $16.1$ & 4.36$\pm$0.27 & $1.8$ & 5.94$\pm$3.25 & 14.75$\pm$0.87 & $-0.12\pm0.05$ & $-0.10\pm0.27$ & $-0.14\pm0.27$ & 10.26 & $0.043\pm0.054$ \\
4 & J0009+109 & $2.462$ & $10.938$ & 0.146 & $9.5$ & 8.53$\pm$0.90 & $7.9$ & 2.10$\pm$0.27 & $4.5$ & 15.75$\pm$3.48 & 10.63$\pm$0.94 & $0.14\pm0.08$ & $0.45\pm0.10$ & $0.47\pm0.09$ & 14.78 & $0.067\pm0.069$ \\
5 & J0010+323 & $2.621$ & $32.390$ & 0.177 & $7.3$ & 7.63$\pm$1.04 & $8.4$ & 2.58$\pm$0.31 & $...$ & 0.39($<$4.49) & 10.21$\pm$1.09 & $-0.02\pm0.09$ & $-0.86\pm1.33$ & $-0.86\pm1.32$ & 5.64 & $-0.071\pm0.080$ \\
6 & J0011$-$113 & $2.804$ & $-11.311$ & 0.089 & $12.7$ & 10.24$\pm$0.81 & $10.7$ & 2.48$\pm$0.23 & $...$ & 0.58($<$3.69) & 12.72$\pm$0.84 & $0.15\pm0.06$ & $-0.85\pm0.74$ & $-0.84\pm0.78$ & 8.93 & $0.009\pm0.030$ \\
7 & J0011$-$292 & $2.962$ & $-29.231$ & 0.137 & $9.6$ & 8.02$\pm$0.84 & $12.1$ & 2.93$\pm$0.24 & $2.1$ & 6.83$\pm$3.28 & 10.95$\pm$0.87 & $-0.06\pm0.07$ & $0.10\pm0.24$ & $0.08\pm0.24$ & 9.36 & $0.034\pm0.066$ \\
8 & J0012$-$313 & $3.132$ & $-31.351$ & 0.235 & $7.6$ & 6.46$\pm$0.85 & $5.5$ & 1.32$\pm$0.24 & $2.6$ & 8.49$\pm$3.29 & 7.78$\pm$0.88 & $0.23\pm0.11$ & $0.30\pm0.19$ & $0.35\pm0.18$ & 6.98 & $-0.020\pm0.061$ \\
9 & J0012$-$154 & $3.139$ & $-15.433$ & 0.177 & $7.2$ & 5.60$\pm$0.78 & $9.7$ & 2.16$\pm$0.22 & $1.5$ & 4.60$\pm$3.06 & 7.76$\pm$0.81 & $-0.08\pm0.09$ & $0.08\pm0.34$ & $0.06\pm0.34$ & 6.81 & $-0.038\pm0.097$ \\
10 & J0016$-$825 & $4.145$ & $-82.587$ & 0.171 & $7.2$ & 5.29$\pm$0.73 & $3.4$ & 0.82$\pm$0.24 & $2.2$ & 5.49$\pm$2.48 & 6.11$\pm$0.77 & $0.36\pm0.14$ & $0.19\pm0.23$ & $0.26\pm0.22$ & 5.02 & $-0.072\pm0.087$ \\
11 & J0021+286 & $5.319$ & $28.601$ & 0.119 & $10.1$ & 9.92$\pm$0.98 & $8.9$ & 2.58$\pm$0.29 & $2.0$ & 7.94$\pm$3.92 & 12.50$\pm$1.02 & $0.11\pm0.07$ & $0.07\pm0.25$ & $0.09\pm0.25$ & 10.20 & $0.076\pm0.075$ \\
12 & J0022$-$193 & $5.733$ & $-19.356$ & 0.150 & $10.6$ & 8.25$\pm$0.78 & $10.5$ & 2.39$\pm$0.23 & $3.7$ & 11.66$\pm$3.12 & 10.64$\pm$0.81 & $0.06\pm0.07$ & $0.34\pm0.13$ & $0.35\pm0.12$ & 5.37 & $-0.034\pm0.041$ \\
13 & J0027+075 & $6.762$ & $7.597$ & 0.193 & $7.4$ & 6.51$\pm$0.88 & $2.9$ & 0.74$\pm$0.25 & $...$ & 1.00($<$4.30) & 7.25$\pm$0.91 & $0.48\pm0.14$ & $-0.64\pm0.97$ & $-0.59\pm1.08$ & 9.65 & $0.024\pm0.098$ \\
14 & J0029+127 & $7.269$ & $12.736$ & 0.206 & $7.7$ & 7.09$\pm$0.93 & $5.6$ & 1.50$\pm$0.27 & $2.8$ & 10.05$\pm$3.58 & 8.59$\pm$0.97 & $0.21\pm0.11$ & $0.34\pm0.17$ & $0.38\pm0.16$ & 4.48 & $-0.059\pm0.075$ \\
15 & J0035$-$791 & $8.974$ & $-79.178$ & 0.121 & $13.0$ & 8.83$\pm$0.68 & $13.4$ & 3.03$\pm$0.23 & $3.8$ & 9.19$\pm$2.38 & 11.86$\pm$0.72 & $-0.02\pm0.05$ & $0.19\pm0.13$ & $0.19\pm0.13$ & 24.40 & $0.092\pm0.057$ \\
16 & J0039+231 & $9.850$ & $23.117$ & 0.170 & $6.6$ & 6.02$\pm$0.92 & $4.3$ & 1.21$\pm$0.27 & $2.4$ & 8.45$\pm$3.55 & 7.23$\pm$0.96 & $0.24\pm0.13$ & $0.33\pm0.20$ & $0.38\pm0.19$ & 3.41 & $-0.090\pm0.089$ \\
17 & J0041$-$093 & $10.455$ & $-9.318$ & 0.026 & $54.1$ & 46.94$\pm$0.87 & $54.2$ & 15.02$\pm$0.28 & $7.2$ & 22.56$\pm$3.12 & 61.96$\pm$0.91 & $0.01\pm0.01$ & $-0.19\pm0.07$ & $-0.18\pm0.07$ & 7.99 & $-0.000\pm0.002$ \\
18 & J0042+412 & $10.668$ & $41.238$ & 0.044 & $31.6$ & 36.63$\pm$1.16 & $32.4$ & 11.97$\pm$0.37 & $4.6$ & 19.84$\pm$4.30 & 48.60$\pm$1.22 & $0.00\pm0.02$ & $-0.13\pm0.11$ & $-0.13\pm0.11$ & 6.38 & $-0.004\pm0.005$ \\
19 & J0048+319 & $12.051$ & $31.900$ & 0.057 & $20.4$ & 21.47$\pm$1.05 & $7.9$ & 2.42$\pm$0.31 & $6.0$ & 24.01$\pm$4.04 & 23.89$\pm$1.09 & $0.49\pm0.05$ & $0.23\pm0.08$ & $0.31\pm0.08$ & 15.54 & $0.009\pm0.014$ \\
20 & J0048$-$732 & $12.057$ & $-73.205$ & 0.094 & $14.6$ & 17.22$\pm$1.92 & $13.9$ & 3.71$\pm$0.27 & $3.9$ & 11.07$\pm$2.83 & 20.93$\pm$1.94 & $0.21\pm0.06$ & $-0.04\pm0.14$ & $-0.00\pm0.14$ & 47.97 & $0.227\pm0.096$   
\enddata
\thispagestyle{empty}
\tablecomments{
  (1) Source number.
  (2) MAXI name determined from the source coordinates.
  (3)-(4) Right ascension and declination in units of degree.
  (5) 1$\sigma$ statistical error of the position in units of degrees. Note that the systematic error is not taken into consideration.
  (6)-(7), (8)-(9), and (10)-(11) Detection significance and 7-year averaged flux in units of 10$^{-12}$ erg cm$^{-2}$ s$^{-1}$ for each energy band.
   When a 1$\sigma$ error is larger than the best-fit value, the 1$\sigma$ upper limit is represented.
  The conversion factor from Crab units into erg cm$^{-2}$ s$^{-1}$ units in the 4--10 keV, 3--4 keV,
  and 10--20 keV bands is respectively $1.21\times10^{-11}$ erg cm$^{-2}$ s$^{-1}$ mCrab$^{-1}$, $3.96\times10^{-12}$ erg cm$^{-2}$ s$^{-1}$
  mCrab$^{-1}$, and $8.51\times10^{-12}$ erg cm$^{-2}$ s$^{-1}$ mCrab$^{-1}$. 
  (12) The flux in the 3--10 keV band. Note that the factor of $1.61\times10^{-11}$ erg cm$^{-2}$ s$^{-1}$ mCrab$^{-1}$ is usable for the conversion. 
  (13) Hardness ratios calculated from the fluxes in the 3--4 keV and  4--10 keV bands. Note that they are derived from the flux in units of Crab. 
  (14)--(15) The same as (13) but for the 4--10 keV and 10--20 keV bands, and the 3--10 keV and 10--20 keV bands.
  (16) Time variability index (see the text for the definition). The hypothesis that the flux does not vary can be
  ruled out for TS$_{\rm var} > 18.48$ at the 99\% confidence level.
  (17) Excess variance. 
  (This entire table is published in the machine-readable format.)
}
\end{deluxetable*}
\clearpage 
\end{longrotatetable}



\begin{turnpage}
\clearpage
\renewcommand{\arraystretch}{1.1}
\begin{deluxetable}{p{0.2cm}ccccccccccccccccccccccccc}
  \thispagestyle{empty}
  \tablefontsize{\footnotesize}
  \tabletypesize{\scriptsize}
  \tablecaption{Possible counterparts of the third MAXI/GSC sources. \label{tab:catalog2}}
  \tablehead{ \colhead{(1)} & \colhead{(2)} & \colhead{} & \colhead{(3)} &  \colhead{(4)}  & \colhead{(5)} & \colhead{(6)} & \colhead{(7)} & \colhead{(8)} & \colhead{(9)} \\ 
              \colhead{} & \multicolumn{1}{c}{MAXI}  & \colhead{} & \multicolumn{5}{c}{Counterpart} & \\ 
              \cline{2-2} \cline{4-8} \vspace{-2mm} \\ 
              \colhead{No.} & \colhead{3MAXI} & \colhead{}     & \colhead{Name} & 
              \colhead{R.A.} & \colhead{Decl.} & \colhead{$z$} & \colhead{Type}  & \colhead{Flag} & \colhead{Note}  
  }
  \startdata
  1 & J0001$-$684 & & ... & $...$ & $...$ & ... & ... & ... & ... \\
2 & J0004$-$348 & & ... & $...$ & $...$ & ... & ... & ... & ... \\
3 & J0006+727 & & ... & $...$ & $...$ & ... & ... & ... & ... \\
4 & J0009+109 & & Mrk 1501 & $2.629$ & $10.975$ & 0.0893 & Blazar & BR & ... \\
5 & J0010+323 & & RXC J0011.7+3225 & $2.935$ & $32.417$ & 0.1073 & Galaxy Cluster & MR & ... \\
6 & J0011$-$113 & & 1RXS J001124.6$-$112843 & $2.853$ & $-11.479$ & ... & Dwarf Nova & R & ... \\
7 & J0011$-$292 & & ... & $...$ & $...$ & ... & ... & ... & ... \\
8 & J0012$-$313 & & ... & $...$ & $...$ & ... & ... & ... & ... \\
9 & J0012$-$154 & & MACS J0011.7$-$1523 & $2.928$ & $-15.389$ & 0.378 & Galaxy Cluster & M & ... \\
10 & J0016$-$825 & & ... & $...$ & $...$ & ... & ... & ... & ... \\
11 & J0021+286 & & RXC J0020.6+2840 & $5.17$ & $28.675$ & 0.094 & Galaxy Cluster & MR & ... \\
12 & J0022$-$193 & & ... & $...$ & $...$ & ... & ... & ... & ... \\
13 & J0027+075 & & ... & $...$ & $...$ & ... & ... & ... & ... \\
14 & J0029+127 & & ... & $...$ & $...$ & ... & ... & ... & ... \\
15 & J0035$-$791 & & 2MASX J00341665$-$7905204 & $8.57$ & $-79.089$ & 0.074 & Sy1 & BR & ... \\
16 & J0039+231 & & ... & $...$ & $...$ & ... & ... & ... & ... \\
17 & J0041$-$093 & & ABELL 85 & $10.408$ & $-9.342$ & 0.0555 & Galaxy Cluster & BMR & ... \\
18 & J0042+412 & & SWIFT J0042.7+4111 & $10.668$ & $41.2$ & ... & confused source & B & ... \\
19 & J0048+319 & & Mrk 348 & $12.196$ & $31.957$ & 0.015 & Blazar & BX & ... \\
20 & J0048$-$732 & & ... & $...$ & $...$ & ... & ... & ... & ...
\enddata
\thispagestyle{empty}
\tablecomments{
   (1) Source number.
    (2) MAXI name.
    (3)-(7) Available information of the counterpart (name, right ascension and declination in units of degree, redshift, and source type)
    (8) Cross-matching flag: B, U, X, M, N, and R represents that the source has more than one counterparts in BAT105, 4U, XTEASMLONG, MCXC,
    XMMSL2, and 1RXS, respectively. The counterpart name listed in column (3) is quoted from the catalog expressed by the leftmost letter in this column.
    (9) Note related to the identification of the counterpart.
    (A) The auto-matching identifies a galaxy cluster, but this is likely inconsistent with the significant variability (TS$_{\rm var} = 87$).
    Then, we select a star as a plausible candidate of the counterpart.
    (B) We automatically find an obscured AGN from BAT105 as the counterpart, but the very soft property is not so common to such objects.
    Then, we find a bright object, CXO J092418.2$-$314217, in a Swift/XRT soft X-ray image. According to a study by \cite{Tom17},
    it can be classified as the LMXB or the CV, and here we representatively adopt the LMXB.
    (C) One of the brightest galaxies is selected for the most plausible counterpart.
    (This entire table is published in the machine-readable format.)
}
\end{deluxetable}
\clearpage
\end{turnpage}

\begin{figure*}
\centering
\subfigure{
\resizebox{7.3cm}{!}{\includegraphics[scale=0.9,angle=-90]{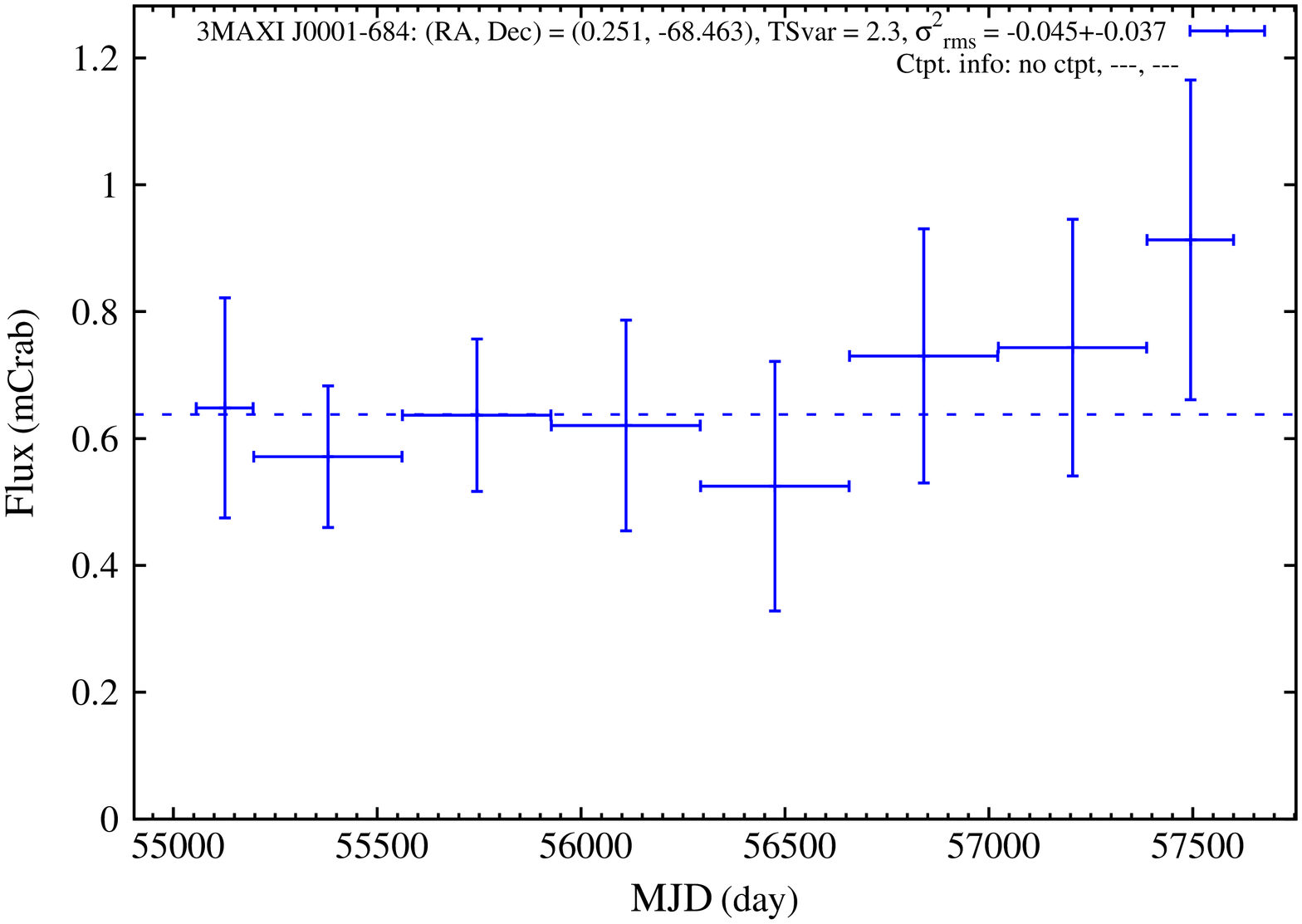}}
\resizebox{7.3cm}{!}{\includegraphics[scale=0.9,angle=-90]{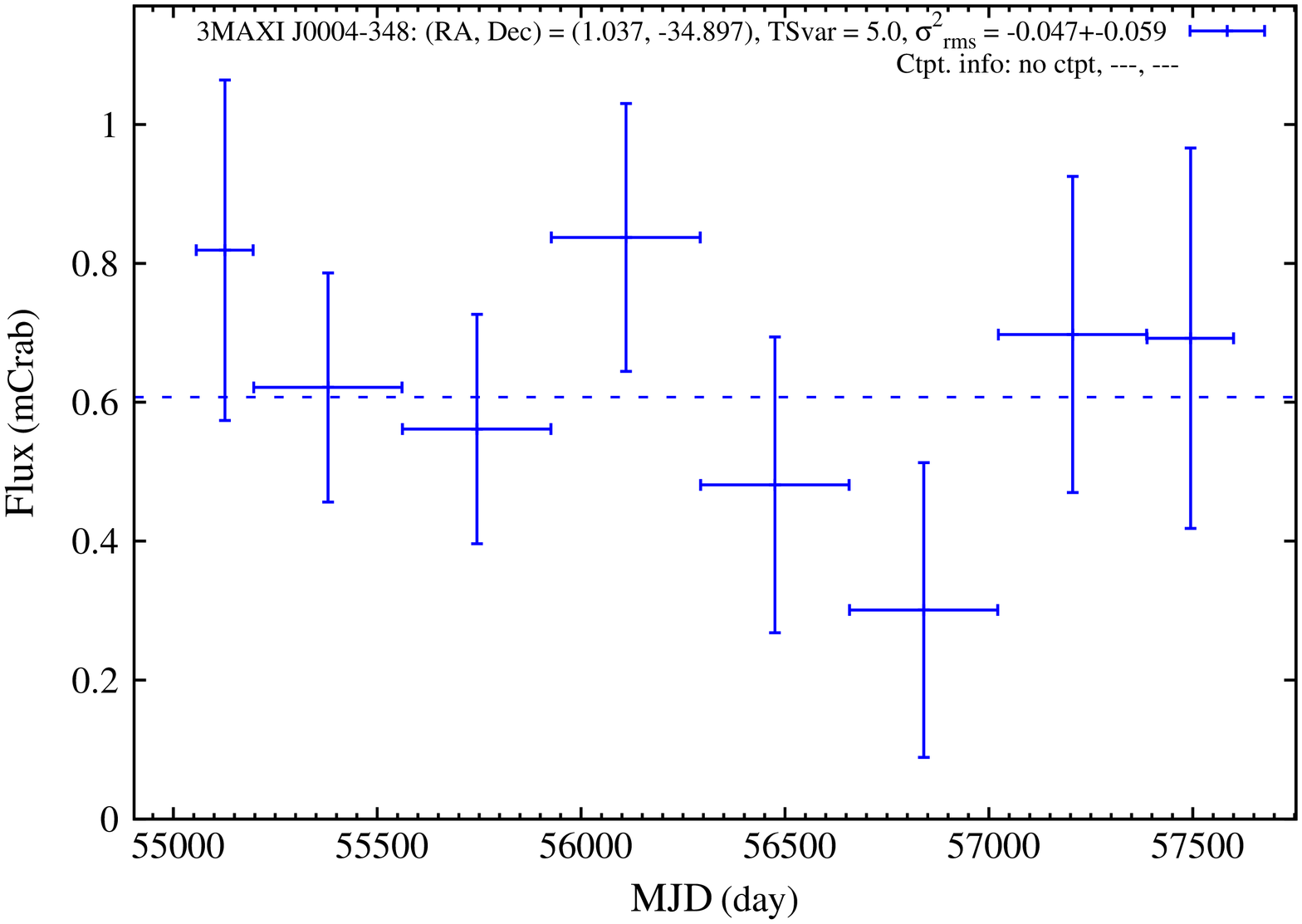}}
}
\subfigure{
\resizebox{7.3cm}{!}{\includegraphics[scale=0.9,angle=-90]{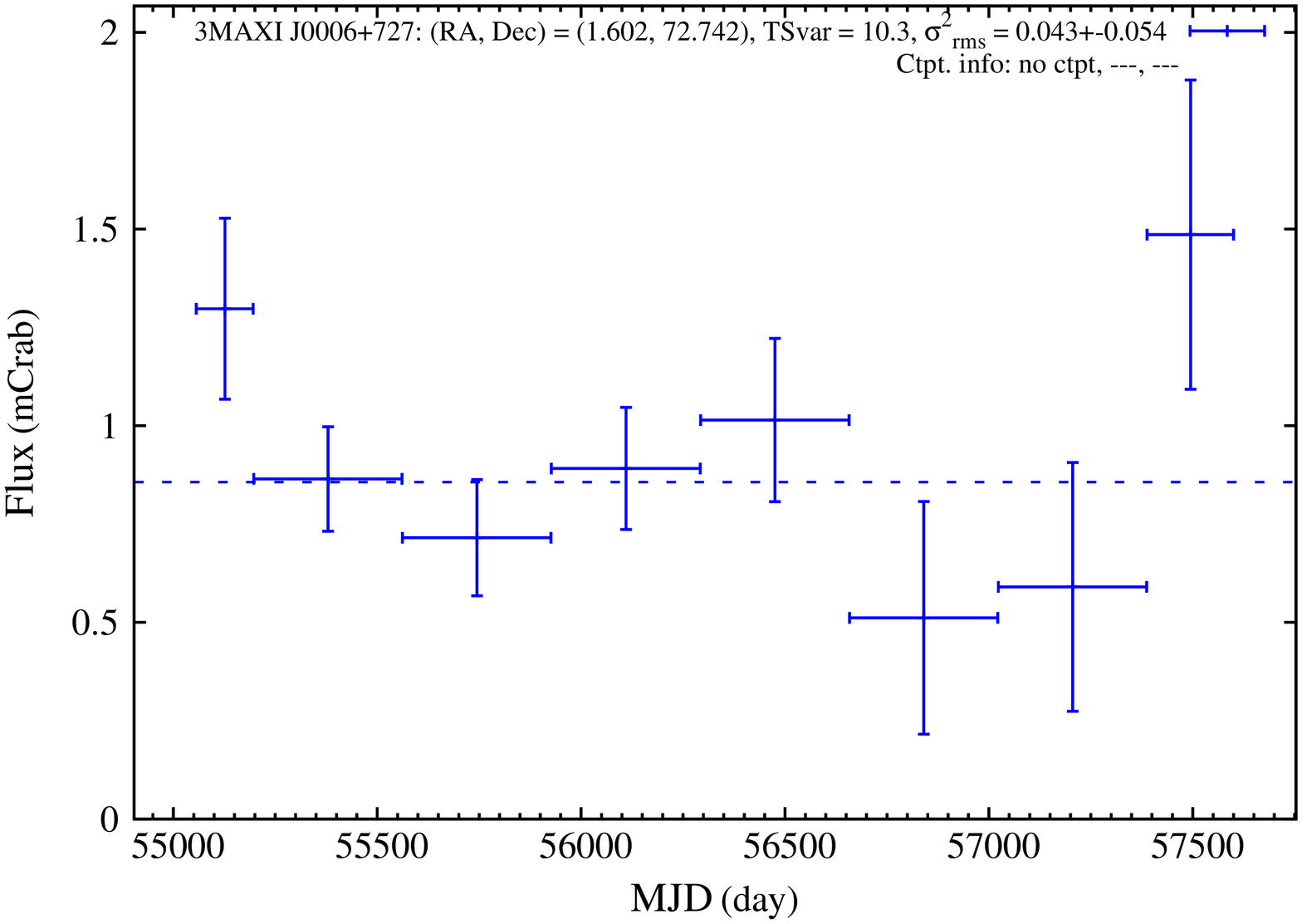}}
\resizebox{7.3cm}{!}{\includegraphics[scale=0.9,angle=-90]{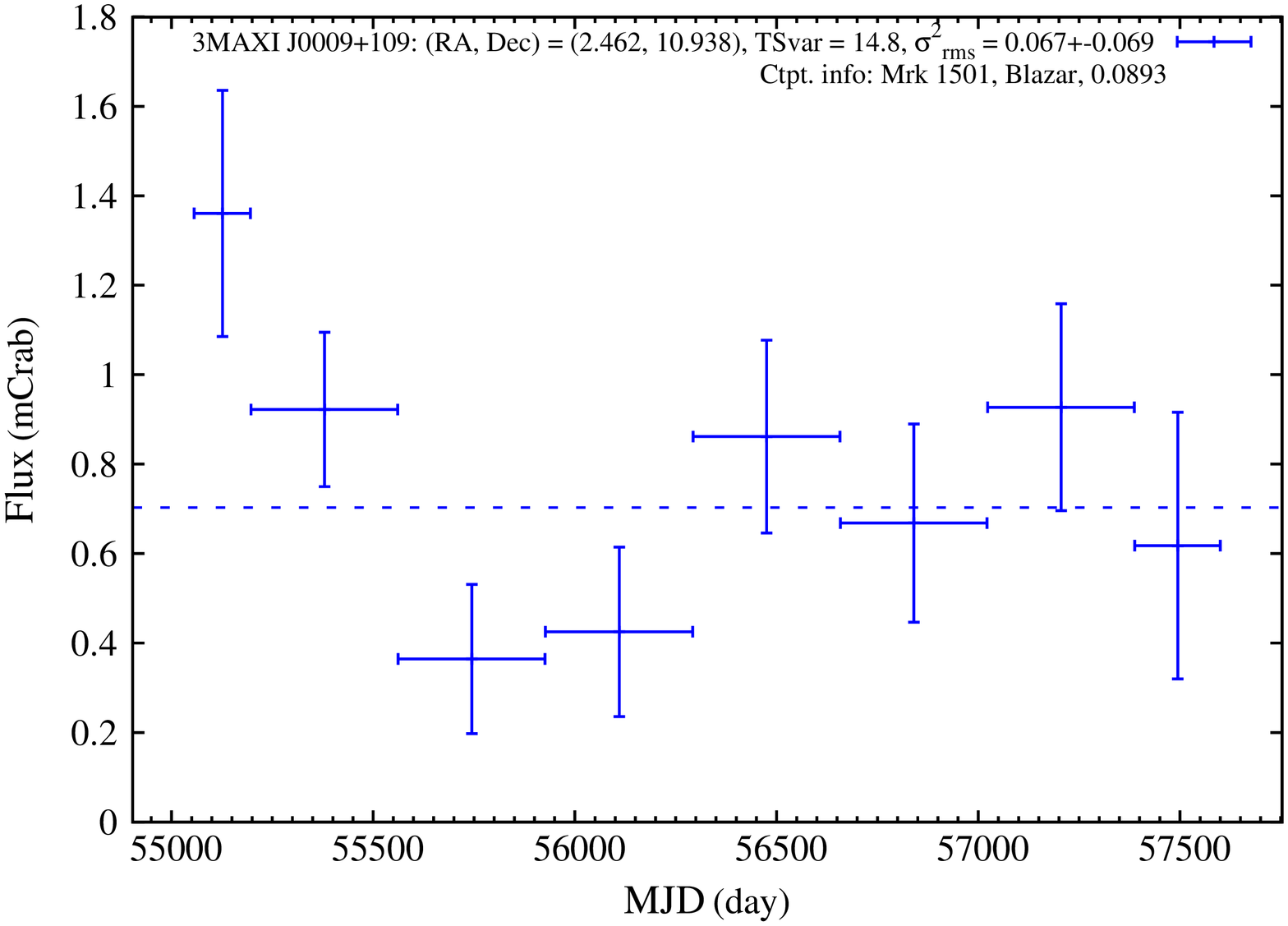}}
}
\subfigure{
\resizebox{7.3cm}{!}{\includegraphics[scale=0.9,angle=-90]{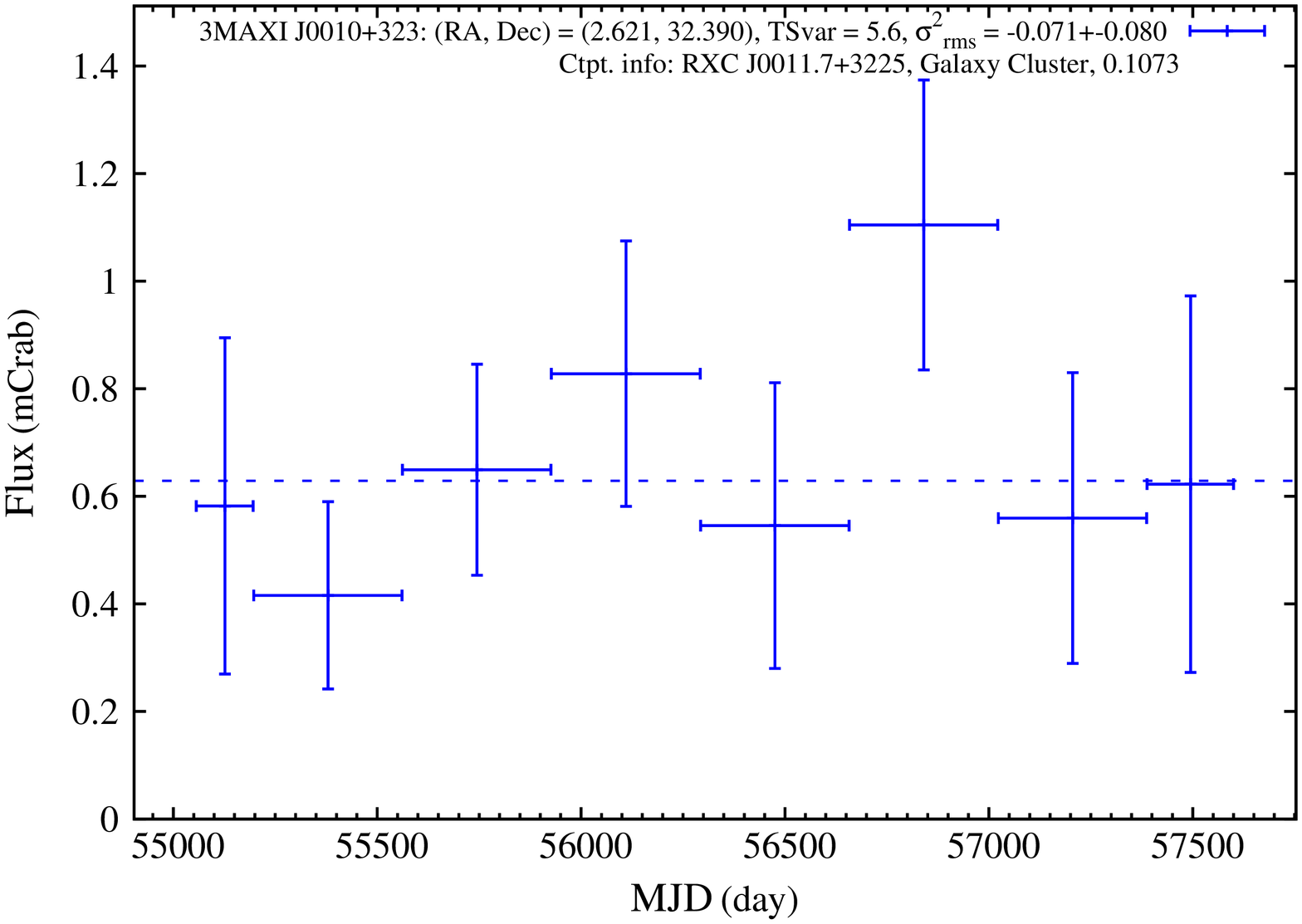}}
\resizebox{7.3cm}{!}{\includegraphics[scale=0.9,angle=-90]{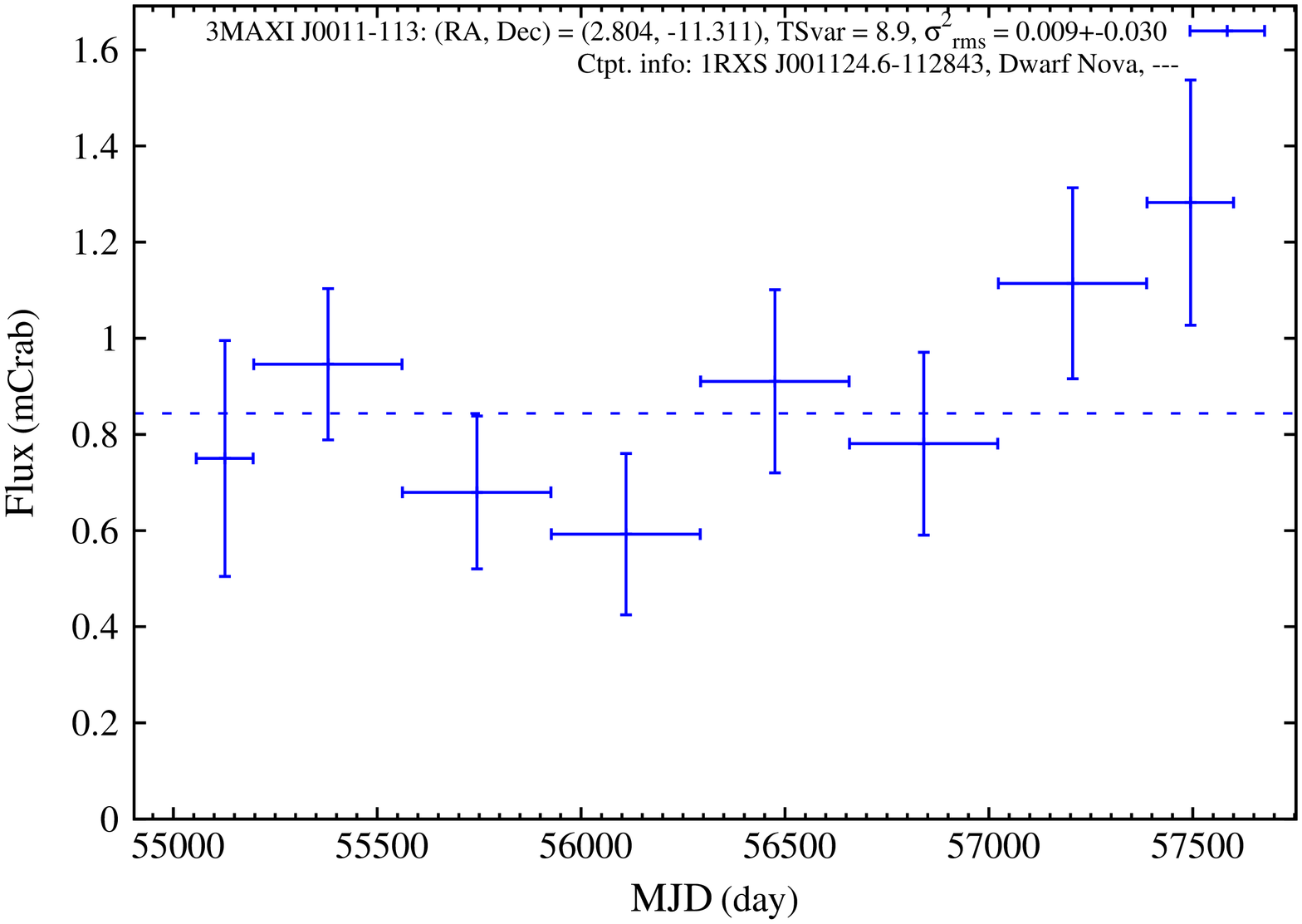}}
}
\subfigure{
\resizebox{7.3cm}{!}{\includegraphics[scale=0.9,angle=-90]{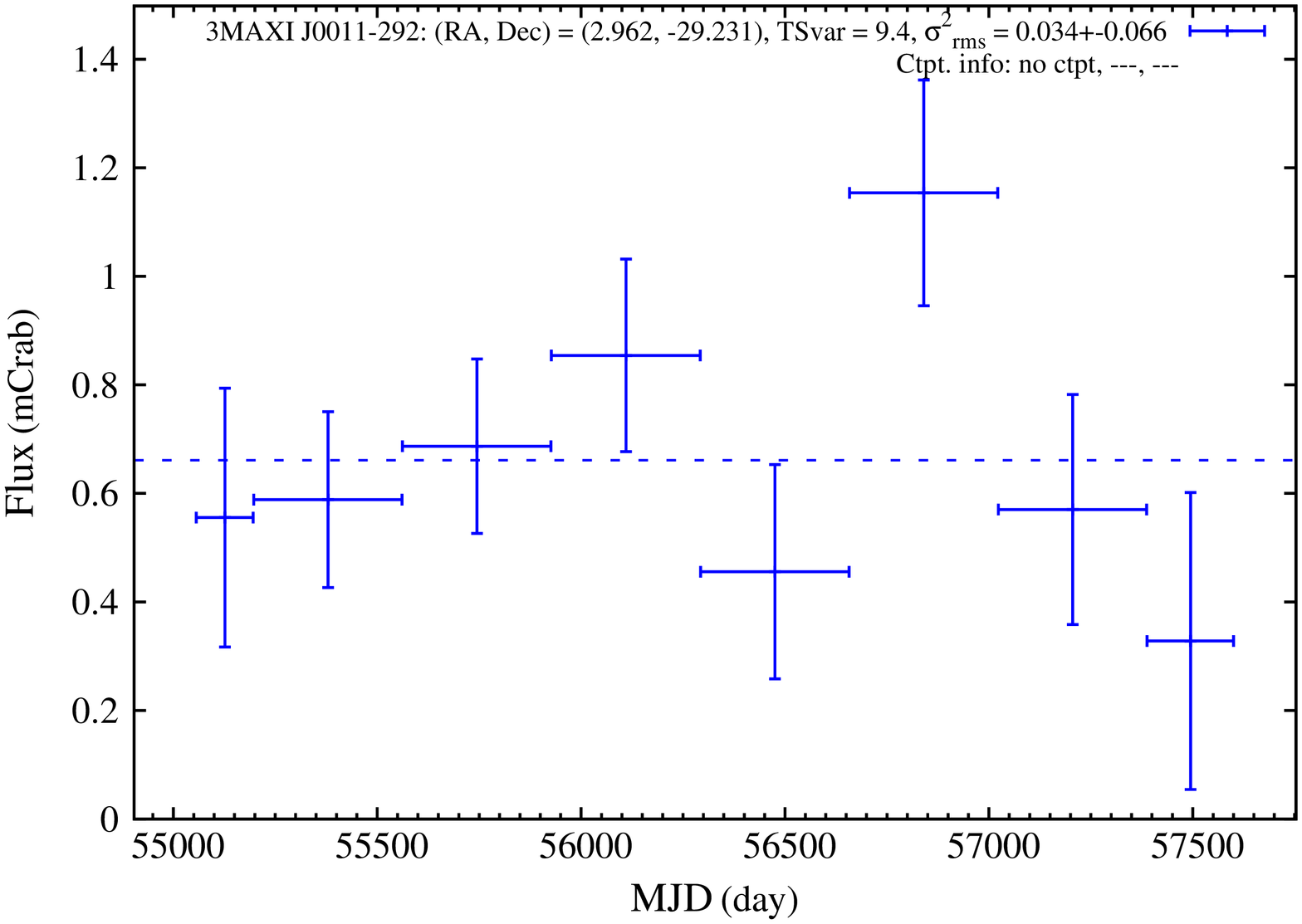}}
\resizebox{7.3cm}{!}{\includegraphics[scale=0.9,angle=-90]{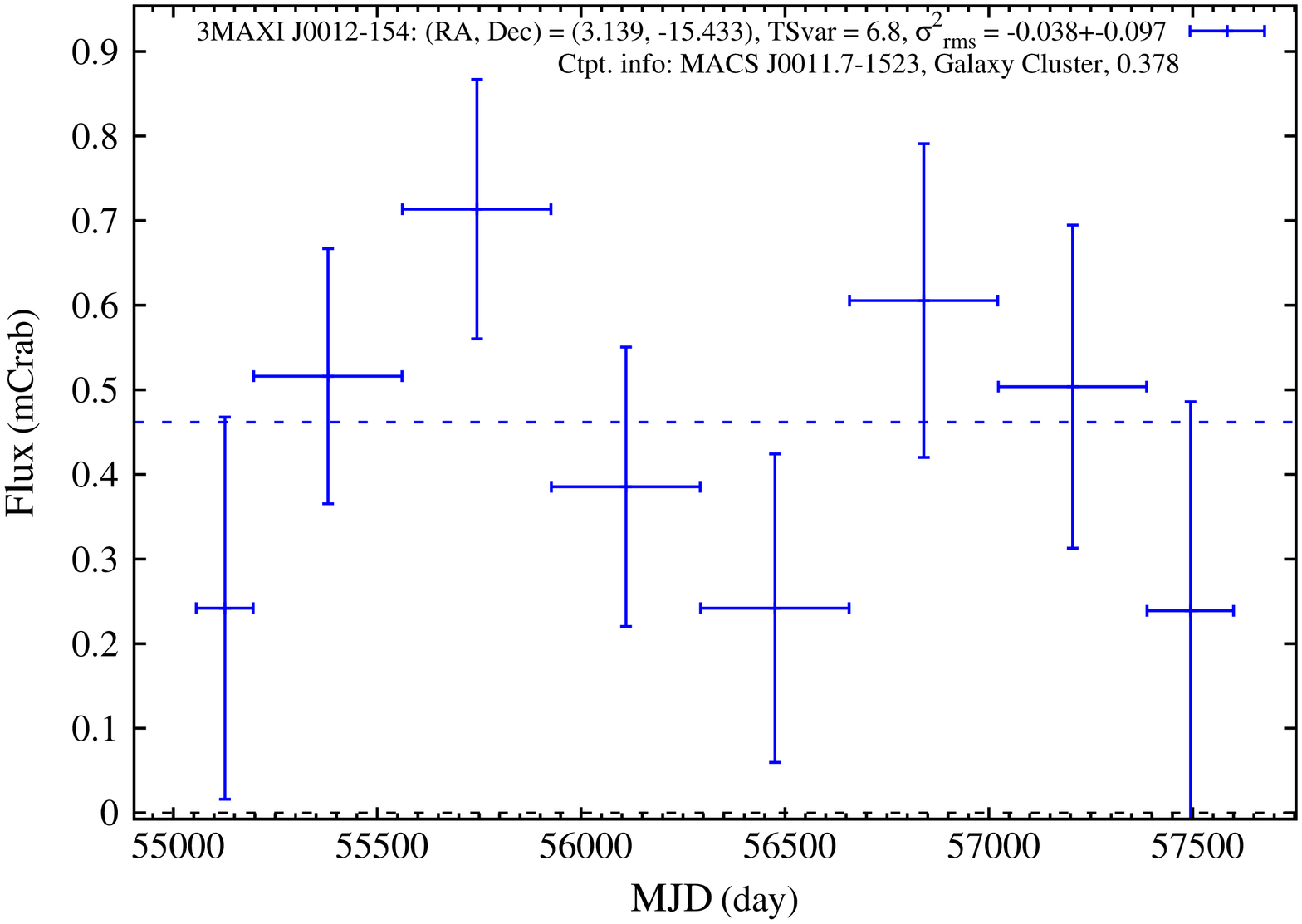}}
}
\caption{
  Lightcurves in the 4--10 keV band during the 7 years in units of mCrab. The blue dashed line denotes the 7-year averaged flux.
    The counterpart information (name, type, and redshift if available) is denoted in each figure.
  (All the figures are published in online.)
}
\label{fig:all_lcs}
\end{figure*}

 \setlength{\topmargin}{1.cm}
 \begin{longrotatetable}
  \clearpage
\renewcommand{\arraystretch}{1.1}
\begin{deluxetable*}{p{0.2cm}ccccccccccccccccccccccccc}
  \thispagestyle{empty}
  \tablefontsize{\footnotesize}
  \tabletypesize{\scriptsize}
  \tablecaption{Properties of transient events. \label{tab:trans_catalog}} 
  \tablehead{ \colhead{(1)} & \colhead{(2)}    & \colhead{(3)}  & \colhead{(4)} & \colhead{(5)} & \colhead{(6)} & \colhead{(7)} & \colhead{(8)} 
            & \colhead{(9)} &  \colhead{}  & \colhead{(10)}   & \colhead{(11)} & \colhead{(12)} & \colhead{(13)} & \colhead{(14)} \\ 
              \colhead{} & \multicolumn{8}{c}{MAXI}  & \colhead{} & \multicolumn{7}{c}{Counterpart} & \\
              \cline{2-9} \cline{11-15} \vspace{-2mm} \\
              \colhead{No.} & \colhead{3MAXIt} & \colhead{R.A.} & \colhead{Decl.} & \colhead{$\sigma_{\rm stat}$} & 
              \colhead{$s_{\rm D, 4-10 keV}$} & \colhead{$f_{\rm 4-10 keV}$} & \colhead{TS$_{\rm var}$} & \colhead{$\sigma^2_{\rm rms}$} & 
              \colhead{} &  \colhead{Name} & \colhead{R.A.}  & \colhead{Decl.} & \colhead{$z$} & \colhead{Type} 
  }
  \startdata
1 & J0051$-$735 & $12.954$ & $-73.562$ & 0.080 & 11.7 & 48.4$\pm$2.8 & 1721.87  & 5.395$\pm$0.782 & & RX J0052.1$-$7319 & $13.058$ & $-73.322$ & ... & HMXB \\
2 & J0912$-$649 & $138.047$ & $-64.911$ & 0.060 & 20.8 & 72.3$\pm$3.5 & 3486.07 & 5.652$\pm$0.329 & & MAXI J0911$-$655 & $138.010$ & $-64.868$ & ... & millisecond X-ray pulsar \\
3 & J1257+014 & $194.471$ & $1.489$ & 0.118 & 9.4 & 18.6$\pm$1.9 & 367.453 & 3.253$\pm$0.849 & & NGC 4845 & $194.505$ & $1.576$ & 0.004110 & Sy2/tidal disruption event \\
4 & J1357$-$095 & $209.257$ & $-9.509$ & 0.139 & 13.6 & 25.5$\pm$1.8 & 758.631 & 5.708$\pm$0.835 & & Swift J1357.2$-$0933 & $209.320$ & $-9.544$ & ... & LMXB
\enddata
\tablecomments{
(1) Source number. 
(2) \maxi transient source name.
(3)-(4) Right ascension and declination in units of degree.
(5) 1$\sigma$ statistical error of the position in units of degrees. Note that the systematic error is not taken into consideration. 
(6)-(7) Detection significance and flux in units of 10$^{-12}$ erg cm$^{-2}$ s$^{-1}$.
(8) Time variability index (see the text for the definition). 
(9) Excess variance.
(10)-(14) Available information of the counterpart (name, right ascension and declination in units of degree, redshift, and source type), 
which is based on the \textit{Swift}/BAT Hard X-ray Transient Monitor catalog (\texttt{https://swift.gsfc.nasa.gov/results/transients/}).
} 
\end{deluxetable*}
\clearpage
 \end{longrotatetable}


\begin{figure*}
\includegraphics[scale=0.36,angle=-90]{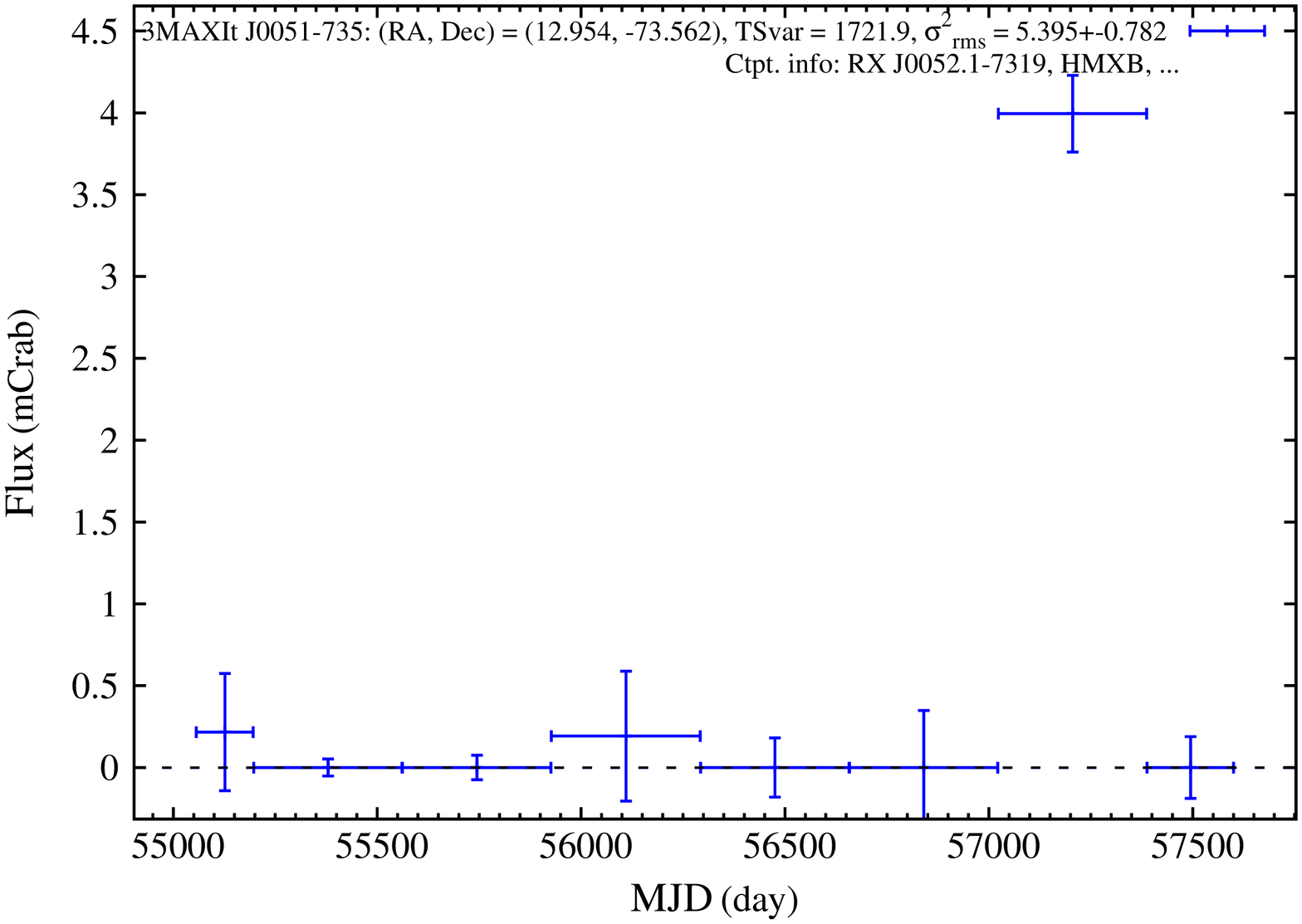} \hspace{-0.3cm}
\includegraphics[scale=0.36,angle=-90]{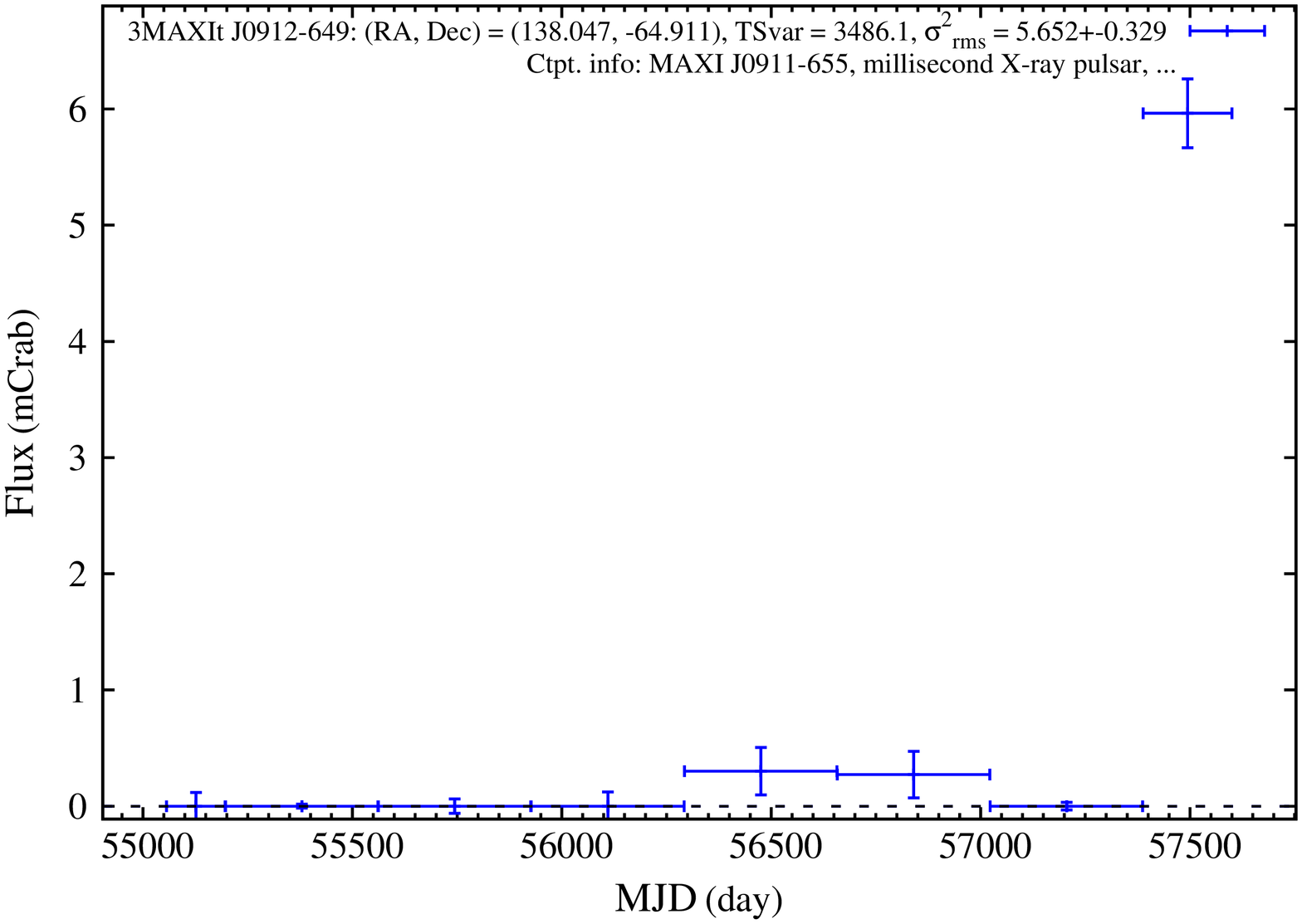} \hspace{-0.3cm}
\includegraphics[scale=0.36,angle=-90]{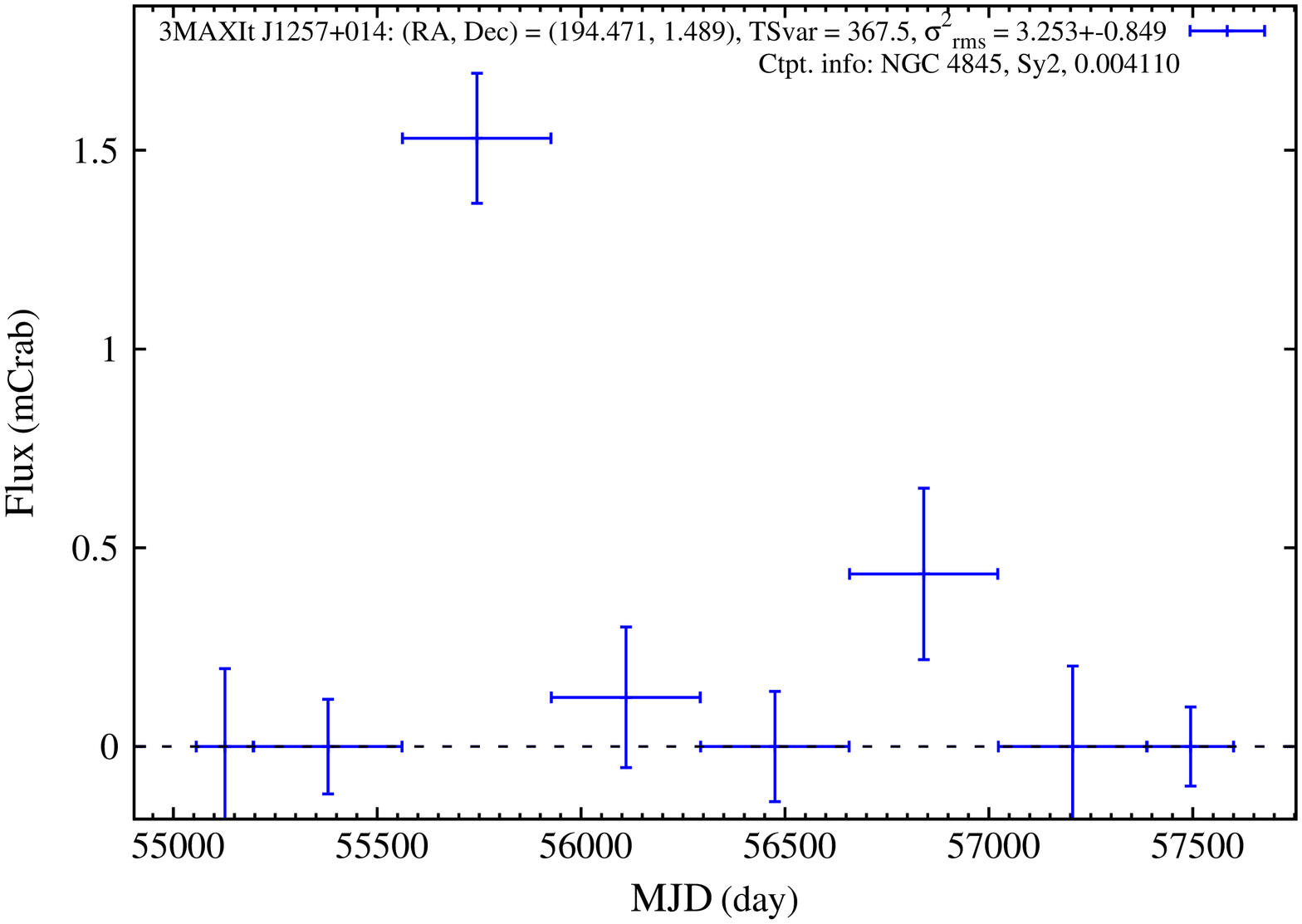} \hspace{-0.3cm}
\includegraphics[scale=0.36,angle=-90]{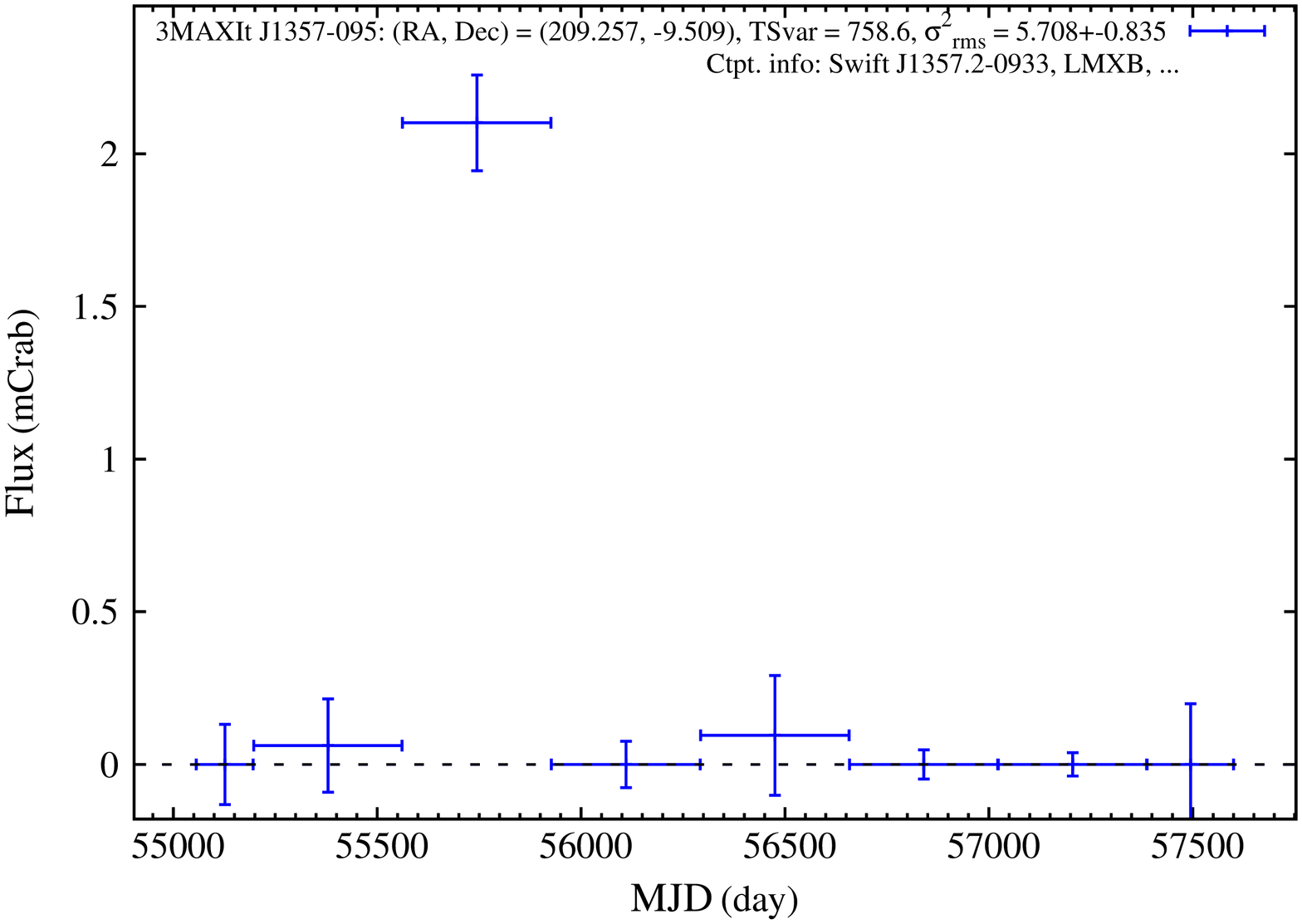}
\caption{\label{fig:trans_lc}
Lightcurves of the transient sources in the 4--10 keV band.
The counterpart information (name, type, and redshift if available) is 
denoted in each figure.
}
\end{figure*}


\end{document}